\documentclass{report}

\usepackage{microtype}
\usepackage{amsmath,amssymb,amsthm}
\usepackage{graphicx}
\usepackage{url}
\usepackage[style = ieee]{biblatex}
\usepackage{rotating}
\usepackage{ifpdf}
\ifpdf
  \usepackage{hyperref}
\fi

\usepackage[dvipsnames]{xcolor}
\usepackage{caption,subcaption}
\usepackage{algorithmic}
\usepackage[ruled,vlined,lined,commentsnumbered]{algorithm2e}
\usepackage{graphicx}
\usepackage{mathtools,bm,array}
\usepackage{enumitem}
\usepackage{multirow}

\newcommand{\here}[1]{\href{#1}{here}}

\newlength\mylen

\author{Josiah W. Smith}
\title{Dual Radar SAR Controller \\ User Guide}
\begin{document}



\maketitle

\begin{abstract}
The following is a user guide for the Dual Radar SAR Controller graphical user interface (GUI) to operate the dual radar synthetic aperture radar (SAR) scanner. 
The scanner was designed in the Spring semester of 2022 by Josiah Smith (RA), Yusef Alimam (UG), and Geetika Vedula (UG) with multiple axes of motion for the radar and target under test. 
The system is operated by a personal computer (PC) running MATLAB.
An AMC4030 motion controller is employed to control the mechanical system. 
An ESP32 microcontroller synchronizes the mechanical motion and radar frame firing to achieving precise positioning at high movement speeds; the software was designed by Josiah Smith (RA) and Benjamin Roy (UG).  
A second system is designed that employs 3-axes of motion (X-Y + rotation) for fine control over the location of the target under test. 
The entire system is capable of efficiently collecting data from colocated and non-colocated radars for multiband fusion imaging in addition to simple single radar imaging. 
\end{abstract}

\tableofcontents

\chapter{Getting Started Guide}
\label{ch:getting_started}

In this chapter, we detail the steps for getting started with the Dual Radar SAR Controller MATLAB application, referred to as the Dual Radar GUI. 
This software is capable of connection with multiple Texas Instruments (TI) millimeter-wave (mmWave) frequency-modulated continuous-wave (FMCW) single-chip radars. 
A simple introduction to FMCW radar signaling is provided in Appendix \ref{ch:fmcw_signal_model} and near-field synthetic aperture radar (SAR) imaging algorithms are briefly introduced in Appendix \ref{ch:reconstruction_algos}. 

\section*{System Requirements}
The following is a set of system requirements for the Dual Radar GUI, where essential requirements are denoted with an asterisk. 
\begin{itemize}
    \item *Windows 10/11 PC (Dual Radar GUI is not currently compatible with MacOS or Linux operating systems).
    \item *Multiple Ethernet ports (required for use with multiple radars). 
    \item High speed USB 3.X to connect with large number of devices of single USB connection. 
    \item 16 GB RAM, decent CPU, enough storage to save files from scans (on the order of 100s of MB per scan for a typical synthetic aperture dimension). 
    \item NVIDIA GPU (optional, for image reconstruction using machine learning-based algorithms). 
\end{itemize}

\section*{Hardware Set Up}
Prior to using the Dual Radar GUI application, the following hardware must be properly configured. 
The remainder of this guide assumes that the hardware has been configured and is ready. 
\begin{itemize}
    \item Scanner stand must be assembled. 
    \item Mechanical system (linear actuators, stepper motors, stepper drivers, motion controller, and ESP32 synchronizer) must be assembled and connected (See Section \ref{gs:connections} for more details on the connections). 
    \item 2 DCA1000EVM data capture cards must be configured at unique IP addresses and mounted to the mechanical system. (See Chapter \ref{ch:dca1000evm_IP_address} for details to set up each DCA1000EVM). 
    \item The 60 GHz radar (TI IWR6843ISK + MMWAVEICEBOOST) and 77 GHz radar (TI IWR1642BOOST) must be modified using the steps detailed in Chapters \ref{ch:hw_trigger_6843} and \ref{ch:hw_trigger_1642}, respectively. 
    \item 60 GHz radar (TI IWR6843ISK + MMWAVEICEBOOST) and 77 GHz radar (TI IWR1642BOOST) must be mounted to their corresponding DCA1000EVM data capture cards and connected. (See Section \ref{gs:connections} for more details on the connections). Physical alignment of the two mounted radars is important to recover images as the relative position of each radar must be taken into account. 
    \item Optionally, the base scanner must be assembled and connected. 
\end{itemize}

\section*{Software Set Up}
The following software is necessary for the set up of the hardware and operation of the Dual Radar GUI. 
\begin{itemize}
    \item \href{https://www.ti.com/tool/MMWAVE-SDK}{TI mmWave SDK 3.5.0.4}
    \item \href{https://www.ti.com/tool/UNIFLASH}{TI Uniflash}
    \item \href{https://software-dl.ti.com/ra-processors/esd/MMWAVE-STUDIO/latest/index_FDS.html}{TI mmWave Studio 2.1.1.0}
    \item \href{https://www.mathworks.com/}{MATLAB 2021b} (recommended, may be compatible with earlier/later releases)
\end{itemize}

\section{Opening the Dual Radar GUI}
After downloading the Dual Radar GUI software, the following steps are required to open the application. 
\begin{itemize}
    \item Open MATLAB and navigate to the downloaded folder containing the application file: \verb$dual_radar_gui.mlapp$. 
    \item Open the \verb$dual_radar_gui.mlapp$ file, which will open the MATLAB Application Designer in a separate window. 
    \item In the MATLAB Application Designer window, press ``Run'' at the top to run the Dual Radar GUI application. 
    \begin{itemize}
        \item If the MATLAB path is set to the proper path (the main folder, containing the application file: \verb$dual_radar_gui.mlapp$), the application will initialize without any issues. Otherwise, if the MATLAB path is incorrect, a selection window will open and the user will be asked to find the folder containing \verb$dual_radar_gui.mlapp$.
        \item If the proper version of TI mmWave Studio (2.1.1.0) is installed to the default location \verb$C:\ti\mmwave_studio_02_01_01_00\$, the application will initialize without any issues. Otherwise, if a different version of mmWave Studio is installed or it is installed at a different location, a selection window will open and the user will be asked to find the \texttt{mmWaveStudio} folder (typically at \\ \verb$...\mmwave_studio_xx_xx_xx_xx\mmWaveStudio\$). 
    \end{itemize}
    \item After initialization (the correct folders are added to the path, etc.), the GUI will open and the indicator lamps will be red, as none of the devices are connected to the application or configured. 
\end{itemize}

The following sections detail the basic operation of the Dual Radar SAR Controller GUI. 
Although the GUI is designed for dual radar operation, it can be applied for single radar use cases by ignoring either the 60 GHz or 77 GHz radar, referred to in the GUI as ``Radar 1'' and ``Radar 2,'' respectively. 
For single radar use, only one radar must be connected, configured, etc.; however, even if two radars are connected and configured, the scan can be performed using only one radar, if specified by the user. 

\section{Connections for Interface with Devices}
\label{gs:connections}
Since the Dual Radar GUI connects to many serial USB devices, identifying each serial device is crucial to proper operation of the application. 
Specifically, difference instances of TI radars or AMC4030 motion controllers will appear under an identical name at different COM ports in Device Manager and it is important to distinguish between the two radars and two AMC4030s (in the case of the base scanner). 

\subsection{Identifying the TI Radars and AMC4030 Motion Controllers}
To identify each TI radar, first open Device Manager by searching ``Device Manager'' in the start menu. 
Then, find the most convenient USB connection (usually at a USB hub) for each radar and disconnect both radars. 
Connect the first radar and two COM ports will appear labeled ``XDS110 Class Application/User UART Port'' and ``XDS110 Class Application/User Data Port.'' 
Record the COM port number corresponding to the ``XDS110 Class Application/User UART Port'' associated with the first radar for later use. 
Similarly, disconnect the first radar and connect the second radar. 
Then, Record the COM port number corresponding to the ``XDS110 Class Application/User UART Port'' associated with the second radar for later use. 

To identify the AMC4030 motion controllers, follow the same procedure, which is summarized below:
\begin{enumerate}
    \item[(1)] Disconnect both devices.
    \item[(2)] Connect one device and record its COM port inside Device Manager for later use. 
    \item[(3)] Disconnect that device, connect the next device, and repeat step (2) until all devices are recorded. 
\end{enumerate}

\section{Connecting the Radars}
\label{sec:connecting_the_radars}
After noting the COM port for each radar, return to the Dual Radar SAR Controller GUI and navigate to the ``Radar Setup'' tab. 
This section follows the numerical order of the various panels in this tab, starting with the Serial Connections panel. 
Pressing ``Connect Radar 1'' opens a window displaying a list of the connected COM ports. 
Select the ``XDS110 Class Application/User UART Port'' corresponding to the 60 GHz IWR1643ISK radar. 
Similarly, select ``Connect Radar 2'' and choose the ``XDS110 Class Application/User UART Port'' corresponding to the 77 GHz xWR1642BOOST radar. 
If the serial connection is successful, the radar connection lamps in the top right of the GUI will be green. 

\section{Configuring the DCA1000EVM Settings}
Assuming both DCA1000EVMs have been configured properly following the instructions in Chapter \ref{ch:dca1000evm_IP_address}, the IP addresses and ports for each DCA1000EVM can be entered into the corresponding fields in the GUI. 
Then, pressing the ``Prepare DCA 1'' button prepares the DCA1000EVM configuration and creates a .json file at \\ \verb$...\mmwave_studio_xx_xx_xx_xx\mmWaveStudio\PostProc\cf1.json$ with the proper configuration and checks the connection of the DCA1000EVM.
If the DCA1000EVM is connected at the specified IP addresses and ports, the message ``System is Connected'' will appear in the MATLAB terminal. 
If the message ``System is Disconnected'' appears in the MATLAB terminal, the following may be causing the issue:
\begin{itemize}
    \item The DCA1000EVM is disconnected from either the USB cable, Ethernet cable, or power cable.
    \item The DCA1000EVM is configured to a different IP address. (To change the DCA1000EVM IP address, follow the steps in Chapter \ref{ch:dca1000evm_IP_address}). 
    \item The Ethernet connection on the PC is not properly configured to the DCA1000EVM IP address. (Follow the steps in Section \ref{general:dca1000evm_pc_connection} to ensure the Ethernet port is properly configured). 
    \item \textbf{In some cases, the system appears disconnected, despite being properly configured, when multiple DCA1000EVMs are connected. In this case, disconnect the other DCA1000EVM and attempt to prepare the DCA1000EVM again. Usually, it will succeed. Then, connect the other DCA1000EVM and attempt the connection of both DCA1000EVMs again. Typically, both DCA1000EVMs will now connect properly.}
\end{itemize}

\section{Configuring the Radars}
\label{sec:configuring_the_radars}
The radar configuration panels are straightforward and follow the corresponding fields in TI mmWave Studio. 
The user can enter the desired chirp parameters for each radar and press the respective Configuration button for each radar to send the configuration to each radar. 
The Serial Number for each radar is the last four digits of the serial number, which is used to store the calibration data for each unique radar. 

After a configuration is sent to the radar, if a HW trigger capture is made, the radar must be power cycled before another configuration can be sent. 
This is a limitation of operating the radars with a HW trigger in addition to command line interface (CLI) configuration. 

Alternatively, if the Hardware Trigger checkbox is unchecked, a software (SW) trigger used, which automatically ends when the number of frames are sent (or if the capture is stopped for No of Frames = 0 for infinite transmission). 

The calibration functionality for each radar is detailed in Chapter \ref{ch:calibration}. 
Once the radar has been calibrated, its calibration data are saved for reuse in future scans. 

At this point, the radars and corresponding DCA1000EVMs have been connected and configured, and the scan can be set up and performed.

\section{Connecting and Configuring the Scanner}
First, the user must switch to the ``Scanner Setup'' tab at the top of the Dual Radar GUI. 

The user must connect the AMC4030 motion controller by pressing the ``Connect AMC4030'' button. 
A window will appear listing the connected serial devices and their COM ports. 
If using the base scanner, it is important to select the correct the proper COM port labeled ``USB-SERIAL CH340'' corresponding to the dual radar scanner.
\textbf{One way to test if the proper AMC4030 is connected is to press the ``Home'' button in the ``Single Movement'' panel on the bottom right of this tab. If the vertical dual radar scanner moves to the home position, then the correct scanner is connected. If the base scanner moves, then you have selected the incorrect COM port. You can disconnect the AMC4030 and select the proper COM port.}

The ESP32 must be connected using the ``Connect ESP32'' button. 
The user should select the COM port corresponding to the ESP32. 

The AMC4030 must be configured with the appropriate parameters for the given linear actuators and settings of each stepper driver, as detailed in Sections \ref{subsec:linear_actuators} and \ref{subsec:stepper_drivers}, respectively. 
After entering the proper settings, the user can configure the AMC4030 by pressing the ``Configure AMC4030'' button, which send the configuration to the AMC4030.

\subsection{Move to Initial Position of Scan}
For many cases, the home position is not the ideal starting position for a given scan. 
Thus, after the AMC4030 is configured, it is recommended to move the platform to a different starting position using the ``Single Movment'' functionality in this tab.
The user can enter a step size (both positive or negative) to move the platform and press the ``Send'' button to send the command which moves the platform. 
The ``Home'' button sends the platform to the home position, effectively resetting the position. 
It is recommended to reset the platform the home position every time the system is turned on. 
After the scanner is moved to the initial/starting position for the scan, the configuration can continue. 

\section{Configuring the SAR Scan Parameters}
Under the ``Configure Scan'' panel, the user can enter the parameters of the scanner and desired SAR aperture. 
The parameters are detailed in Table \ref{tab:scan_params}. 
After the parameters are entered, the user can configure the scan, which updates the scan dimensions and estimated scan time fields, by pressing the ``Configure Scan'' button. \\

\begin{minipage}{\textwidth}
Table \ref{tab:scan_params}: Parameters of SAR scan. 
\centering
\begin{tabular}{c|c}
    Parameter & Meaning \\
     \hline
    X-Max Size (mm) & Maximum dimension of linear actuator along the $x$-direction in mm \\
    Y-Max Size (mm) & Maximum dimension of linear actuator along the $y$-direction in mm \\
    File Name & Name of the scan\footnote{The scan is saved to a folder with name corresponding to the date of the scan, so names can be reused. The GUI will check if the scan name already exists before performing the scan.} \\
    X-Step Size (mm) & Step size between each sample along the $x$-direction in mm \\
    Y-Step Size (mm) & Step size between each sample along the $y$-direction in mm \\
    Periodicity & Approximate periodicity along the $x$-direction \\
    X-Offset (mm) & Offset from the starting location to start collecting samples in mm\footnote{Allows the effect of up-ramp of the acceleration to be diminished.} \\
    $\Delta$X (mm) & Distance between the radars along the $x$-direction in mm \\
    Num X-Steps & Number of samples along the $x$-direction \\
    Num Y-Steps & Number of samples along the $y$-direction \\ 
    Radar 1 Checkbox & To use radar 1 for scan \\
    Radar 2 Checkbox & To use radar 2 for scan
\end{tabular}
\label{tab:scan_params}
\end{minipage} \\

Additionally, the ``Scan Notes'' text area is provided for the user to record notes about the specified scan. 
We have found this helpful for describing the target scene, scanning parameters, etc. for later use. 
The scan notes are recorded into a text file and saved with the scan data for reference. 

\section{Starting the Scan}
After the scan is configured and all devices are connected and configured, all the indicator lamps should be green. 
In this case, the scan is ready to commence and the user can start the command by pressing the ``Start Scan'' button. 

During the scan, the platform will perform a raster back-and-forth scanning motion during which the radars will be triggered at the specified locations to create a uniform grid synthetic aperture elements from which to reconstruct a high-resolution image. 
The DCA1000EVM must be reset for each horizontal scan, which is currently the largest inefficiency in the scanning process. 
For each horizontal motion, HW triggers are sent by the ESP32 to each radar. 
After each radar receives a HW pulse, it sends a frame (typically of 4 chirps). 
The beat signals are streamed to the DCA1000EVM over LVDS and then to the PC over UDP via Ethernet. 
The reason we need to reset for each horizontal scan is currently unknown, but after the HW triggers stop at the end of the horizontal motion, either the radar or DCA1000EVM stops. 
The radar may stop streaming and need to be reset or the DCA1000EVM may reset since the data stops streaming. 
We have attempted different streaming modes for the DCA1000EVM that are supposed to change the stopping criterion of the DCA1000EVMs, but we have not seen any difference and this is likely a bug or due to operating the hardware in an unsupported manner. 
Additional system details can be found in \cite{yanik2020development}. 

The following are potential known issues that may occur during the scanning process (typically when multiple DCA1000EVMs are uses simultaneously):
\begin{itemize}
    \item Too many packets are received to a DCA1000EVM: this is a common issue with no known cause or solution. If too many packets are received by the DCA1000EVM, we typically truncate the received data and assume that the extra data can be ignored. This does not seem to degrade image quality substantially. 
    \item Not enough packets are received to a DCA1000EVM: this occurs rarely and also does not have a known cause or solution. If too few packets are received, we terminate the scan and reset the process. Typically, the scan can be attempted again with success after one or two attempts. 
\end{itemize}

\section{Loading the SAR Data}
After the scan completes successfully, the user can press the ``Load Data'' button to load the data and store it to a convenient format for later processing. 
Data are saved to \verb$data\<date>\<fileName>\$ with MATLAB files to load the data and reconstruct the images using the radar imaging toolbox, as detailed in Chapter \ref{ch:image_reconstruction}. 
\chapter{Reconstructing Images from Dual Radar SAR Scans}
\label{ch:image_reconstruction}

Using the dual radar GUI, scans can be performed using 1 or 2 radars as specified in the previous chapter.
In the following sections, we detail how to reconstruct images from each of the radars or using multiband signal fusion algorithms. 
The following steps assume that the MATLAB path remains located in the main \texttt{dual-radar-gui} folder regardless of which file is opened. 

\section{Reconstruct Image from Radar 1 (60 GHz)}
\label{sec:image_radar1}
To reconstruct the image using radar 1, open the file:

\verb$data\<date>\<fileName>\<fileName>_radar1.m$. 

The script first attempts to add the proper image reconstruction toolbox and then proceeds to load the data. 
When calling \texttt{DualRadarLoadAll()}, the second argument indicates the method for loading the radar data.
When the value is $1.5$, the algorithm interpolates the data to a uniform grid with $\lambda/4$ spacing for 77 GHz along the vertical direction. 
For a value of $1$, the algorithm leaves the spacing with the original 60 GHz $\lambda/4$ vertical spacing. 

The remaining sections of the script are self-explanatory to set the imaging parameters (volume of interest), number of voxels, display parameters, etc. 

\section{Reconstruct Image from Radar 2 (77 GHz)}
\label{sec:image_radar2}
To reconstruct the image using radar 2, open the file:

\verb$data\<date>\<fileName>\<fileName>_radar2.m$. 

The script first attempts to add the proper image reconstruction toolbox and then proceeds to load the data. 
When calling \texttt{DualRadarLoadAll()}, the second argument indicates the method for loading the radar data.
For a value of $2$, the algorithm leaves the spacing with the original 77 GHz $\lambda/4$ vertical spacing. 

The remaining sections of the script are self-explanatory to set the imaging parameters (volume of interest), number of voxels, display parameters, etc. 

\section{Reconstruct Image with Both Radars using Matrix Fourier Transform (MFT)}
\label{sec:image_mft}
To reconstruct the image with both radars using the matrix Fourier transform (MFT) method detailed in \cite{li2008mft} and discussed in the matrix pencil algorithm paper \cite{wang2018wavenumber}, open the file:

\verb$data\<date>\<fileName>\<fileName>_dual_radar.m$. 

The script first attempts to add the proper image reconstruction toolbox and then proceeds to load the data. 
When calling \texttt{DualRadarLoadAll()}, the second argument indicates the method for loading the radar data.
For a value of $3$, the algorithm loads in both the 60 and 77 GHz radar data, applies multistatic-to-monostatic conversion \cite{yanik2019sparse,yanik2019cascaded} to both sets of data, interpolates the 60 GHz data to an identical sampling grid as the 77 GHz radar, and zero-pads between the samples. 

The remaining sections of the script are self-explanatory to set the imaging parameters (volume of interest), number of voxels, display parameters, etc. 

\section{Reconstruct Image using Super-Resolution Algorithm}
\label{sec:image_sr}

Software requirements for Josiah Smith super-resolution algorithms:
\begin{enumerate}
    \item[1)] Cuda 11.6
    \item[2)] PyTorch 1.11
    \item[3)] Install requirements.txt from the \texttt{dual-radar-fusion-SR} repository
\end{enumerate}

After loading in the dual radar data as for the MFT method, a super-resolution algorithm can be applied. 
A typical call to the super-resolution algorithm with recommended settings is included below:

\verb$DualRadarSR(target,"./saved/fftnet1055.tar",4096,"range","min-max","norm3",1.2);$
\chapter{Calibrating a mmWave Radar}
\label{ch:calibration}
In this chapter, the calibration process using the dual radar GUI is discussed. 
Mathematical details of the calibration procedure can be found in \cite{yanik2020development,yanik2019sparse,yanik2019cascaded}. 

After the radar and DCA1000EVM are connected and configured, following Sections \ref{sec:connecting_the_radars} through \ref{sec:configuring_the_radars}, each radar can be calibrated using the dual radar GUI. 
Each radar must be calibrated before it can be used for scanning to compensate for constant phase errors and range bias. 
The serial number of the radar is used in the GUI to store the calibration data associated with that radar. 
It is recommended to calibrate the radar every time the system is power cycled; however, this may be unnecessary. 

To calibrate the radar, a corner reflector is required. 

Calibration steps:
\begin{enumerate}
    \item[1)] Connect and configure radar.
    \item[2)] Move radar to the correct position ($x$ and $y$) using the Single Command functionality of the scanner.
    \item[3)] Place corner reflector in front of radar (recommended distance of 300 mm from the radar boresight) such that the center of the corner reflector is aligned vertically with the lowest RX element. 
    \item[4)] Follow calibration steps in GUI (calibrate empty scene (optional): moves radar to different position to capture samples of an empty scene / background clutter of surrounding area). 
\end{enumerate}
\chapter{Mechanical System}
\label{ch:mech_system}
In this chapter, we detail the mechanical system on which the radars are mounted. 
The system scans both radars in a raster pattern and synchronizes the triggers such that a uniform grid is achieved by both radars. 

\section{Physical Scanner Stand}
\label{sec:physical_scanner_stand}

\section{Scanner Motion}
\label{sec:scanner_motion}
The mechanical scanner moves the radars along a 2-D $x$-$y$ plane using linear actuators and stepper motors, which are driven by stepper drivers. 

\subsection{MJUnit Linear Actuators}
\label{subsec:linear_actuators}
\begin{figure}[ht]
    \centering
    \includegraphics[width=0.4\textwidth]{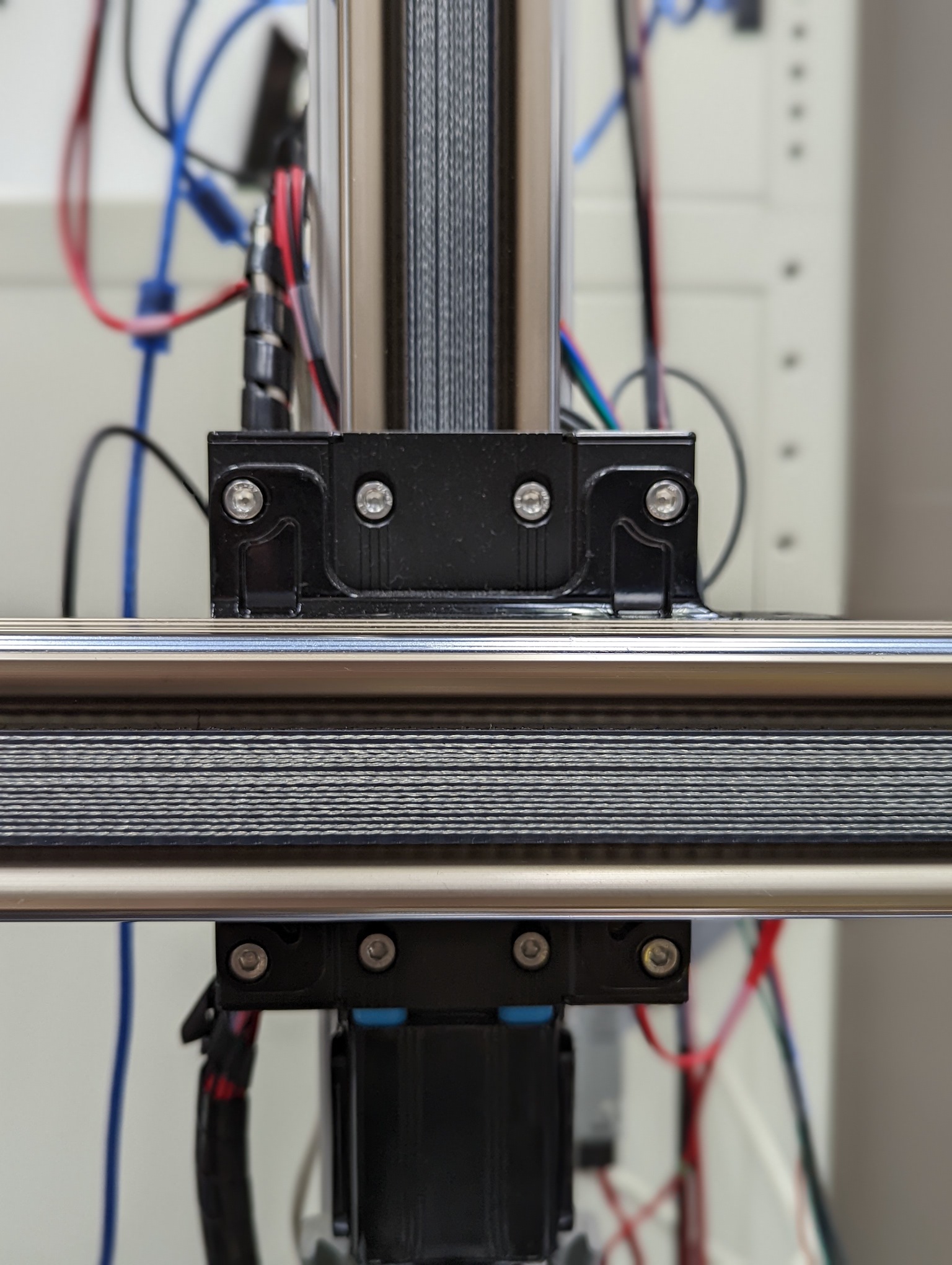}
    \caption{MJUnit MJ50 $x$-$y$ platform mounted to scanner stand.}
    \label{fig:mjunit}
\end{figure}

The 8020 stand discussed in Section \ref{sec:physical_scanner_stand} holds the linear actuators. 
The linear actuators used are MJUnit MJ45 or MJ50 rails, which are belt driven and are capable of speeds upward of 200 mm/s. 
The large dual-radar scanner uses MJ50 rails, which are shown in Fig. \ref{fig:mjunit}. 

\subsubsection{Linear Actuator AMC4030 Settings}
\label{subsubsec:amc4030_actuator_DIST}
Each linear actuator type has a different gear ratio that varies the ratio of millimeters per revolution. 
This ratio is important for the AMC4030 motion controller for converting the rotational motion accurately to linear (millimeters). 
Table \ref{tab:amc4030_actuator_DIST} details the required settings that must be applied for each linear actuator type.
See \cite{yanik2020development} for more details and comparison of various actuator types. 

\begin{table}[ht]
    \centering
    \caption{AMC4030 Software Settings for Each Linear Actuator Type}
    \begin{tabular}{c|c}
        \hline
        Actuator Type & DIST Parameter \\
        \hline\hline
        MJ50 & 110 mm/rev \\
        MJ45 & 110 mm/rev \\
        Rotator & 36 mm/rev (converts mm to deg) \\
        Old Screw Rail (purchased by M. E. Yanik) & 5 mm/rev \\
        \hline
    \end{tabular}
    \label{tab:amc4030_actuator_DIST}
\end{table}

\subsection{Stepper Drivers}
\label{subsec:stepper_drivers}
Two varieties of stepper drivers are used for the dual radar scanner. 
A high-power stepper driver is employed for the larger stepper motor detailed in the following section and a smaller stepper driver is used with the smaller stepper motor.

\begin{figure}[ht]
\centering
    \begin{subfigure}[b]{0.4\textwidth}
         \centering
         \includegraphics[width=\textwidth]{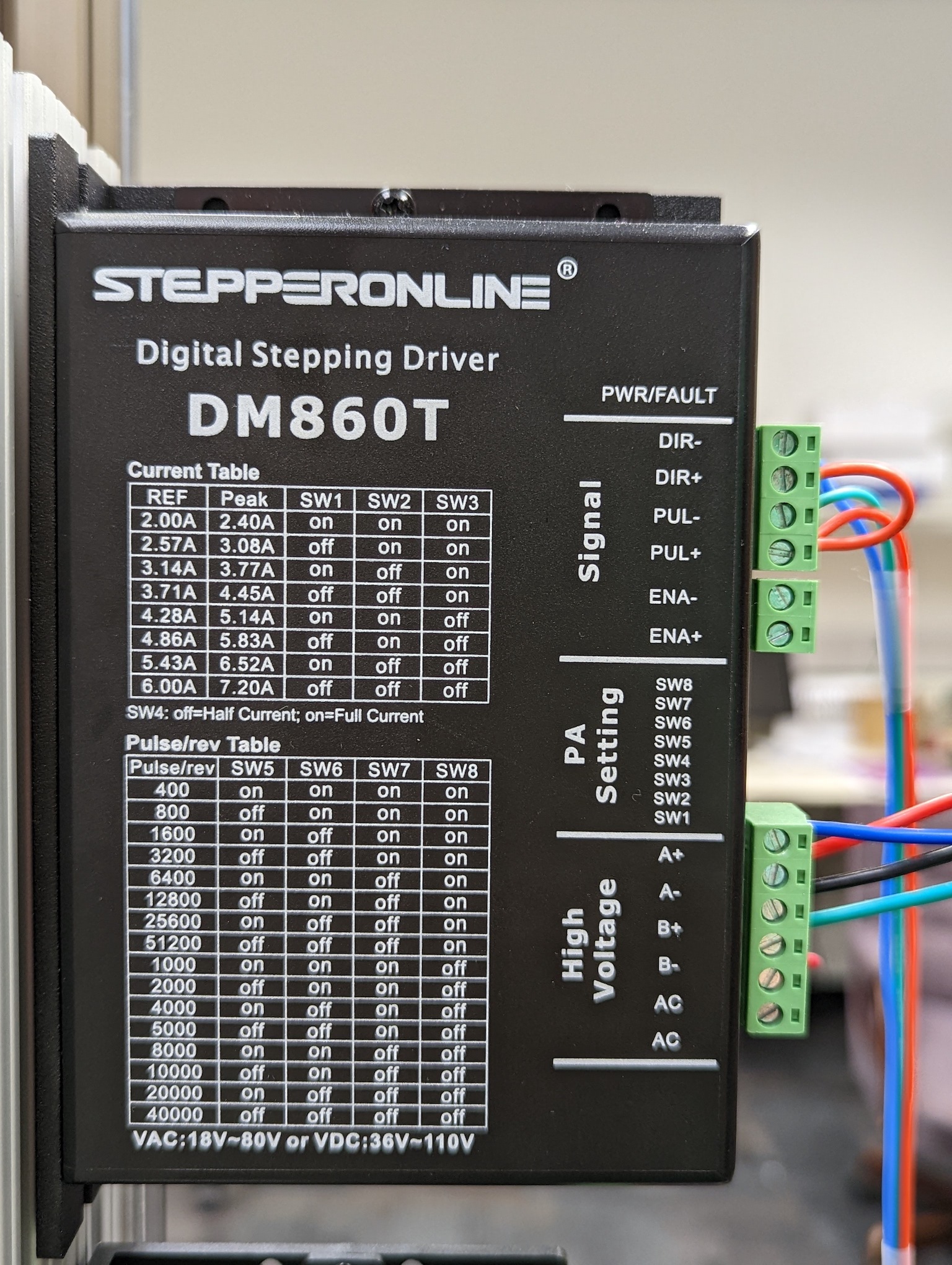}
         \caption{}
         \label{fig:driver_large}
    \end{subfigure}
    \begin{subfigure}[b]{0.4\textwidth}
         \centering
         \includegraphics[width=\textwidth]{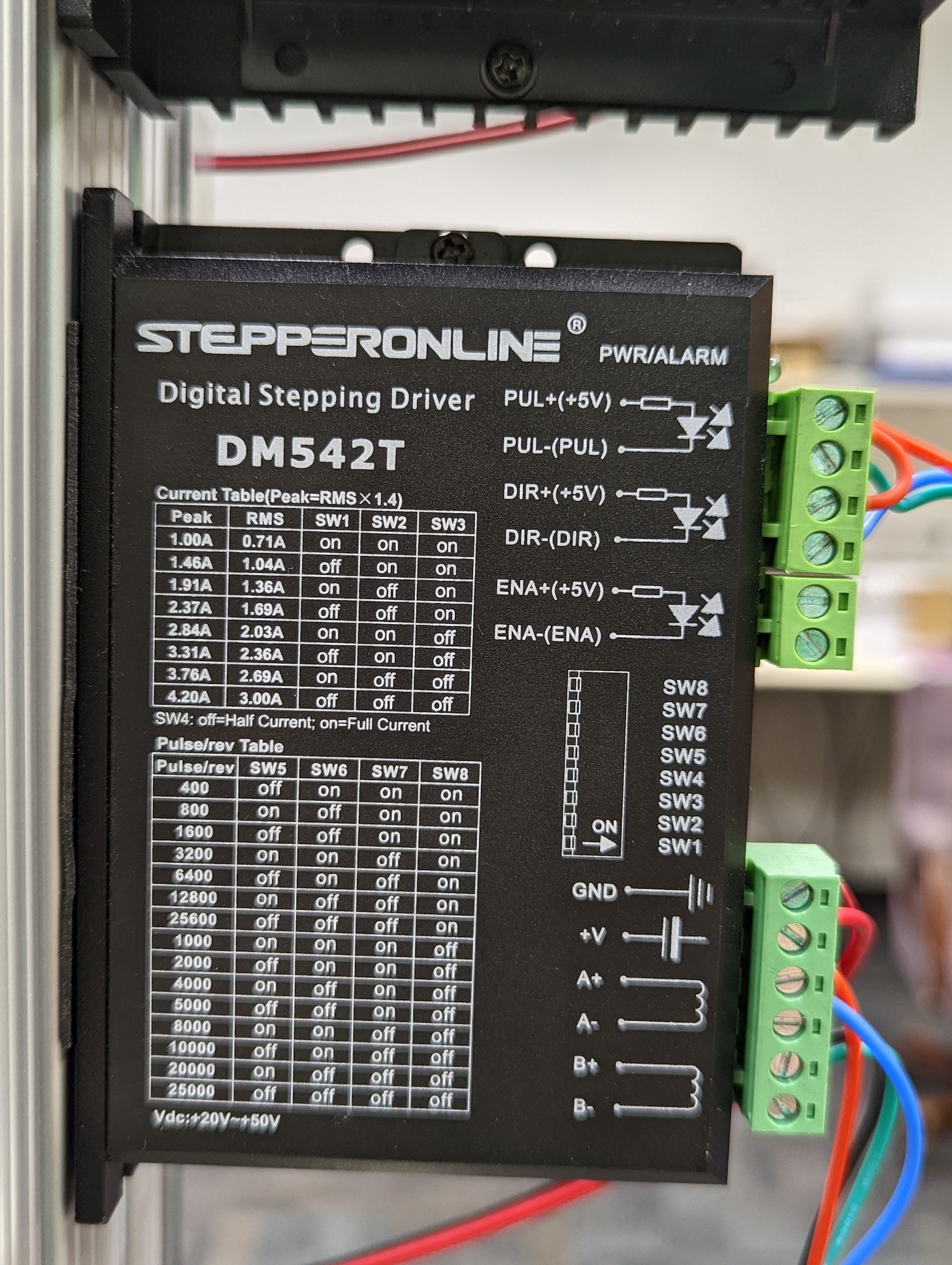}
         \caption{}
         \label{fig:driver_small}
    \end{subfigure}
\caption{Stepper drivers: (a) DM860T high-power stepper driver for driving the NEMA34 motor at the vertical $y$-axis. (b) DM542T stepper driver for driving the NEMA23 motor at the horizontal $x$-axis.}
\label{fig:drivers}
\end{figure}

The larger stepper driver used to drive the larger NEMA34 motor operating the vertical $y$-axis is the DM860T, as shown in Fig. \ref{fig:driver_large}. 
A 50 V power supply is employed to power the stepper driver and stepper motor. 
The larger motor must be used for the vertical axis due to the load of the horizontal linear actuator. 
The weight of the radars is small compared to the heavy horizontal linear rail. 

The NEMA23 motor operating the horizontal $x$-axis is driven by a DM542T stepper driver, as shown in Fig. \ref{fig:driver_small}. 
A 24 V power supply is employed to power this stepper driver and stepper motor. 
The smaller motor can be used for the horizontal axis because the required torque is lower for the horizontal motion.

\subsubsection{Stepper Driver Settings}
\label{subsubsec:stepper_driver_settings}
The stepper driver settings for each driver are listed on the back of the stepper drivers and are pictured in Fig. \ref{fig:drivers}. 
There are two primary settings for the stepper drivers: 1) available current and 2) pulses per revolution. 
Both settings are controlled by a set of DIP switches on the side of the stepper driver between the wiring connections (to the right in the images in Fig. \ref{fig:drivers}) and are detailed as follows:
\begin{itemize}
    \item[1)] \textbf{Current:} We recommend setting the stepper driver such that the maximum available current is provided to the stepper motor. This is not thoroughly tested; however, we find that, particularly for the vertical $y$-axis, low current can cause incorrect positioning due to the heavy load. Typically, this setting requires setting switches SW1, SW2, and SW3 to the off position. \textit{NOTE: Please check with the stepper motor documentation to ensure that the current setting is compatible and will not damage the stepper motor.}
    \item[2)] \textbf{Pulses Per Revolution:} the pulses per revolution (pulse/rev) setting controls the precision of the motion. If the stepper driver is set to $N$ pulses/rev, the stepper motor will complete exactly 1 rotation when the AMC4030 sends $N$ pulses. Using the analysis from Section \ref{subsubsec:amc4030_actuator_DIST}, the conversion from pulses to linear motion (millimeters) can be computed as follows. Let $M$ be the mm/rev setting corresponding to the linear actuator being used, as shown in Table \ref{tab:amc4030_actuator_DIST}. The linear distance per pulse can be computed as $M/N$. For example, if an MJ50 linear actuator is used, such as for the large dual radar scanner, $M = 110$ mm/rev and suppose $N = 20000$ pulse/rev. Hence, the conversion is $M/N = 0.0055$ mm/pulse. This computation is necessary for synchronizing the motion using the synchronizer software discussed in Section \ref{sec:synchronizer} \cite{yanik2020development}. We recommend using a high value of number of pulses to achieve highly precise positioning since the wavelength is on the order of millimeters and typical spacing is $\lambda/4$, where $\lambda$ is the wavelength of the center frequency (62 GHz or 79 GHz). 
\end{itemize}

\subsection{Stepper Motors}
\label{subsec:stepper_motors}

\begin{figure}[ht]
\centering
    \begin{subfigure}[b]{0.4\textwidth}
         \centering
         \includegraphics[width=\textwidth]{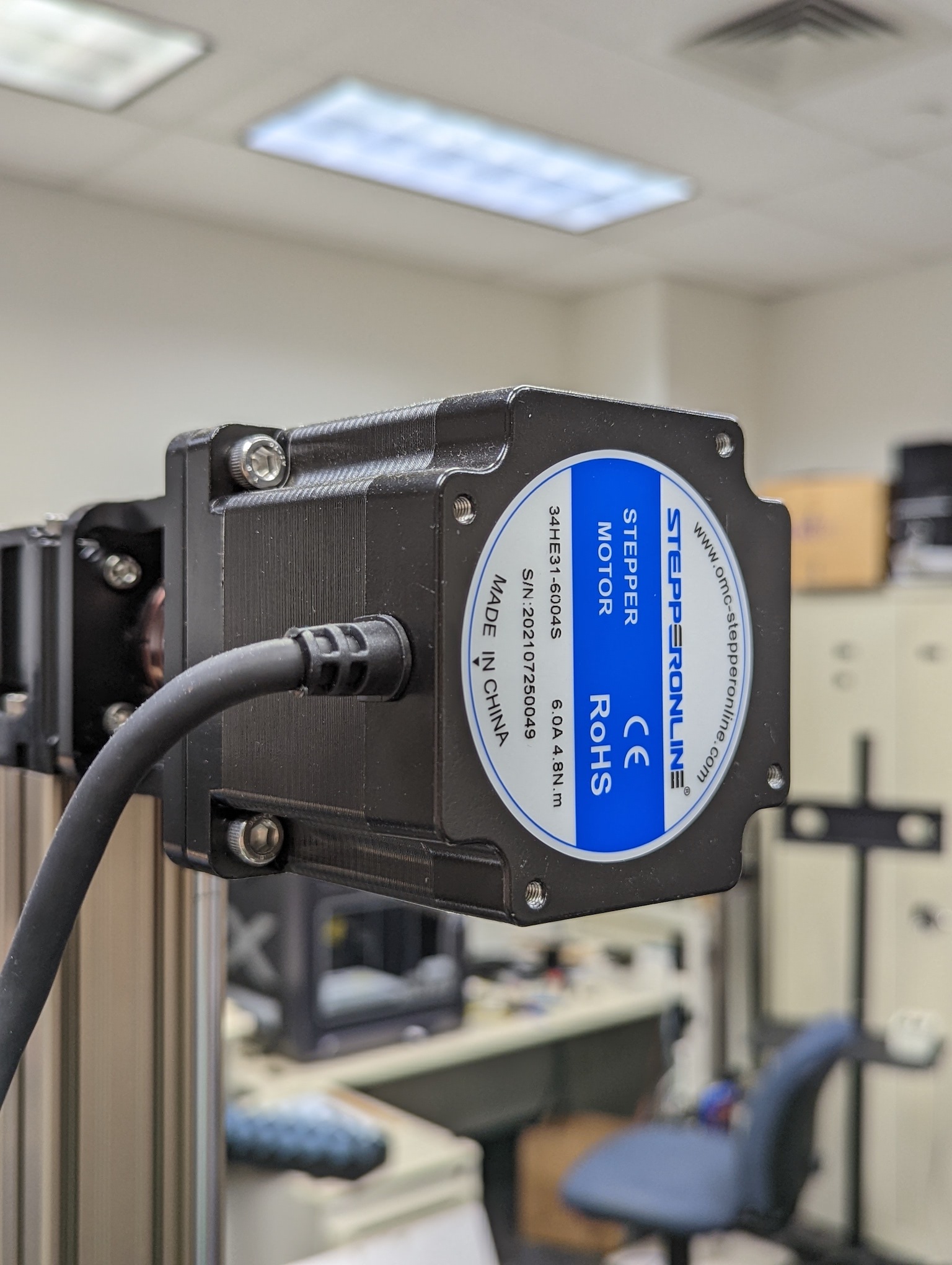}
         \caption{}
         \label{fig:motor_large}
    \end{subfigure}
    \begin{subfigure}[b]{0.4\textwidth}
         \centering
         \includegraphics[width=\textwidth]{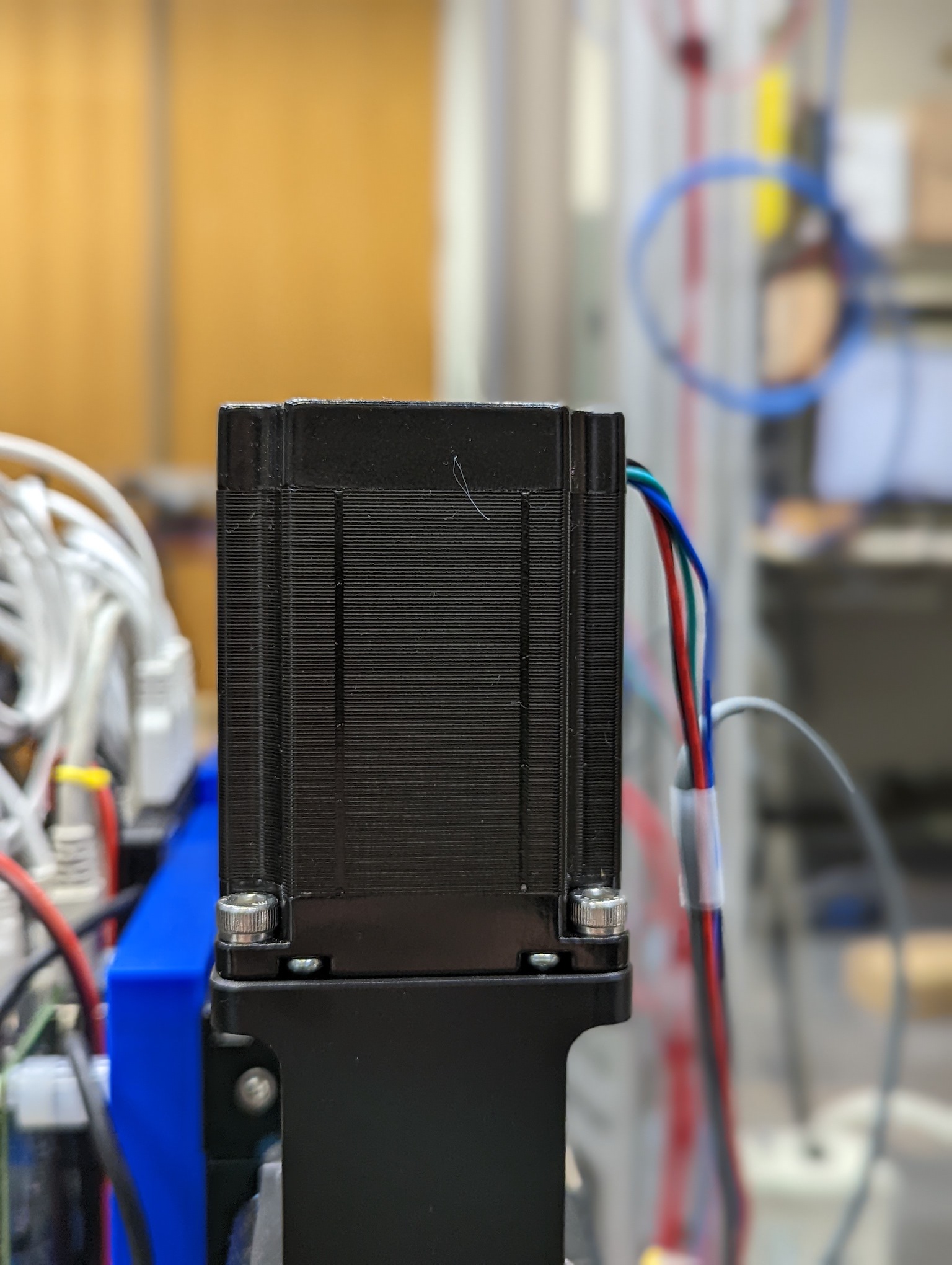}
         \caption{}
         \label{fig:motor_small}
    \end{subfigure}
\caption{Stepper motors: (a) NEMA34 motor at the vertical $y$-axis. (b) NEMA23 motor at the horizontal $x$-axis.}
\label{fig:motors}
\end{figure}

The stepper motors are shown in Fig. \ref{fig:motors}. 
A NEMA34 motor is used for the vertical $y$-axis and a NEMA23 motor is used for the horizontal $x$-axis. 
The motors are connected to the linear actuators using the black metal housing shown in Fig. \ref{fig:motors}. 
Within the housing is a cylindrical connector that tightens around each shaft (of the linear actuator and stepper motor). 
There are many sizes of these cylindrical connectors corresponding to different stepper motor shaft diameters. 
\textit{NOTE: The connector must be tightened very tightly around each shaft or the connection will not be strong enough for the motion, particularly for the vertical $y$-axis. However, the connectors are somewhat fragile, so be careful when tightening not to strip out the bolt heads or damage the connector.}

The wiring of the stepper motors and stepper drivers is discussed later in Section \ref{sec:wiring}. 

\subsection{AMC4030 Motion Controller}
\label{subsec:amc4030}
The motion of the entire system is controlled by a FUYU AMC4030 motion controller. 
The AMC4030 motion controller may be obscure, but it has an excellent MATLAB API that allows for direct control from MATLAB, enabling automation of scanning patterns and synchronization. 
The AMC4030 connections are detailed later in Section \ref{sec:wiring}. 

\subsubsection{AMC4030 Software}
\label{subsubsec:amc4030_software}
The AMC4030 software can be downloaded from the FUYU website \url{https://www.fuyumotion.com/manual/}. 
We have had difficulty opening the software on certain computers and are not sure why that is. 
However, even though the AMC4030 software GUI does not open, the API interface does work and the dual radar GUI and still control the scanner.

The AMC4030 shows up in Windows' Device Manager as a CH340. 
The driver included by FUYU does not always work for some reason, but we have had success with the driver at this website \url{https://sparks.gogo.co.nz/ch340.html}. 

\subsection{ESP32 Microcontroller}
\label{subsec:esp32}
The ESP32 microcontroller is used to synchronize the radar transmissions along the horizontal motion. 
The synchronizer, detailed in Section \ref{sec:synchronizer}, overcomes two primary issues: 1) uniform spacing of radar transmissions without constant velocity and 2) synchronizing both radars to transmit at the same physical locations although being horizontally displaced. 

\begin{figure}[ht]
    \centering
    \includegraphics[width=0.4\textwidth]{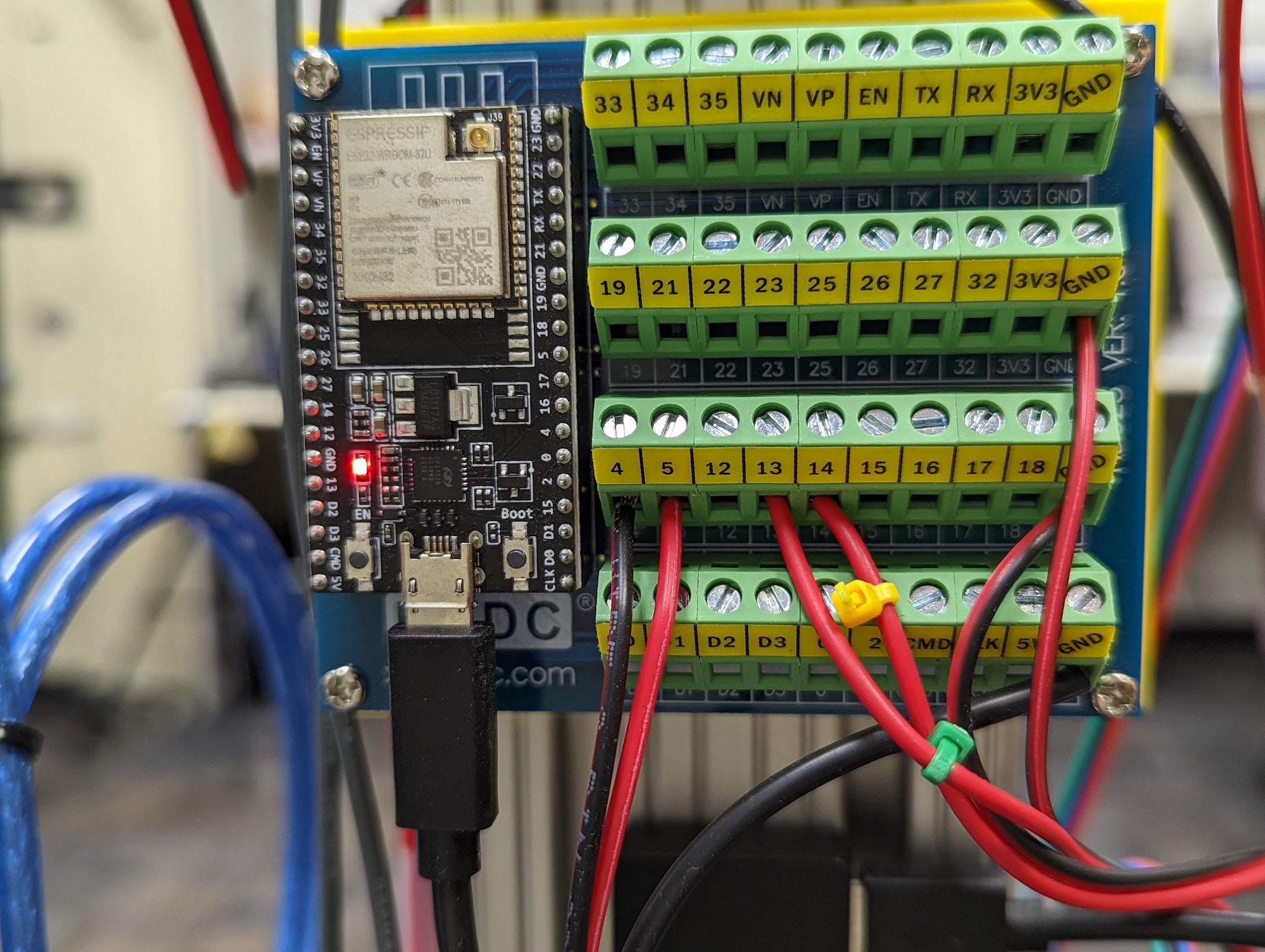}
    \caption{ESP32 microcontroller wired for dual radar synchronization.}
    \label{fig:esp32}
\end{figure}

The ESP32 is mounted to a breakout board, as shown in Fig. \ref{fig:esp32} to make the connections to accessible for each GPIO pin. 
The wiring of the ESP32 is discussed in Section \ref{sec:wiring} and the synchronizer software is detailed in Section \ref{sec:synchronizer}. 

\section{Wiring}
\label{sec:wiring}
The wiring of the mechanical system is crucial for proper operation and should be one of the first steps in debugging when an issue arises. 
This section details the various connections required for the entire system. 

\begin{figure}[ht]
    \centering
    \includegraphics[width=0.9\textwidth]{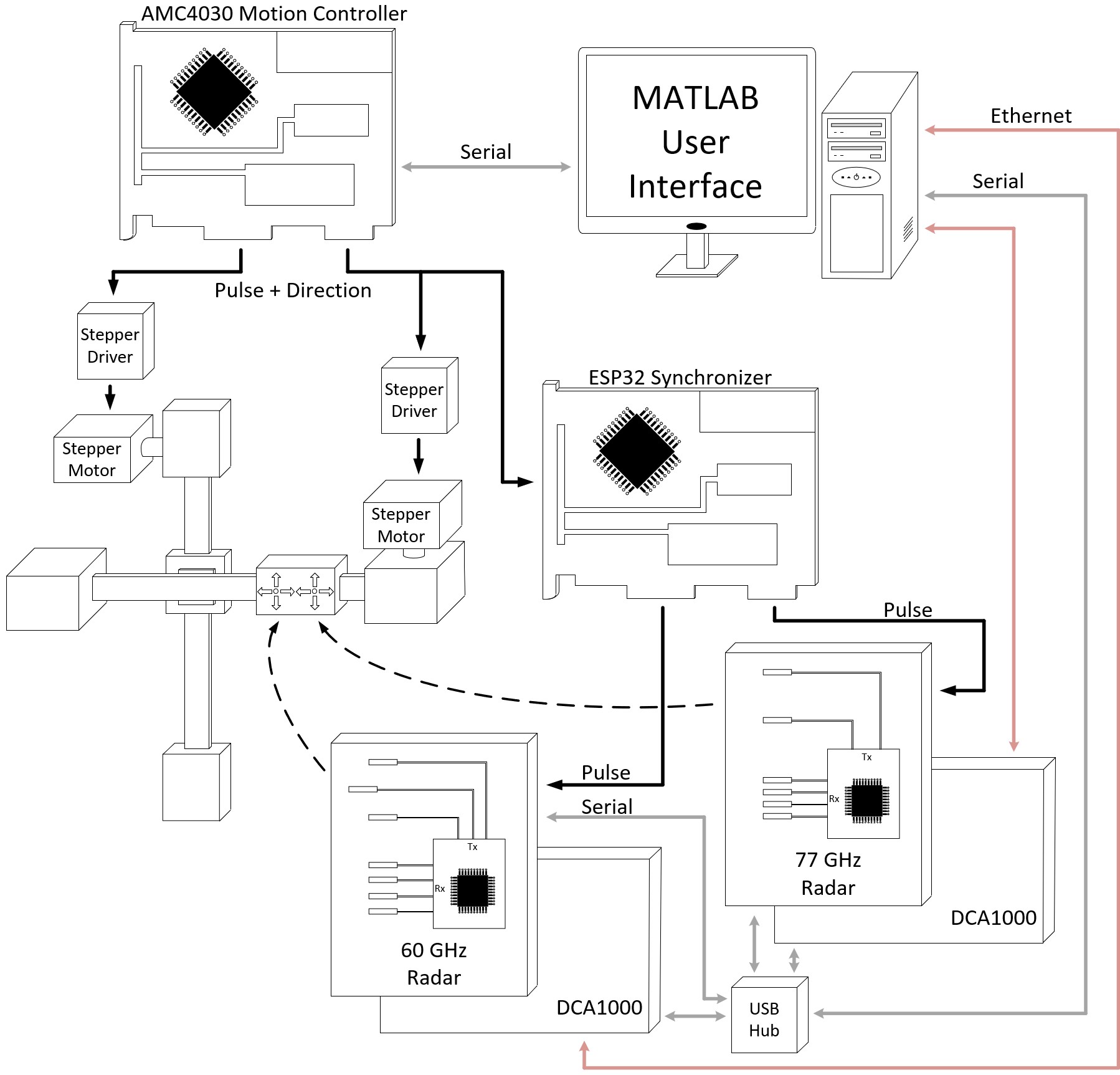}
    \caption{High-level diagram of the wiring required for the dual radar system. This illustration neglects power required for each device.}
    \label{fig:system_overview}
\end{figure}

A high-level view of the wiring is shown in Fig. \ref{fig:system_overview}. 
Although this figure does not include the power connections for each element, the required power connections are detailed for each component in the following sections. 

\subsection{Radar and DCA1000EVM Wiring}
\label{subsec:radar_wiring}
\begin{figure}[ht]
\centering
    \begin{subfigure}[b]{0.45\textwidth}
         \centering
         \includegraphics[width=\textwidth]{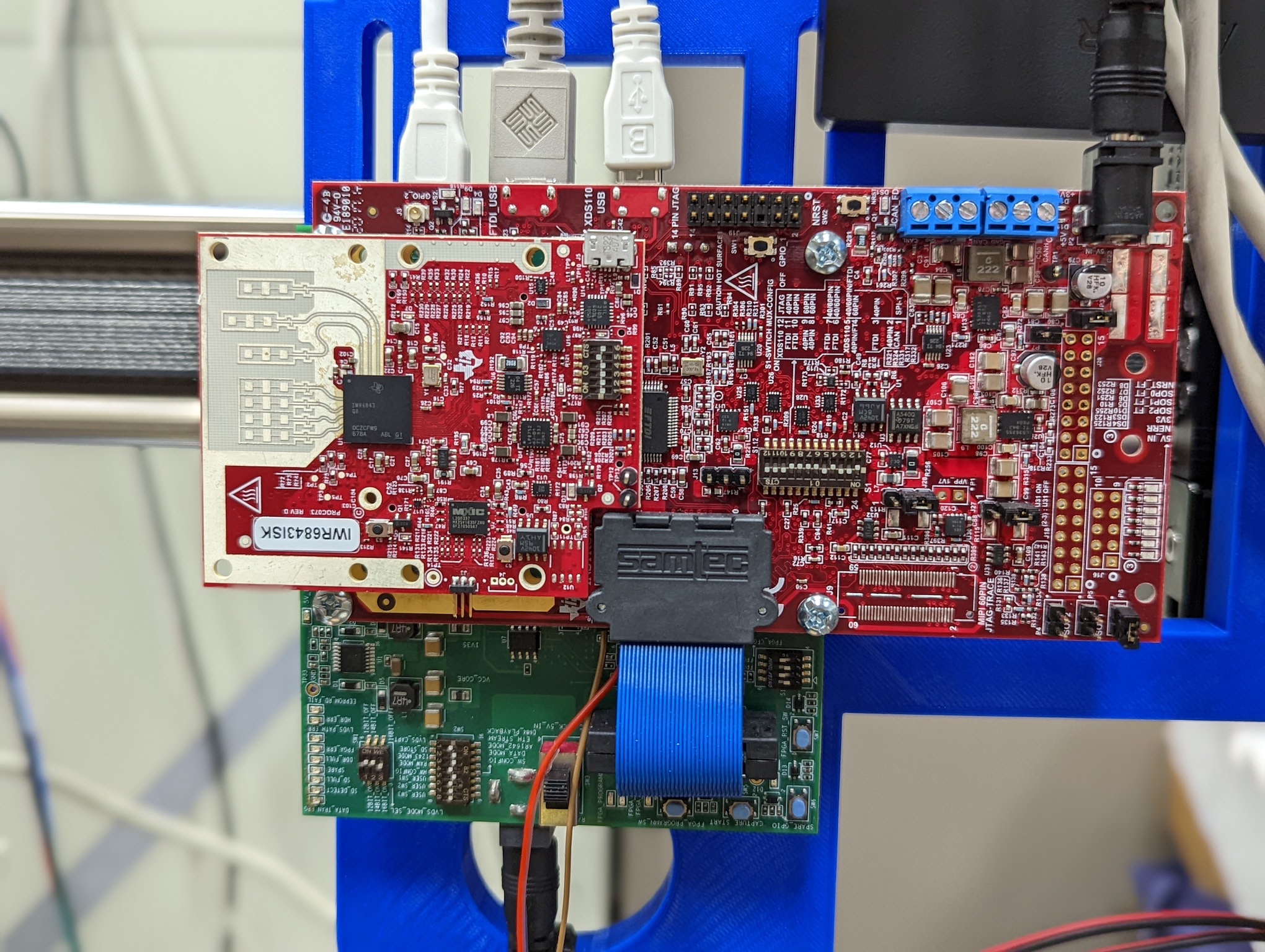}
         \caption{}
         \label{fig:radar60}
    \end{subfigure}
    \begin{subfigure}[b]{0.45\textwidth}
         \centering
         \includegraphics[width=\textwidth]{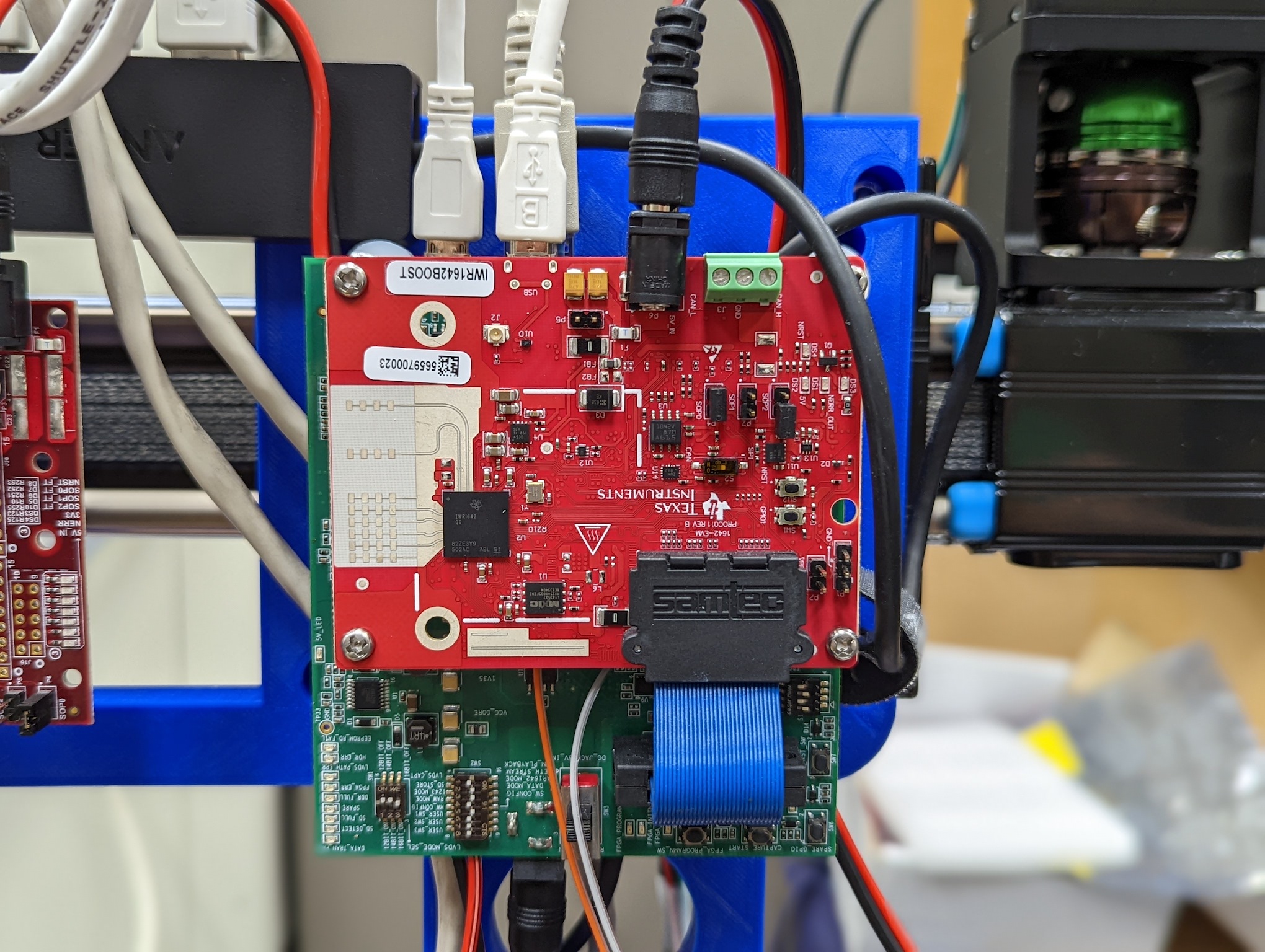}
         \caption{}
         \label{fig:radar77}
    \end{subfigure}
\caption{Radars: (a) 60 GHz radar (considered radar 1). (b) 77 GHz radar (considered radar 2). Both radars are mounted to DCA1000EVM boards via the 60-pin blue cables shown. Each DCA1000EVM is programmed to a different IP address following Chapter \ref{ch:dca1000evm_IP_address}. }
\label{fig:radars}
\end{figure}

As shown in Fig. \ref{fig:radars}, both radars are mounted to DCA1000EVMs, which are mounted to the back plate (blue 3-D printed PLA plastic). 
Each DCA1000EVM must be set to a different IP address, as discussed in Chapter \ref{ch:dca1000evm_IP_address}. 

The required connections are detailed as follows:
\begin{enumerate}
    \item \textbf{Connection between radars and DCA1000EVMs:} the radars are connected to the DCA1000EVMs using the 60-pin blue cables. 
    \item \textbf{Connection between radars and PC:} the radars are connected to the PC via a serial connection. The white USB cables shown are connected to a USB hub attached to the blue back plate. A USB extension cable connects that USB hub to another USB hub mounted to the dual radar scanner stand. The second USB hub is connected to the PC. The correct USB connection on each radar is detailed in Chapters \ref{ch:hw_trigger_6843} and \ref{ch:hw_trigger_1642}. 
    \item \textbf{Connection between DCA1000EVMs and PC:} the DCA1000EVMs are connected to the PC via a serial connection and Ethernet connection. The USB connection is identical to that between the radars and PC as the DCA1000EVMs connect via USB to the same USB hub attached to the blue back plate. The Ethernet connection is made directly to the PC. Two long Ethernet cables are used to make this connection. 
    \item \textbf{Connection between radars and ESP32:} the radars are connected to the ESP32 synchronizer via simple cables capable of a 3.3 V or 5 V signal. The HW pulse is sent from the ESP32 to trigger the radar at the proper spatial location to create a uniform grid despite the non-uniform motion. The connection of each radar is different and are detailed in Chapters \ref{ch:hw_trigger_6843} and \ref{ch:hw_trigger_1642}. \textit{NOTE: Connecting the ESP32 to the radars is EXTREMELY important. We recommend testing the connection with all the cables prior to assembling any system.}
    \item \textbf{Power connection for the radars and DCA1000EVMs:} the radars and DCA1000EVMs each require a 5 V ~3 A power supply. The power is provided to the blue back plate by a 24 V power supply, detailed in Section \ref{subsec:power_supply_wiring}, which is connected to a downconverter to provide a 5 V power supply. The connection is split off as necessary and connectors are provided for each radar and DCA1000EVM. 
\end{enumerate}

\subsection{ESP32 Wiring}
\label{subsec:esp32_wiring}
\begin{figure}[ht]
    \centering
    \includegraphics[width=0.4\textwidth]{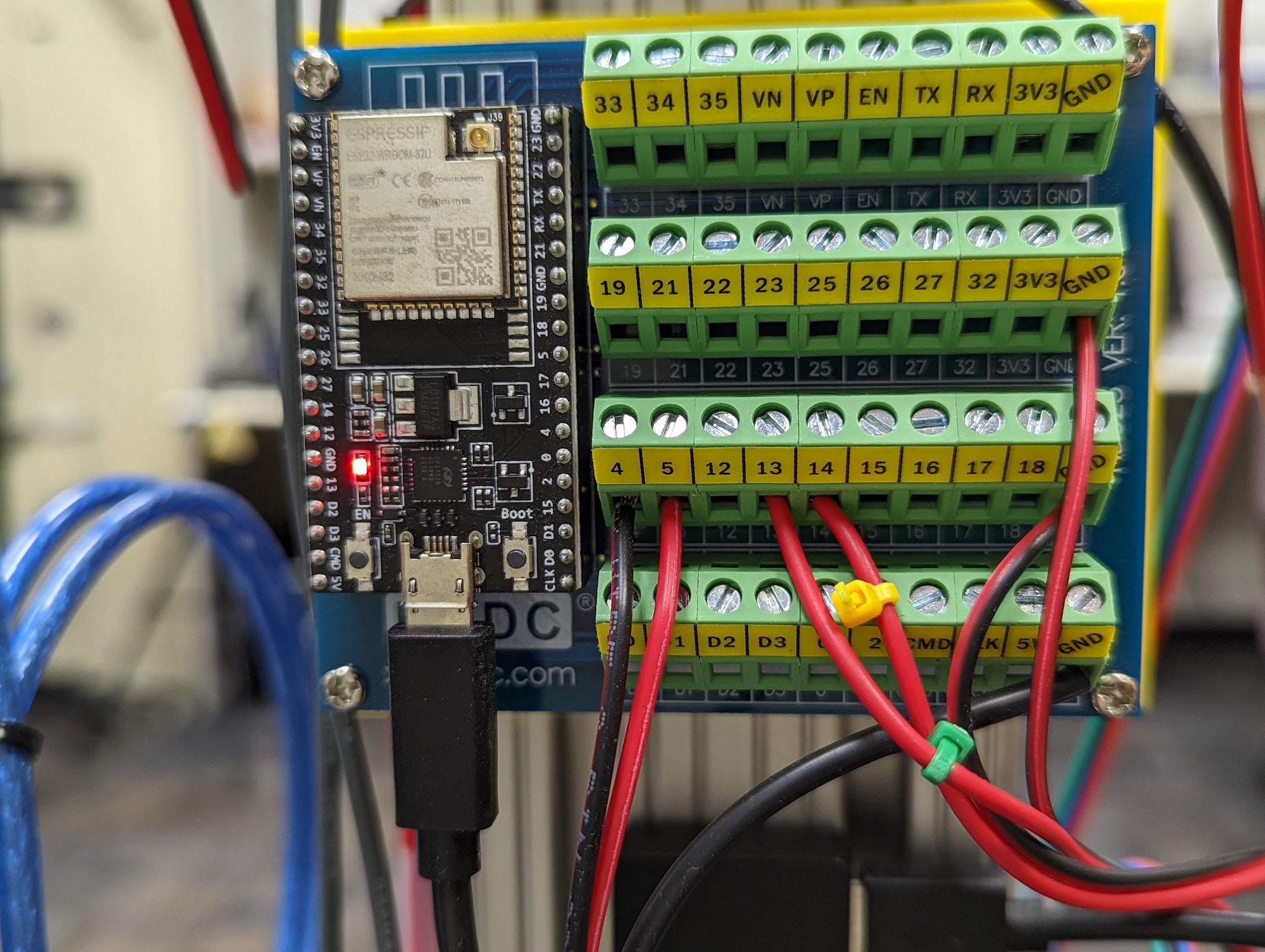}
    \caption{ESP32 microcontroller wired for dual radar synchronization.}
    \label{fig:esp32_wiring}
\end{figure}

A properly wired ESP32 is shown in Fig. \ref{fig:esp32_wiring}. 
The ESP32 is connected to the AMC4030 to count the direction and pulses sent to the stepper driver and is connected to the radars to send the HW trigger pulses. 

The required connections are detailed as follows:
\begin{enumerate}
    \item \textbf{Connection between ESP32 and AMC4030:} the ESP32 is connected to the AMC4030 to monitor the direction and pulse pins in order to count the pulses sent to the stepper driver. The connections are shown in Fig. \ref{fig:amc4030_wiring_close}. The ESP32 is connected with the black and dark red wires to the DIR1 and PUL1 pins on the AMC4030. The black wire (connected to DIR1 on the AMC4030) is connected to GPIO pin 4 on the ESP32 and the red wire (connected to PUL1 on the AMC4030) is connected to the GPIO pin 5 on the ESP32, as shown in Fig. \ref{fig:esp32_wiring}. Additionally, the ESP32 must be connected to the ground of the AMC4030 to create a common ground. 
    
    \begin{figure}[h]
        \centering
        \includegraphics[width=0.4\textwidth]{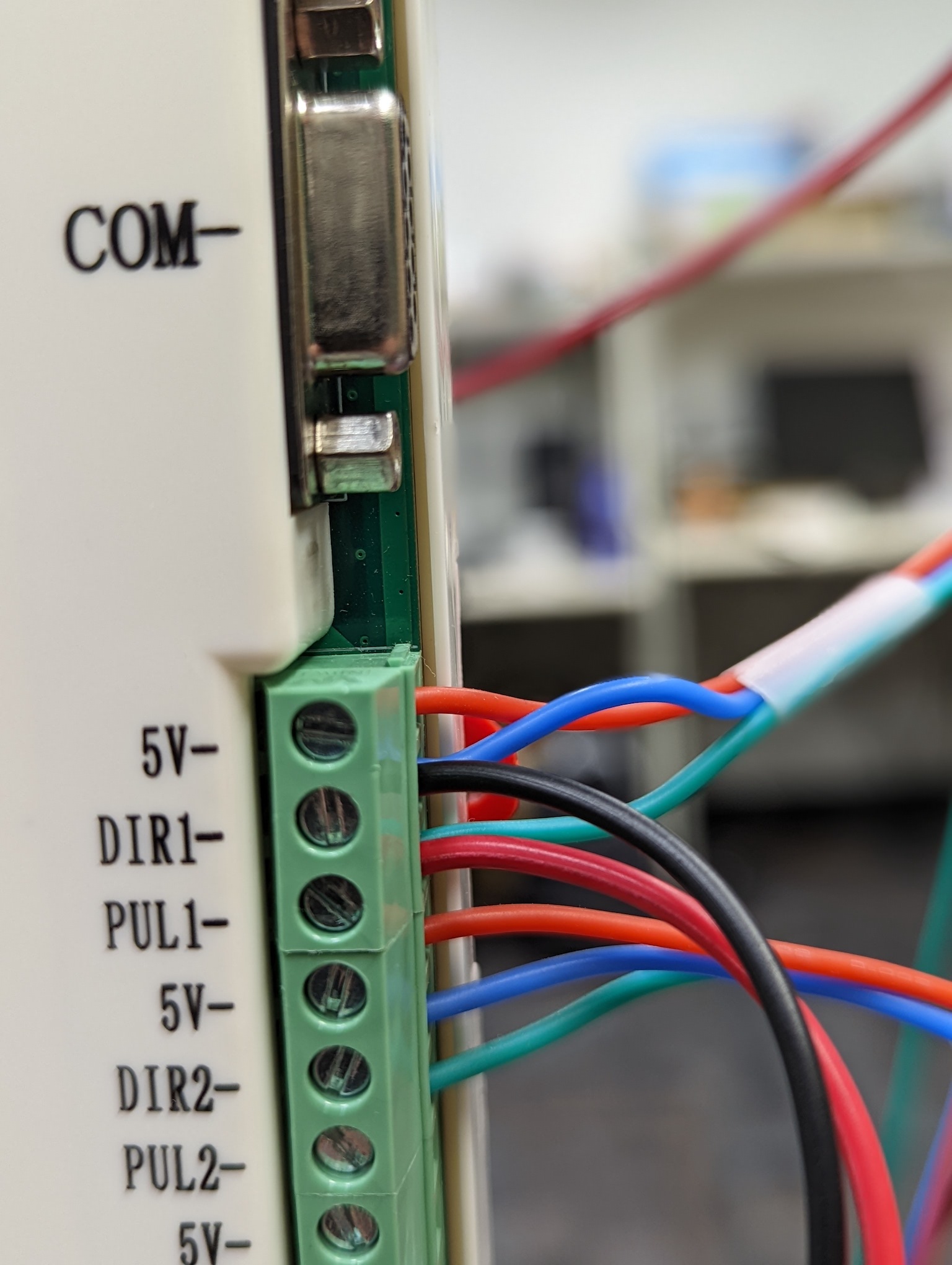}
        \caption{ESP32 connection with the AMC4030. The ESP32 is connected with the black and dark red wires to the DIR1 and PUL1 pins on the AMC4030. The black wire (connected to DIR1 on the AMC4030) is connected to GPIO pin 4 on the ESP32 and the red wire (connected to PUL1 on the AMC4030) is connected to the GPIO pin 5 on the ESP32, as shown in Fig. \ref{fig:esp32_wiring}.}
        \label{fig:amc4030_wiring_close}
    \end{figure}
    
    \item \textbf{Connection between ESP32 and radars:} the radars are connected to the ESP32 synchronizer via simple cables capable of a 3.3 V or 5 V signal. The HW pulse is sent from the ESP32 to trigger the radar at the proper spatial location to create a uniform grid despite the non-uniform motion. The connection of each radar is different and are detailed in Chapters \ref{ch:hw_trigger_6843} and \ref{ch:hw_trigger_1642}. As shown in Fig. \ref{fig:esp32_wiring}, radar 1 (60 GHz radar indicated with the green zip tie) is connected to GPIO pin 13 on the ESP32 and radar 2 (77 GHz radar indicated with the yellow zip tie) is connected to GPIO pin 13 on the ESP32. Additionally, both radars must be connected to the ground of the ESP32 to create a common ground. \textit{NOTE: Connecting the ESP32 to the radars is EXTREMELY important. We recommend testing the connection with all the cables prior to assembling any system.}
    
    \item \textbf{Connection between the ESP32 and PC:} the ESP32 is powered by the PC and receives commands from the PC for the various motion states of the scanner. The connection is made over a serial USB cable that is attached from the ESP32 to the USB hub mounted on the scanner stand and then to the PC. 
    
    \item \textbf{Power connection for the ESP32:} the ESP32 is powered by the USB connection with the PC.
\end{enumerate}

\subsection{AMC4030 Wiring}
\label{subsec:amc4030_wiring}
The AMC4030 receives commands from the PC, sends pulses to the stepper drivers, is connected to the ESP32 for synchronization, and is connected to homing switches to set the home position of the scanner.

\begin{figure}[ht]
\centering
    \begin{subfigure}[b]{0.45\textwidth}
         \centering
         \includegraphics[width=\textwidth]{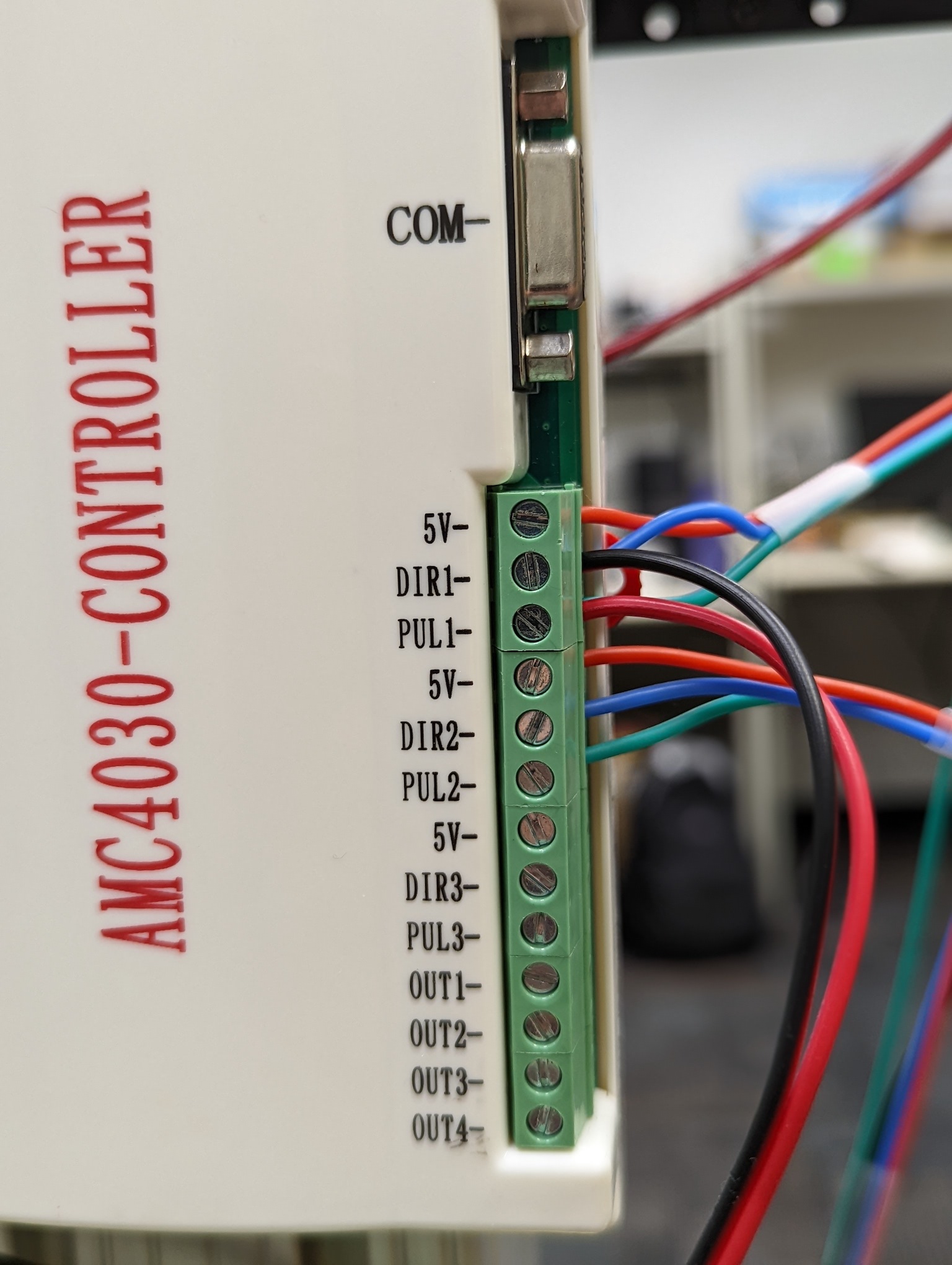}
         \caption{}
         \label{fig:amc4030_wiring_sub}
    \end{subfigure}
    \begin{subfigure}[b]{0.45\textwidth}
         \centering
         \includegraphics[width=\textwidth]{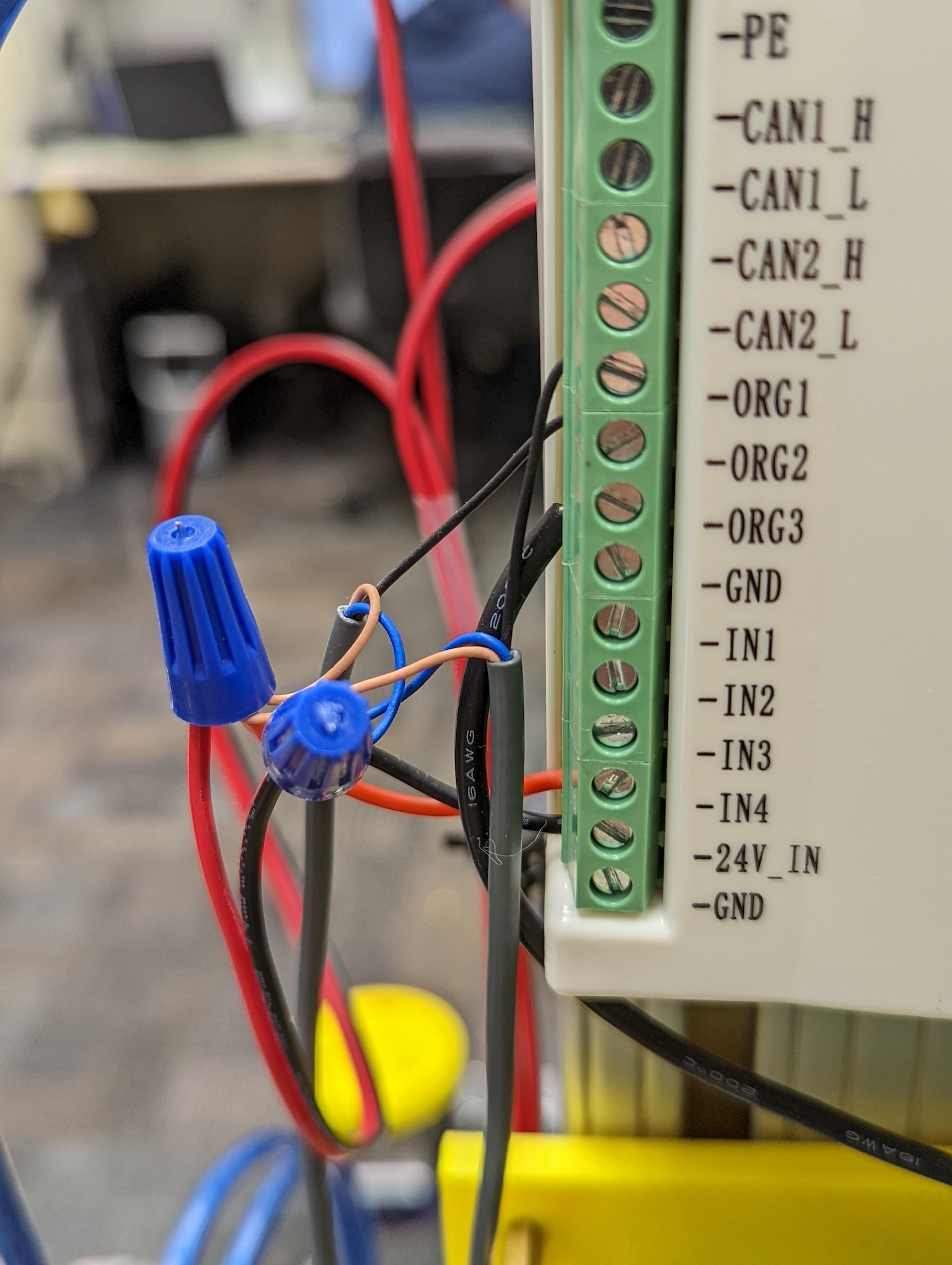}
         \caption{}
         \label{fig:homing_switch_wiring}
    \end{subfigure}
\caption{AMC4030 wiring: (a) AMC4030 connections with the stepper drivers and ESP32. See Figs. \ref{fig:driver_amc_connection} and \ref{fig:esp32_wiring} for corresponding connections on the stepper drivers and ESP32, respectively. (b) AMC4030 connection to the homing switches. The homing switch blue wire is connected to a 24 V power supply. The light brown wire on the homing switch is connected to ground. The black wire on the homing switch is connected to the ORG1 or ORG2 pin on the AMC4030 corresponding to the horizontal $x$-axis or vertical $y$-axis, respectively. }
\label{fig:amc4030_wiring}
\end{figure}

The required connections are detailed as follows:
\begin{enumerate}
    \item \textbf{Connection between AMC4030 and stepper drivers:} the stepper drivers are connected to AMC4030 using 3 wires (red, blue, and green), as shown in Figs. \ref{fig:amc4030_wiring_sub} and \ref{fig:driver_amc_connection}. The red wire is for a constant high (5 V) signal and is connected to the stepper driver's PUL+ and DIR+ pins and to the AMC4030's 5 V pin. The blue wire is for the direction pin, which indicates which direction the motion is to take, and is connected to the DIR- pin on the stepper driver and the DIR pin on the AMC4030. The green wire is for the pulses sent from the AMC4030 to the stepper driver and is connected to the PUL- pin on the stepper driver and the PUL pin on the AMC4030. The horizontal $x$-axis is connected to axis 1 on the AMC4030 and the vertical $y$-axis is connected to axis 2 on the AMC4030. 
    
    \begin{figure}[ht]
        \centering
        \includegraphics[width=0.4\textwidth]{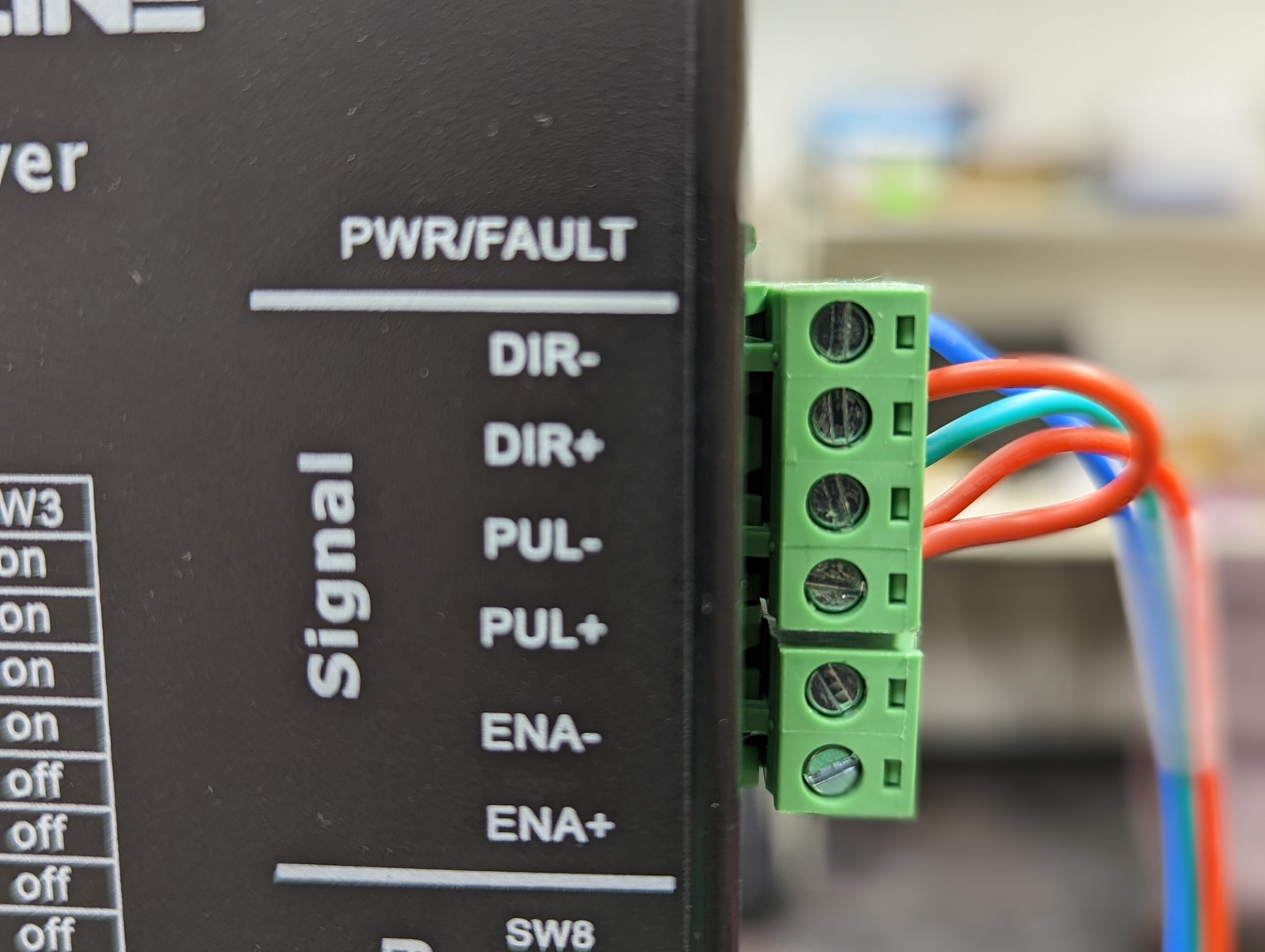}
        \caption{Stepper driver connection to the AMC4030. The red wire is for a constant high (5 V) signal and is connected to the stepper driver's PUL+ and DIR+ pins and to the AMC4030's 5 V pin. The blue wire is for the direction pin, which indicates which direction the motion is to take, and is connected to the DIR- pin on the stepper driver and the DIR pin on the AMC4030. The green wire is for the pulses sent from the AMC4030 to the stepper driver and is connected to the PUL- pin on the stepper driver and the PUL pin on the AMC4030. See Fig. \ref{fig:amc4030_wiring_sub} for the corresponding connections on the AMC4030. The horizontal $x$-axis is connected to axis 1 on the AMC4030 and the vertical $y$-axis is connected to axis 2 on the AMC4030. }
        \label{fig:driver_amc_connection}
    \end{figure}
    
    \item \textbf{Connection between AMC4030 and homing switches:} two homing switches are connected to the AMC4030, 1 for each of the for the $x$- and $y$-axes. The wiring is shown in Fig. \ref{fig:homing_switch_wiring} and the homing switch is shown in Fig. \ref{fig:homing_switch} properly mounted to the linear actuator. The homing switch blue wire is connected to a 24 V power supply. The light brown wire on the homing switch is connected to ground. The black wire on the homing switch is connected to the ORG1 or ORG2 pin on the AMC4030 corresponding to the horizontal $x$-axis or vertical $y$-axis, respectively. 
    
    \begin{figure}[ht]
        \centering
        \includegraphics[width=0.4\textwidth]{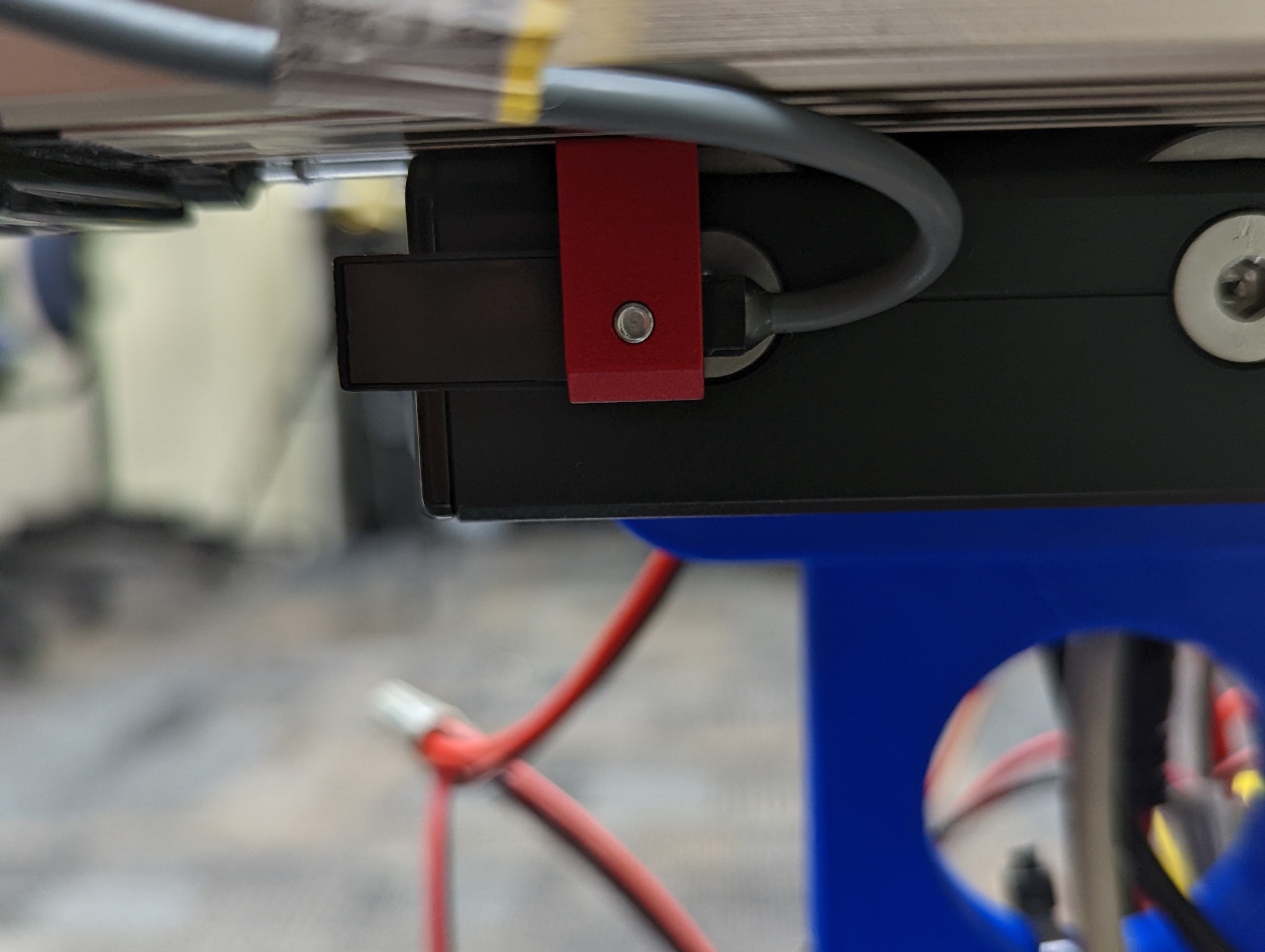}
        \caption{Homing switch connected to AMC4030 and mounted to linear actuator.}
        \label{fig:homing_switch}
    \end{figure}
    
    \item \textbf{Connection between ESP32 and AMC4030:} the ESP32 is connected to the AMC4030 to monitor the direction and pulse pins in order to count the pulses sent to the stepper driver. The connections are shown in Fig. \ref{fig:amc4030_wiring_close}. The ESP32 is connected with the black and dark red wires to the DIR1 and PUL1 pins on the AMC4030. The black wire (connected to DIR1 on the AMC4030) is connected to GPIO pin 4 on the ESP32 and the red wire (connected to PUL1 on the AMC4030) is connected to the GPIO pin 5 on the ESP32, as shown in Fig. \ref{fig:esp32_wiring}. Additionally, the ESP32 must be connected to the ground of the AMC4030 to create a common ground. 
    
    \item \textbf{Connection between AMC4030 and PC:} the AMC4030 is connected to the PC over a serial USB cable connected to the USB hub mounted to the scanner stand. 
    
    \item \textbf{Power connection for the AMC4030:} the AMC4030 is powered by a 24 V power supply, as shown in Fig. \ref{fig:homing_switch_wiring}. The power supply is connected to the 24V\_IN and GND pins on the AMC4030. 
\end{enumerate}

\subsection{Stepper Driver Wiring}
\label{subsec:stepper_driver_wiring}
The stepper driver is connected to the AMC4030 and stepper motors. 

The required connections are detailed as follows:
\begin{enumerate}
    \item \textbf{Connection between AMC4030 and stepper drivers:} the stepper drivers are connected to AMC4030 using 3 wires (red, blue, and green), as shown in Figs. \ref{fig:amc4030_wiring_sub} and \ref{fig:driver_amc_connection}. The red wire is for a constant high (5 V) signal and is connected to the stepper driver's PUL+ and DIR+ pins and to the AMC4030's 5 V pin. The blue wire is for the direction pin, which indicates which direction the motion is to take, and is connected to the DIR- pin on the stepper driver and the DIR pin on the AMC4030. The green wire is for the pulses sent from the AMC4030 to the stepper driver and is connected to the PUL- pin on the stepper driver and the PUL pin on the AMC4030. The horizontal $x$-axis is connected to axis 1 on the AMC4030 and the vertical $y$-axis is connected to axis 2 on the AMC4030. 
    
    \item \textbf{Connection between the stepper drivers and stepper motors:} the stepper drivers are connected the stepper motors using a standard 4 wire connection, as shown in Fig. \ref{fig:driver_motor_connection}. The wiring is shown with two different configurations. The A+ and A- pins on the stepper driver should be connected to the red and blue wires of the stepper motor, and the B+ and B- pins on the stepper driver should be connected to the black and green wires of the stepper motor. The direction of the stepper motor can be reversed by switching the connection of either the A or B pair of wires, as shown in Figs. \ref{fig:driver_motor_connection2} and \ref{fig:driver_motor_connection3}. 

    \begin{figure}[ht]
    \centering
        \begin{subfigure}[b]{0.45\textwidth}
             \centering
             \includegraphics[width=\textwidth]{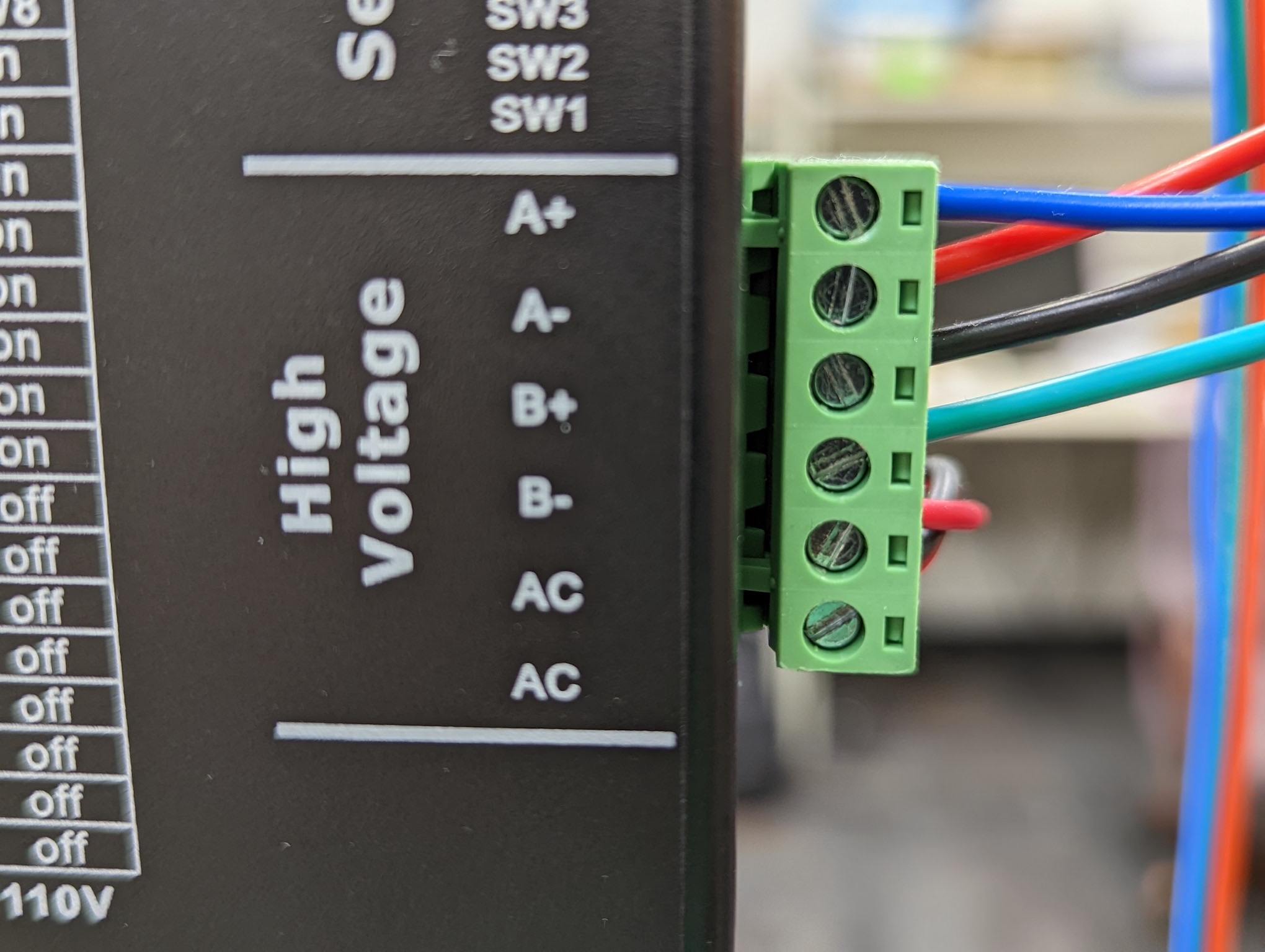}
             \caption{}
             \label{fig:driver_motor_connection2}
        \end{subfigure}
        \begin{subfigure}[b]{0.45\textwidth}
             \centering
             \includegraphics[width=\textwidth]{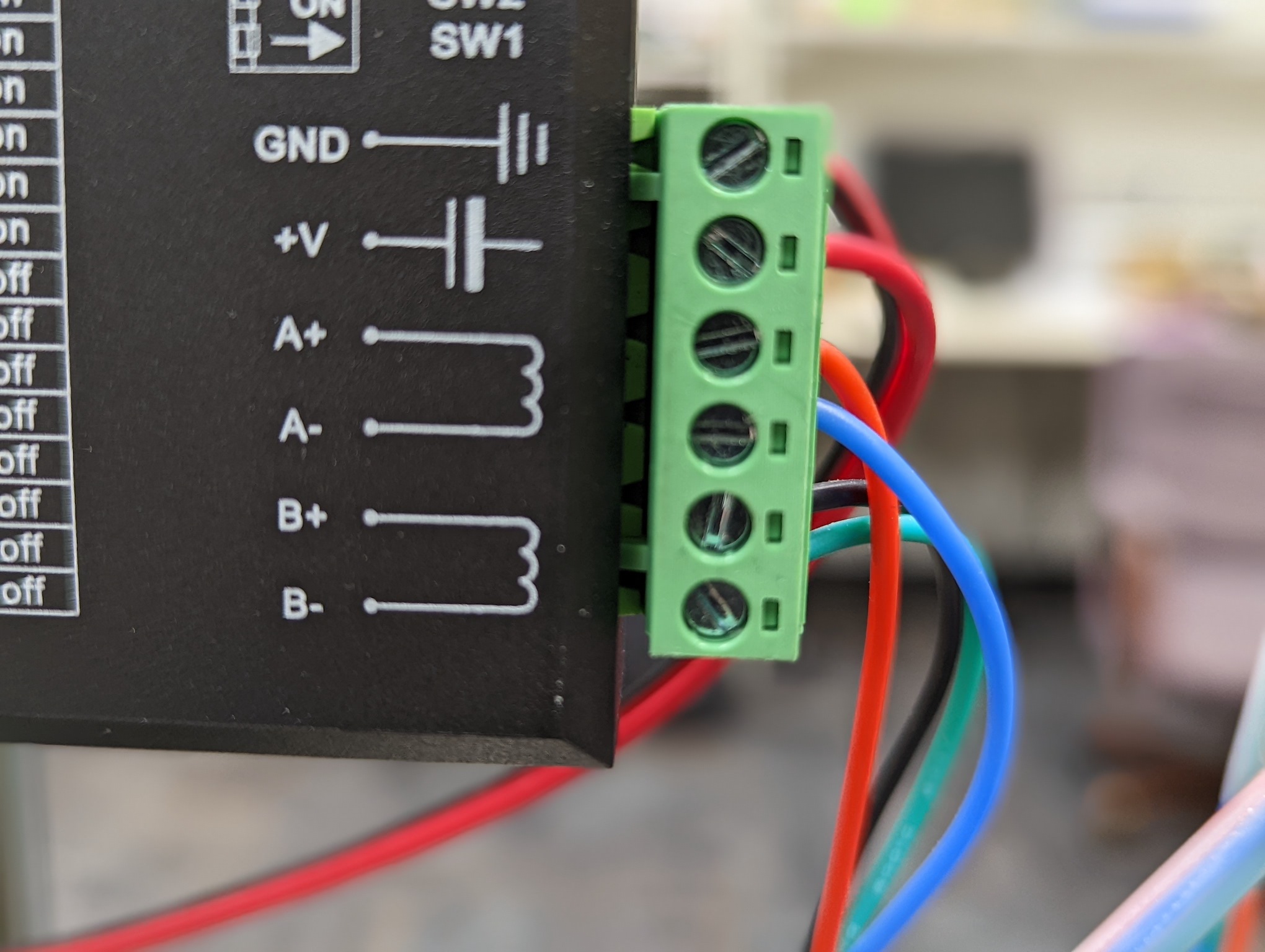}
             \caption{}
             \label{fig:driver_motor_connection3}
        \end{subfigure}
    \caption{Stepper driver connection with stepper motors: (a) DM860T connection with NEMA34 stepper motor at the vertical $y$-axis. (b) DM542T connection with the NEMA23 stepper motor at the horizontal $x$-axis. The wiring is shown with two different configurations. The A+ and A- pins on the stepper driver should be connected to the red and blue wires of the stepper motor, and the B+ and B- pins on the stepper driver should be connected to the black and green wires of the stepper motor. The direction of the stepper motor can be reversed by switching the connection of either the A or B pair of wires, as shown in (a) and (b).}
    \label{fig:driver_motor_connection}
    \end{figure}
    
    \item \textbf{Power connection for the stepper drivers:} The stepper drivers are connected to either a 24 V or 50 V power supply on the pins as shown in Fig. \ref{fig:driver_motor_connection}. 
\end{enumerate}

\subsection{Stepper Motor Wiring}
\label{subsec:stepper_motor_wiring}
The stepper motors are connected to the stepper drivers and mounted to the linear actuators, as shown in Fig. \ref{fig:motors}. 
See \ref{subsec:stepper_motors} for more details regarding the mounting. 

The required connections are detailed as follows:
\begin{enumerate}
    \item \textbf{Connection between the stepper drivers and stepper motors:} the stepper drivers are connected the stepper motors using a standard 4 wire connection, as shown in Fig. \ref{fig:driver_motor_connection}. The wiring is shown with two different configurations. The A+ and A- pins on the stepper driver should be connected to the red and blue wires of the stepper motor, and the B+ and B- pins on the stepper driver should be connected to the black and green wires of the stepper motor. The direction of the stepper motor can be reversed by switching the connection of either the A or B pair of wires, as shown in Figs. \ref{fig:driver_motor_connection2} and \ref{fig:driver_motor_connection3}. 
    \item \textbf{Power connection for the stepper motors:} the power for the stepper motors is provided from the stepper drivers. There is no need for any additional power connections. 
\end{enumerate}

\subsection{PC Wiring}
\label{subsec:pc_wiring}
The PC controls the entire scan and must be connected to the radars, DCA1000EVMs, ESP32, and AMC4030. 
The PC must be powered on, running Windows (10 is recommended, 11 is not tested), and MATLAB 2021 or 2022 recommended. 

The required connections are detailed as follows:
\begin{enumerate}
    \item \textbf{Connection between radars and PC:} the radars are connected to the PC via a serial connection. The white USB cables shown are connected to a USB hub attached to the blue back plate. A USB extension cable connects that USB hub to another USB hub mounted to the dual radar scanner stand. The second USB hub is connected to the PC. The correct USB connection on each radar is detailed in Chapters \ref{ch:hw_trigger_6843} and \ref{ch:hw_trigger_1642}. 
    
    \item \textbf{Connection between DCA1000EVMs and PC:} the DCA1000EVMs are connected to the PC via a serial connection and Ethernet connection. The USB connection is identical to that between the radars and PC as the DCA1000EVMs connect via USB to the same USB hub attached to the blue back plate. The Ethernet connection is made directly to the PC. Two long Ethernet cables are used to make this connection. 
    
    \item \textbf{Connection between the ESP32 and PC:} the ESP32 is powered by the PC and receives commands from the PC for the various motion states of the scanner. The connection is made over a serial USB cable that is attached from the ESP32 to the USB hub mounted on the scanner stand and then to the PC. 
    
    \item \textbf{Connection between AMC4030 and PC:} the AMC4030 is connected to the PC over a serial USB cable connected to the USB hub mounted to the scanner stand. 
\end{enumerate}

\subsection{Power Supply Wiring}
\label{subsec:power_supply_wiring}
The dual radar scanner uses 4 power supplies: 3x 24 V power supplies and 1x 50 V power supply. 
The power supplies mounted to the base of the scanner stand. 
The power supplies are connected to the radars, DCA1000EVMs, AMC4030, and stepper drivers as discussed in the Sections above. 

\section{Dual Radar SAR Position Synchronizer}
\label{sec:synchronizer}
The dual radar synchronizer is a crucial element to the system as it overcomes two integral challenges for dual radar scanning: 
\begin{enumerate}
    \item At high speeds, the linear actuators do not follow a simple constant velocity model, but rather accelerate at the start of each motion and decelerate at the end of each motion \cite{yanik2020development}. Hence, if the radar is programmed to trigger at regular time intervals (uniform periodicity), the positions of each trigger will not follow a regular uniform grid. Rather, they will be spaced unevenly and the position of each sample is unknown. The synchronizer tracks the position of the platform regardless of the velocity and acceleration and sends HW triggers to the radars so they trigger at regular spatial intervals to create an ideal uniform grid.
    
    \item To recover an image, the position of both radars must be known at each sample. The synchronizer tracks the position of the platform and sends a HW trigger to each radar when that radar reaches the next breakpoint so as to align the captures horizontally. 
\end{enumerate}

\subsection{Summary of Functionality}
The dual radar synchronizer is written in FreeRTOS and implemented on an ESP32. 

\textbf{Position Estimation via Pulse Counting:} The synchronizer counts the pulses sent from the AMC4030 to track the exact position of the platform across the horizontal motion. 
The AMC4030 sends pulses unevenly following the acceleration profile, as discussed in \cite{yanik2020development}. 
The synchronizer can convert the number of pulses to spatial distance using the analysis in Section \ref{subsubsec:stepper_driver_settings}. 

\textbf{HW Trigger Sent to Radars:} The synchronizer sends a HW pulse to the radars when the platform reaches a breakpoint. 
Each breakpoint is specified by the user to denote the spatial sampling interval. 
Knowing the relative distance between the radars, the synchronizer tracks the location of each radar independently and sends the HW trigger when each radar meets the next breakpoint. 
Hence, the scanner can perform two-direction scanning where the synchronizer counts up and then down and sends the HW trigger pulse at the specified breakpoints.

\chapter{Using the Base Scanner}
\label{ch:base_scanner}

The base scanner GUI can be accessed under the Tools menu in the dual radar GUI. 
Functionality for operating the base scanner is simple and detailed in the base scanner GUI. 
At the time of writing this manual, the base scanner is being built and the functionality is limited. 
Over the summer of 2022, Yusef Alimam will be attempting to automate data collection using the base scanner and will improve the functionality. 
\chapter{Changing DCA1000EVM IP Addresses and Ports}
\label{ch:dca1000evm_IP_address}

The following is a simple guide for changing the configuration of a DCA1000EVM. 
To operate multiple DCA1000EVM devices on the same PC, the IP addresses and ports associated with each DCA1000EVM must be unique. 
Hence, these steps must be carefully followed to ensure each DCA1000EVM is properly configured for multi-radar coordination. 

\section{If the Current DCA1000EVM IP Address is {Unknown}}
If the current IP address is unknown, the following steps are necessary to reset the DCA1000EVM to factory defaults prior to setting a new IP address. 

\begin{enumerate}
    \item[1)] Place SW2.6 to the ON position (towards pin 11).
    \item[2)] Power cycle the DCA1000EVM (This loads the default Ethernet settings from the DCA1000EVM’s FPGA).
    \item[3)] The IP address for the DCA1000EVM is now 192.168.33.180.
    \item[4)] Set the IPv4 address on active PC LAN port to 192.168.33.30 (see the \href{https://software-dl.ti.com/ra-processors/esd/MMWAVE-STUDIO/latest/exports/DCA1000_Quick_Start_Guide.pdf}{DCA1000EVM Quick Start Guide}). 
    \begin{enumerate}
        \item[a)] Once the device is properly connected to the PC, open the start menu and search ``View Network Connections,'' as shown in Fig. \ref{fig:view_network_connections}. 
        \item[b)] Inside the ``Network Connections'' of Control Panel, right click on the Ethernet port of choice and select ``Properties.''
        \item[c)] ``Local Area Connection Properties'' window will open. Right click on ``Internet Protocol Version 4 (TCP/IPv4).''
        \item[d)] ``Internet Protocol Version 4 (TCP/IPc4)'' window will open. Set the IP address field to 192.168.33.30.  
        \item[e)] The Subnet mask field can remain the default 255.255.255.0. 
        \item[f)] Press ``OK'' on all the windows and you can close ``Network Connections.''
    \end{enumerate}
    \item[5)] Follow the instructions in the following Section to set a new IP address to the DCA1000EVM. 
\end{enumerate}

\section{Setting New IP Addresses and Ports to the DCA1000EVM}
\begin{enumerate}
    \item[1)] Modify DCA1000EVM configuration file (.json). The typical location is \\ ``C:/ti/mmwave\_studio\_xx\_xx\_xx\_xx/mmWaveStudio/PostProc/.''
    
    \begin{enumerate}
        \item[a)] Lines 9--20 will have the format given in Fig. \ref{fig:dca_change}.
    
        \begin{figure}[th]
            \centering
            \includegraphics[width=0.85\textwidth]{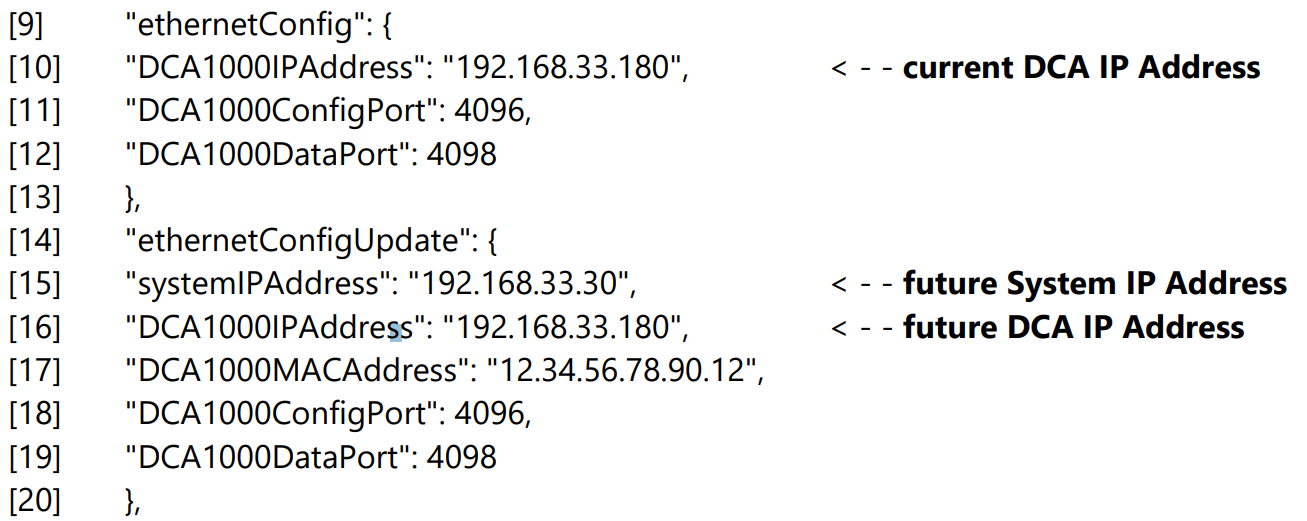}
            \caption{Lines 9--20 of the .json configuration file used to change the DCA1000EVM IP addresses and ports.}
            \label{fig:dca_change}
        \end{figure}
    
        \item[b)] Ensure that the IP address on line 10 matches the current IP address of the DCA1000EVM.
        \item[c)] Change the IP addresses on lines 15 - 16 to the new IP address.
        \item[d)] Save the json file under a new name, such as: newIP.json.\footnote{If you want to use different IP address with mmWave Studio, make sure the json is called \texttt{cf.json}. mmWave Studio will use and modify the \texttt{cf.json} in the \texttt{PostProc} folder. Hence, it is recommended that you use different .json files for each radar named \texttt{cf1.json}, for example.}
    \end{enumerate}
    
    \item[2)] Send the new configuration file to DCA1000EVM. 
    \begin{enumerate}
        \item[a)] Power cycle the DCA1000EVM.
        \item[b)] Open Powershell.
        \item[c)] Use the following commands to enter the correct directory and update the DCA1000EVM’s EEPROM:
        \begin{enumerate}
            \item[] \verb$cd "C:\ti\mmwave_studio_xx_xx_xx_xx\mmWaveStudio\PostProc\"$
            \item[] \verb$.\DCA1000EVM_CLI_Control.exe eeprom newIP.json$
        \end{enumerate}
    \end{enumerate}
    \item[3)] Set the IPv4 address on active PC LAN port to the new System IP Address (see the \href{https://software-dl.ti.com/ra-processors/esd/MMWAVE-STUDIO/latest/exports/DCA1000_Quick_Start_Guide.pdf}{DCA1000EVM Quick Start Guide}). 
        \begin{enumerate}
            \item[a)] Once the device is properly connected to the PC, open the start menu and search ``View Network Connections,'' as shown in Fig. \ref{fig:view_network_connections}. 
            \item[b)] Inside the ``Network Connections'' of Control Panel, right click on the Ethernet port of choice and select ``Properties.''
            \item[c)] ``Local Area Connection Properties'' window will open. Right click on ``Internet Protocol Version 4 (TCP/IPv4).''
            \item[d)] ``Internet Protocol Version 4 (TCP/IPc4)'' window will open. Set the IP address field to 192.168.xxx.xxx (the \textbf{NEW} DCA1000EVM System IP Address).  
            \item[e)] The Subnet mask field can remain the default 255.255.255.0. 
            \item[f)] Press ``OK'' on all the windows and you can close ``Network Connections.''
        \end{enumerate}
    \item[4)] Update the .json file
        \begin{enumerate}
            \item[a)] In the .json file, change line 10 to match the IP address on line 15.
            \item[b)] Save changes.
        \end{enumerate}
    \item[5)] Verify changes
        \begin{enumerate}
            \item[a)] Place SW2.6 to the OFF position (towards pin 6).
            \item[b)] Power cycle the DCA1000EVM. (This loads the Ethernet settings from the DCA1000EVM’s EEPROM).
            \item[c)] Open Powershell. 
            \item[d)] Use the following commands to enter the correct directory and verify system status:
                \begin{enumerate}
                    \item[] \verb$cd "C:\ti\mmwave_studio_xx_xx_xx_xx\mmWaveStudio\PostProc\"$
                    \item[] \verb$.\DCA1000EVM_CLI_Control.exe query_sys_status newIP.json$
                \end{enumerate}
            \item[e)] If response is ``System is connected'' then the device is functioning properly. If the response is ``System is disconnected'' then ensure that the following are true:
                \begin{enumerate}
                    \item[i] SW2.6 is in the OFF position (towards pin 6).
                    \item[ii] PC’s IP Address is set to the correct IP Address for the system and not the IP Address for the DCA.
                    \item[iii] Correct .json file is called when running \verb$query_sys_status$ command.
                \end{enumerate}
        \end{enumerate}
\end{enumerate}

\chapter{Using Hardware Trigger with TI 6843 60 GHz Radar}
\label{ch:hw_trigger_6843}

The following is a simple demo for enabling the hardware (HW) trigger with a Texas Instruments IWR6843ISK 60 GHz radar in conjunction with the Dual Radar SAR Controller application.

\section{Hardware Requirements}
\begin{enumerate}
    \item[1)] \href{https://www.ti.com/tool/DCA1000EVM}{DCA1000EVM} with modifications (see Section \ref{6843:DCA}). 
    \item[2)] \href{https://www.ti.com/tool/MMWAVEICBOOST}{MMWAVEICBOOST} with modifications (see Section \ref{6843:MMWAVEICBOOST}).
    \item[3)] \href{https://www.ti.com/tool/IWR6843ISK}{IWR6843ISK} with modifications (see Section \ref{6843:IWR6843ISK}). 
    \item[4)] ESP32 microcontroller to send HW trigger pulses.
\end{enumerate}

\section{Software Requirements}
\begin{enumerate}
    \item[1)] \href{https://www.ti.com/tool/MMWAVE-SDK}{TI mmWave SDK 3.5.0.4}
    \item[2)] \href{https://www.ti.com/tool/UNIFLASH}{TI Uniflash}
    \item[3)] \href{https://software-dl.ti.com/ra-processors/esd/MMWAVE-STUDIO/latest/index_FDS.html}{TI mmWave Studio 2.1.1.0}
    \item[4)] Dual Radar SAR Controller software (requires MATLAB 2021b (recommended, may be compatible with earlier/later releases)).
\end{enumerate}

\section{Hardware Set Up}

\subsection{DCA1000EVM Hardware Configuration}
\label{6843:DCA}
\begin{enumerate}
    \item[1)] Remove R120 \href{https://e2e.ti.com/support/sensors-group/sensors/f/sensors-forum/710632/awr1243-dca1000-and-1243-hardware-trigger-sts_no_lvds_data}{(suggested here)}. 
    \item[2)] Connect to the MMWAVEICBOOST with 60 pin connector, as shown on \href{https://www.ti.com/lit/ug/swru546d/swru546d.pdf?ts=1620300878197}{page 16 here}.
    \item[3)] Connect to the PC over USB on the RADAR/FTDI connnector (J1 on DCA1000EVM) and over Ethernet.
    \item[4)] Connect 5V/3A power.
\end{enumerate}

\subsection{MMWAVEICBOOST Hardware Configuration}
\label{6843:MMWAVEICBOOST}
\begin{enumerate}
    \item[1)] Remove R346 and short R348 (suggested \href{https://e2e.ti.com/support/sensors-group/sensors/f/sensors-forum/998616/iwr6843isk-custom-binary-for-cli-lvds-hw-trigger/3692785}{here} and \href{https://e2e.ti.com/support/sensors-group/sensors/f/sensors-forum/972613/iwr6843isk-hardware-trigger}{here}).
    \begin{enumerate}
        \item[a)] Follow \href{https://e2e.ti.com/support/sensors-group/sensors/f/sensors-forum/998616/iwr6843isk-custom-binary-for-cli-lvds-hw-trigger/3692785}{Josiah's E2E post}:
        \item[b)] Download xWR6843 EVM Schematic Drawing, Assembly Drawing, and Bill of Materials - SWRR164C.zip from \href{https://www.ti.com/tool/IWR6843ISK#design-files}{here}.
        \item[c)] On page 9 of PROC074B(001)\_Sch.pdf (for rev B of the MMWAVEICBOOST), under ``RNR FOR SYNC IN'', 40PIN\_SYNC\_IN needs to be routed to \\ RADAR\_SYNC\_IN.
        \item[d)] From that diagram, DCA\_SYNC\_IN is shorted via R346 to \ RADAR\_SYNC\_IN.
        \item[e)] Hence, remove R346 and place 0 ohm resistor over R348. Now 40PIN\_SYNC\_IN is routed to RADAR\_SYNC\_IN.
    \end{enumerate}
    \item[2)] Switch Settings
    \begin{enumerate}
        \item[a)] From page 11 \href{https://www.ti.com/lit/ug/swru546d/swru546d.pdf?ts=1620300878197}{here}, use the switch settings on S1 for DCA1000 Mode, as shown in Fig. \ref{fig:MMWAVEICBOOST_switches}. 
    \end{enumerate}
    
    \begin{figure}[th]
        \centering
        \includegraphics[width=0.95\textwidth]{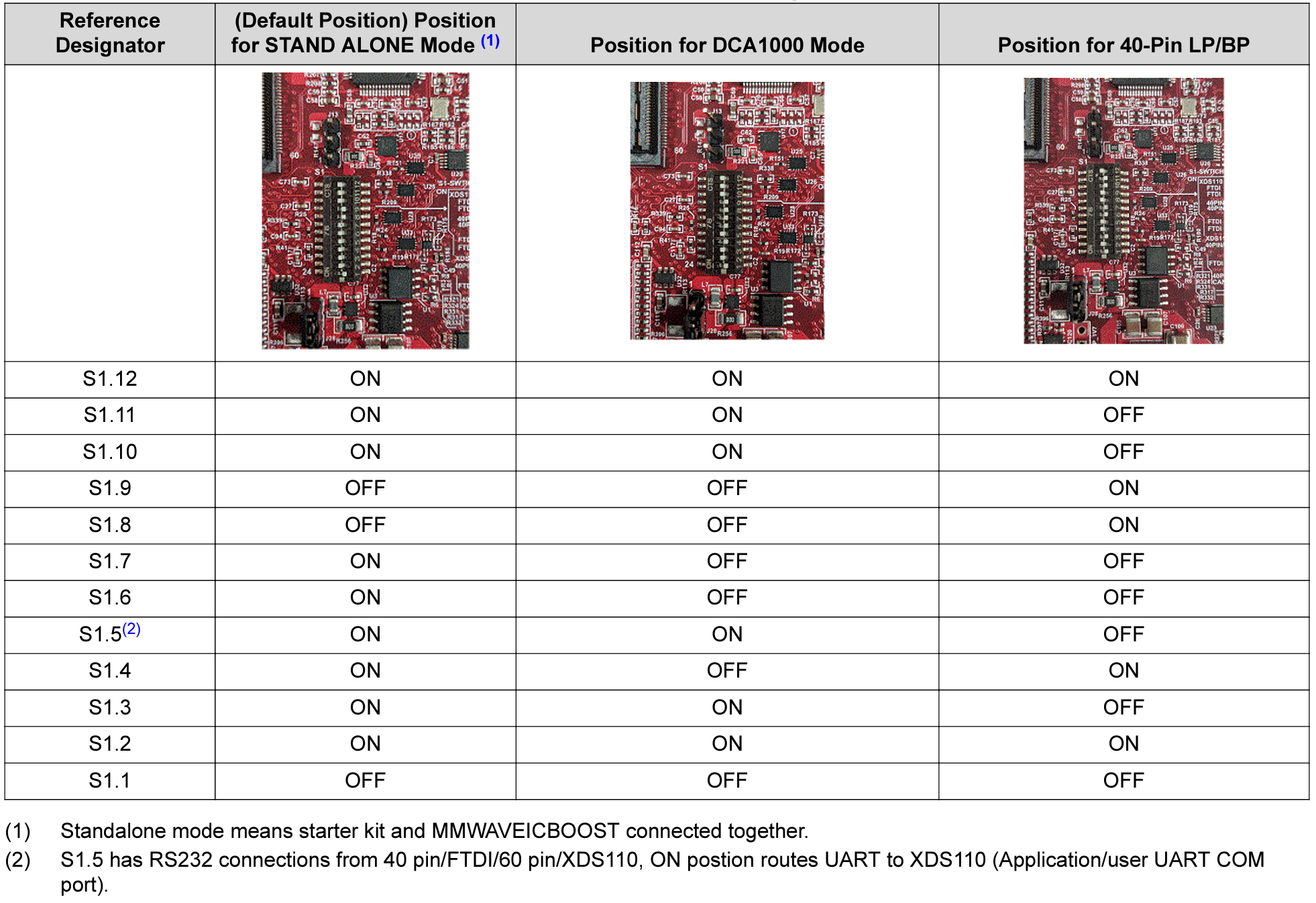}
        \caption{Switch Settings for MMWAVEICBOOST}
        \label{fig:MMWAVEICBOOST_switches}
    \end{figure}
    
    \item[3)] Connect to Microcontroller (MCU)
    \begin{enumerate}
        \item[a)] Follow \href{https://e2e.ti.com/support/sensors-group/sensors/f/sensors-forum/998616/iwr6843isk-custom-binary-for-cli-lvds-hw-trigger/3692785}{Josiah's E2E post}:
        \item[b)] On page 8 of PROC074B(001)\_Sch.pdf and page 19 of here, 40PIN\_SYNC\_IN is pin 9 of J5 \textbf{(IMPORTANT)}. See picture \href{https://e2e.ti.com/support/sensors-group/sensors/f/sensors-forum/972613/iwr6843isk-hardware-trigger}{here} for pin 9 of J5 input. Ground is pin 4 of J5 or pin 2 of J6. 
    \end{enumerate}
    \item[4)] Connect to the PC over USB on XDS110\_USB/J1 and attach IWR6843 as shown on page 16 \href{https://www.ti.com/lit/ug/swru546d/swru546d.pdf?ts=1620300878197}{here}.
    \item[5)] Connect 5V/3A power.
\end{enumerate}

\subsection{IWR6843ISK Hardware Configuration}
\label{6843:IWR6843ISK}
\begin{enumerate}
    \item[1)] Switch Settings
    \begin{enumerate}
        \item[a)] From step 1 \here{https://dev.ti.com/tirex/explore/content/mmwave_industrial_toolbox_4_7_0/labs/common/docs/hardware_setup/hw_setup_mmwaveicboost_mode_flashing.html} or page 45 \here{https://www.ti.com/lit/ug/swru546d/swru546d.pdf?ts=1620300878197}, use the switch settings on S1, as shown in Fig. \ref{fig:IWR6843ISK_switches}.
    \end{enumerate}
    
    \begin{figure}[th]
        \centering
        \includegraphics[width=0.65\textwidth]{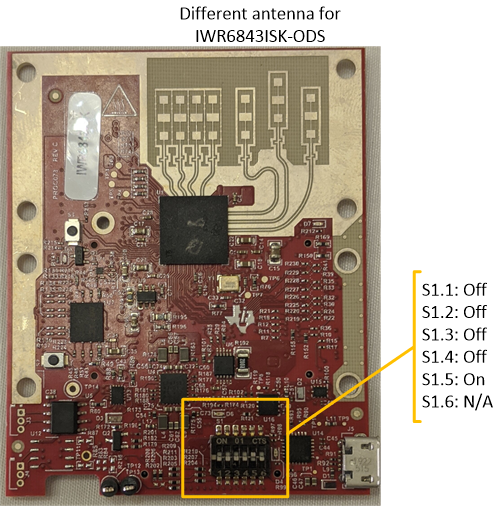}
        \caption{Switch Settings for IWR6843ISK}
        \label{fig:IWR6843ISK_switches}
    \end{figure}
    
    \item[2)] Connect to MMWAVEICEBOOST, as shown on page 16 \here{https://www.ti.com/lit/ug/swru546d/swru546d.pdf?ts=1620300878197}.
\end{enumerate}

\section{Software Set Up}

\subsection{Flash the SDK Demo Binary to the MMWAVEICBOOST (See \href{http://software-dl.ti.com/ra-processors/esd/MMWAVE-SDK/latest/exports/mmwave_sdk_user_guide.pdf}{TI SDK 3.5.0.4 User Guide} Section 4.2)}
\begin{enumerate}
    \item[1)] Set the device to Flash Programming Mode by bridging SOP0 and SOP2 as shown in step 5 \here{https://dev.ti.com/tirex/explore/content/mmwave\_industrial_toolbox\_4\_7\_0/labs/common/docs/hardware\_setup/hw\_setup_mmwaveicboost\_mode\_flashing.html}. 
    \item[2)] Power cycle the MMWAVEICBOOST.
    \item[3)] Once the device is properly connected to the PC, download the demo firmware using Unifash. (Typically under the path: \\ ``C:/ti/mmwave\_sdk\_03\_05\_00\_04/packages/ti/demo/xwr68xx/mmw'').
    \item[4)] Once the download is complete, set the device to Functional Mode by bridging only SOP0 and remove the bridge on SOP2. 
    \item[5)] Power cycle the MMWAVEICBOOST. 
\end{enumerate}

\subsection{Set Up the DCA1000EVM on Proper IP Address (See the \href{https://software-dl.ti.com/ra-processors/esd/MMWAVE-STUDIO/latest/exports/DCA1000_Quick_Start_Guide.pdf}{DCA1000EVM Quick Start Guide})}
\label{general:dca1000evm_pc_connection}
\begin{enumerate}
    \item[1)] Once the device is properly connected to the PC, open the start menu and search ``View Network Connections,'' as shown in Fig. \ref{fig:view_network_connections}. 
    
    \begin{figure}[th]
        \centering
        \includegraphics[width=0.85\textwidth]{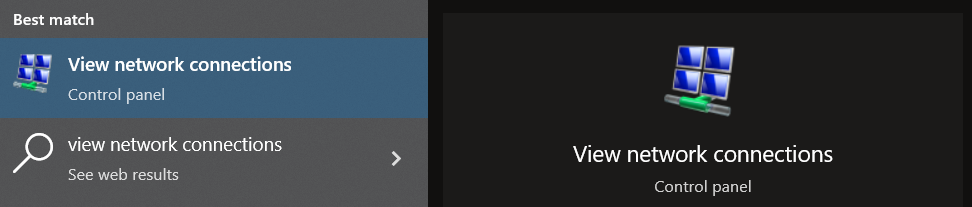}
        \caption{Example of Windows start menu search result for viewing network connections}
        \label{fig:view_network_connections}
    \end{figure}
    
    \item[2)] Inside the ``Network Connections'' of Control Panel, right click on the Ethernet port of choice and select ``Properties.''
    \item[3)] ``Local Area Connection Properties'' window will open. Right click on ``Internet Protocol Version 4 (TCP/IPv4).''
    \item[4)] ``Internet Protocol Version 4 (TCP/IPc4)'' window will open. Set the IP address field to 192.168.33.30, or the desired IP address if different. (See Chapter \ref{ch:dca1000evm_IP_address} on changing the IP address of the DCA1000EVM using the DCA1000 CLI Utility, which is necessary for a dual radar setup).  
    \item[5)] The Subnet mask field can remain the default 255.255.255.0. 
    \item[6)] Press ``OK'' on all the windows and you can close ``Network Connections.''
\end{enumerate}

\subsection{Open the Dual Radar SAR Controller MATLAB Application}
\begin{enumerate}
    \item[1)] Open MATLAB and navigate to the folder containing ``dual\_radar\_gui.mlapp'' and open that file. The MATLAB App Designer window will open the application.
    \item[2)] Press ``Run'' at the top of the page. 
    \item[3)] The app will open and all the indicators will be red as none of the devices are connected to the application or configured. 
    \item[4)] Assuming you have installed mmWave Studio 2.1.1.0 to the typical location, it will open normally. Otherwise, it will ask you to find the installation location of mmWave Studio 2.1.1.0. 
    \item[5)] Press ``Connect Radar 1.''
    \item[6)] A window will appear asking to select a serial COM port. Select the COM port corresponding to the entry in device manager labeled ``XDS110 Class Application/User UART.'' 
    \item[7)] Under the DCA 1 Configuration panel, enter the correct System and DCA IP addresses and ports for the radar. Press ``Prepare DCA 1.'' If a Windows Defender Firewall window opens, give access to private networks and public networks (check both boxes) and select ``Allow access.'' (The optional drop down menu in the bottom right corner of the DCA 1 Configuration panel is used to select from a set of defaults with different patterns and is straightforward). 
    \item[8)] Under the Radar 1 (60 GHz) Configuration panel, enter the desired chirp parameters and press ``Configure Radar 1.'' The Radar 1 Configuration lamp should change to a green color to indicate the configuration was successful. If the configuration stalls and the lamp is stuck on yellow, this indicates a communication error between the PC and radar, and the radar must be disconnected and power cycled. Then, attempt these steps again. 
    \item[9)] Under the Radar 1 (60 GHz) Configuration panel, press ``Start'' to start the 60 GHz radar. 
    \begin{enumerate}
        \item[a)] The DCA1000EVM will start waiting for data over LVDS. 
        \item[b)] The MMWAVEICBOOST and IWR6843ISK will wait for HW trigger from MCU. 
        \item[c)] If everything is working properly at this point, the D7 LED on the IWR6843ISK and the DS2 LED on the MMWAVEICBOOST will turn on. 
    \end{enumerate}
    \item[10)] Start the MCU sending pulses. If everything is working properly the DATA\_TRAIN\_PRG LED on the DCA1000EVM will be flashing while the radar is triggered. 
    \item[11)] Once the MCU stops sending pulses, the radar and DCA1000EVM will automatically stop the capture. (Pressing the ``stop'' button is only necessary if you set No of Frames to 0, for infinite capture mode, with HW trigger disabled, indicating the SW trigger will be used). 
\end{enumerate}

\chapter{Using Hardware Trigger with TI 1642 77 GHz Radar}
\label{ch:hw_trigger_1642}

The following is a simple demo for enabling the hardware (HW) trigger with a Texas Instruments xWR1642BOOST 77 GHz radar in conjunction with the Dual Radar SAR Controller application. 

\section{Hardware Requirements}
\begin{enumerate}
    \item[1)] \href{https://www.ti.com/tool/DCA1000EVM}{DCA1000EVM} with modifications (see Section \ref{1642:DCA}). 
    \item[3)] \href{https://www.ti.com/tool/IWR1642BOOST}{xWR1642BOOST} with modifications (see Section \ref{1642:xWR1642BOOST}). 
    \item[4)] ESP32 microcontroller to send HW trigger pulses.
\end{enumerate}

\section{Software Requirements}
\begin{enumerate}
    \item[1)] \href{https://www.ti.com/tool/MMWAVE-SDK}{TI mmWave SDK 3.5.0.4}
    \item[2)] \href{https://www.ti.com/tool/UNIFLASH}{TI Uniflash}
    \item[3)] \href{https://software-dl.ti.com/ra-processors/esd/MMWAVE-STUDIO/latest/index_FDS.html}{TI mmWave Studio 2.1.1.0}
    \item[4)] Dual Radar SAR Controller software (requires MATLAB 2021b (recommended, may be compatible with earlier/later releases)).
\end{enumerate}

\section{Hardware Set Up}

\subsection{DCA1000EVM Hardware Configuration}
\label{1642:DCA}
\begin{enumerate}
    \item[1)] Remove R120 \href{https://e2e.ti.com/support/sensors-group/sensors/f/sensors-forum/710632/awr1243-dca1000-and-1243-hardware-trigger-sts_no_lvds_data}{(suggested here)}. 
    \item[2)] Connect to the MMWAVEICBOOST with 60 pin connector, as shown on \href{https://www.ti.com/lit/ug/swru546d/swru546d.pdf?ts=1620300878197}{page 16 here}.
    \item[3)] Connect to the PC over USB on the RADAR/FTDI connnector (J1 on DCA1000EVM) and over Ethernet.
    \item[4)] Connect 5V/3A power.
\end{enumerate}

\subsection{xWR1642BOOST Hardware Configuration}
\label{1642:xWR1642BOOST}
\begin{enumerate}
    \item[1)] Connect to Microcontroller (MCU)
    \begin{enumerate}
        \item[a)] Follow \href{https://e2e.ti.com/support/sensors-group/sensors/f/sensors-forum/998616/iwr6843isk-custom-binary-for-cli-lvds-hw-trigger/3692785}{Josiah's E2E post}:
        \item[b)] For IWR1642Boost, on page 10 of PROC049B(002)\_Sch.pdf, AR\_SYNC\_IN is pin 9 of J6 \textbf{(IMPORTANT)}.
        \item[c)] For AWR1642Boost, On page 10 of PROC011C(002)\_Sch.pdf, AR\_SYNC\_IN is pin 9 of J6 \textbf{(IMPORTANT)}.
        \item[d)] For newest revisions, the R165 resistor is already shorted, enabling the HW trigger input.
        \item[e)] Ground is pin 4 of J6 or pin 2 of J5. 
    \end{enumerate}
    \item[2)] Connect to the PC over. 
    \item[3)] Connect 5V/3A power.
\end{enumerate}

\section{Software Set Up}

\subsection{Flash the SDK Demo Binary to the xWR1642BOOST (See \href{http://software-dl.ti.com/ra-processors/esd/MMWAVE-SDK/latest/exports/mmwave_sdk_user_guide.pdf}{TI SDK 3.5.0.4 User Guide} Section 4.2)}
\begin{enumerate}
    \item[1)] Set the device to Flash Programming Mode by bridging SOP0 and SOP2. Similar bridging is shown on the MMWAVEICBOOST in step 5  \here{https://dev.ti.com/tirex/explore/content/mmwave\_industrial_toolbox\_4\_7\_0/labs/common/docs/hardware\_setup/hw\_setup_mmwaveicboost\_mode\_flashing.html}. 
    \item[2)] Power cycle the xWR1642BOOST.
    \item[3)] Once the device is properly connected to the PC, download the demo firmware using Unifash. (Typically under the path: \\ ``C:/ti/mmwave\_sdk\_03\_05\_00\_04/packages/ti/demo/xwr68xx/mmw'').
    \item[4)] Once the download is complete, set the device to Functional Mode by bridging only SOP0 and remove the bridge on SOP2. 
    \item[5)] Power cycle the xWR1642BOOST. 
\end{enumerate}

\subsection{Set Up the DCA1000EVM on Proper IP Address (See the \href{https://software-dl.ti.com/ra-processors/esd/MMWAVE-STUDIO/latest/exports/DCA1000_Quick_Start_Guide.pdf}{DCA1000EVM Quick Start Guide})}
\begin{enumerate}
    \item[1)] Once the device is properly connected to the PC, open the start menu and search ``View Network Connections,'' as shown in Fig. \ref{fig:view_network_connections}. 
    \item[2)] Inside the ``Network Connections'' of Control Panel, right click on the Ethernet port of choice and select ``Properties.''
    \item[3)] ``Local Area Connection Properties'' window will open. Right click on ``Internet Protocol Version 4 (TCP/IPv4).''
    \item[4)] ``Internet Protocol Version 4 (TCP/IPc4)'' window will open. Set the IP address field to 192.168.33.30, or the desired IP address if different. (See Chapter \ref{ch:dca1000evm_IP_address} on changing the IP address of the DCA1000EVM using the DCA1000 CLI Utility, which is necessary for a dual radar setup).  
    \item[5)] The Subnet mask field can remain the default 255.255.255.0. 
    \item[6)] Press ``OK'' on all the windows and you can close ``Network Connections.''
\end{enumerate}

\subsection{Open the Dual Radar SAR Controller MATLAB Application}
\begin{enumerate}
    \item[1)] Open MATLAB and navigate to the folder containing ``dual\_radar\_gui.mlapp'' and open that file. The MATLAB App Designer window will open the application.
    \item[2)] Press ``Run'' at the top of the page. 
    \item[3)] The app will open and all the indicators will be red as none of the devices are connected to the application or configured. 
    \item[4)] Assuming you have installed mmWave Studio 2.1.1.0 to the typical location, it will open normally. Otherwise, it will ask you to find the installation location of mmWave Studio 2.1.1.0. 
    \item[5)] Press ``Connect Radar 2.''
    \item[6)] A window will appear asking to select a serial COM port. Select the COM port corresponding to the entry in device manager labeled ``XDS110 Class Application/User UART.'' 
    \item[7)] Under the DCA 2 Configuration panel, enter the correct System and DCA IP addresses and ports for the radar. Press ``Prepare DCA 2.'' If a Windows Defender Firewall window opens, give access to private networks and public networks (check both boxes) and select ``Allow access.'' (The optional drop down menu in the bottom right corner of the DCA 2 Configuration panel is used to select from a set of defaults with different patterns and is straightforward). 
    \item[8)] Under the Radar 2 (77 GHz) Configuration panel, enter the desired chirp parameters and press ``Configure Radar 2.'' The Radar 2 Configuration lamp should change to a green color to indicate the configuration was successful. If the configuration stalls and the lamp is stuck on yellow, this indicates a communication error between the PC and radar, and the radar must be disconnected and power cycled. Then, attempt these steps again. 
    \item[9)] Under the Radar 2 (77 GHz) Configuration panel, press ``Start'' to start the 77 GHz radar. 
    \begin{enumerate}
        \item[a)] The DCA1000EVM will start waiting for data over LVDS. 
        \item[b)] The MMWAVEICBOOST and IWR6843ISK will wait for HW trigger from MCU. 
        \item[c)] If everything is working properly at this point, the D7 LED on the IWR6843ISK and the DS2 LED on the MMWAVEICBOOST will turn on. 
    \end{enumerate}
    \item[10)] Start the MCU sending pulses. If everything is working properly the DATA\_TRAIN\_PRG LED on the DCA1000EVM will be flashing while the radar is triggered. 
    \item[11)] Once the MCU stops sending pulses, the radar and DCA1000EVM will automatically stop the capture. (Pressing the ``stop'' button is only necessary if you set No of Frames to 0, for infinite capture mode, with HW trigger disabled, indicating the SW trigger will be used). 
\end{enumerate}

\appendix 

\chapter{Preliminaries of FMCW Signaling}
\label{ch:fmcw_signal_model}

\section{Monostatic FMCW Signal Model}

\subsection{FMCW Chirp}
In this section we derive the simple signal model for a monostatic (Tx and Rx are assumed to be at the same position in space) scenario with a single, ideal point scatterer (reflector). 
By definition, an FMCW signal, called an FMCW chirp, has a frequency linearly increasing with time. 
We can express this relationship as
\begin{equation}
    \label{eq:f_vs_time}
    f(t) \triangleq f_0 + K t, \quad 0 \leq t \leq T,
\end{equation}
where $f$ is the instantaneous frequency as a function of $t$, $f_0$ is the start frequency (frequency at time $t = 0$), $K$ is the slope of the chirp, the bandwidth covered by the chirp is given by $B = KT$, and $t$ is called the ``fast time'' variable.

The definition of instantaneous frequency is given by
\begin{equation}
    \label{eq:instantaneous_frequency}
    f(t) \triangleq \frac{1}{2 \pi} \frac{\partial}{\partial t} \phi(t),
\end{equation}
where $\phi(t)$ is the phase of the signal, containing the frequency content, e.g.,
\begin{equation}
    \label{eq:m(t)}
    m(t) \triangleq e^{j\phi(t)}.
\end{equation}
Hence, the phase term $\phi(t)$ can be expressed as
\begin{equation}
    \label{eq:phi}
    \phi(t) = 2 \pi \int_0^t f(t') dt'.
\end{equation}

Substituting (\ref{eq:f_vs_time}) into (\ref{eq:phi}) yields
\begin{align}
    \label{eq:phi2}
    \begin{split}
        \phi(t) &= 2 \pi \int_0^t(f_0 + K t') dt', \\
        &= 2 \pi \left[ f_0 t' + 0.5 K (t')^2 \right]_0^t, \\
        &= 2 \pi (f_0t + 0.5 K t^2).
    \end{split}
\end{align}

\subsection{FMCW Beat Signal}

Thus, by the definition of the FMCW chirp as a sinusoidal signal whose frequency linearly increases with time, we can derive the phase of the signal and express the transmitted FMCW pulse by substituting (\ref{eq:phi2}) into (\ref{eq:m(t)}) as
\begin{equation}
    \label{eq:transmitted_signal}
    m(t) = e^{j\phi(t)} = e^{j 2 \pi (f_0 t + 0.5 K t^2)}, \quad 0 \leq t \leq T,
\end{equation}
where $T$ is the duration of the FMCW pulse. 

Assuming the monostatic radar antenna element is located at $(x',y',z')$ and the point scatterer is located at $(x_0,y_0,z_0)$, the distance between the radar and point target can be expressed as 
\begin{equation}
    \label{eq:R_0}
    R_0 = \sqrt{(x_0-x')^2 + (y_0-y')^2 + (z_0-z')^2}.
\end{equation}
The FMCW pulse, expressed in (\ref{eq:transmitted_signal}) is transmitted from the antenna, propagates through space traveling a distance of $R_0$ to the point scatter, reflects from the point scatterer, travels another $R_0$ back to the radar. 
The round-trip time delay required for this propagation is given by  
\begin{equation}
    \label{eq:tau_0}
    \tau_0 = \frac{2R_0}{c},
\end{equation}
where $c$ is the speed of light.

As a result, the signal received at the radar is a time-delayed and scaled version of the transmitted signal as
\begin{align}
    \label{eq:received_signal}
    \begin{split}
        \hat{s}(t) &= \frac{\sigma}{R_0^2} m(t - \tau_0), \\
        &= \frac{\sigma}{R_0^2} e^{j 2 \pi (f_0(t - \tau_0) + 0.5 K (t - \tau_0)^2)}, \\
        &= \frac{\sigma}{R_0^2} e^{j 2 \pi (f_0t - f_0\tau_0 + 0.5Kt^2 - K \tau_0 t + 0.5 K \tau_0^2)}, \\
        &= \frac{\sigma}{R_0^2} e^{j 2 \pi (f_0t + 0.5Kt^2 - f_0 \tau_0 - K \tau_0 t + 0.5 K \tau_0^2)}, \\
        &= \frac{\sigma}{R_0^2} \underbrace{e^{j 2 \pi (f_0t + 0.5Kt^2)}}_{m(t)} e^{-j 2 \pi (f_0 \tau_0 + K \tau_0 t - 0.5 K \tau_0^2)},
    \end{split}
\end{align}
where $\sigma$ is known as the ``reflectivity'' of the scatterer (how reflective the point target is) and the $1/R_0^2$ term is the round-trip path loss or amplitude decay (the intuition here is that farther targets give weaker reflections). Note that the received signal $\hat{s}(t)$ contains a factor which is the transmitted signal $m(t)$.

The next step in the signal chain is known as ``dechirping'' and removes this factor of $m(t)$ by multiplying the conjugate of the received signal by the transmitted signal. 
After dechirping, the signal is known as the IF signal or beat signal. 
The beat signal can be expressed as
\begin{align}
    \label{eq:s(t)}
    \begin{split}
        s(t) &= m(t) \hat{s}^*(t), \\
        &= e^{j 2 \pi (f_0 t + 0.5 K t^2)} \frac{\sigma}{R_0^2}e^{-j 2 \pi (f_0t + 0.5Kt^2)} e^{j 2 \pi (f_0 \tau_0 + K \tau_0 t - 0.5 K \tau_0^2)}, \\
        &= \frac{\sigma}{R_0^2} e^{j 2 \pi (f_0 t + 0.5 K t^2) -j 2 \pi (f_0t + 0.5Kt^2)} e^{j 2 \pi (f_0 \tau_0 + K \tau_0 t - 0.5 K \tau_0^2)}, \\
        &= \frac{\sigma}{R_0^2} e^{j 2 \pi (f_0 \tau_0 + K \tau_0 t - 0.5 K \tau_0^2)}.
    \end{split}
\end{align}

In radar literature, it is common practice to express the beat signal as a function of $k$ rather than $t$, where $k(t)$ is the instantaneous wavenumber corresponding to the instantaneous frequency $f(t)$ given by 
\begin{equation}
    \label{eq:k(t)}
    k(t) \triangleq \frac{2 \pi}{c}f(t).
\end{equation}
Substituting (\ref{eq:f_vs_time}) into (\ref{eq:k(t)}) yields
\begin{equation}
    \label{eq:k_vs_time}
    k(t) = \frac{2 \pi}{c} (f_0 + K t), \quad 0 \leq t \leq T.
\end{equation}

Hence, let us express $s(t)$ as a function of $k$ by rewriting (\ref{eq:s(t)}) in terms of (\ref{eq:k_vs_time}). 
Note that for short distances $\tau_0^2$ is negligible and the last term in (\ref{eq:s(t)}) can be ignored.
Substituting (\ref{eq:tau_0}) into (\ref{eq:s(t)}) and recalling the definitions in (\ref{eq:f_vs_time}) and (\ref{eq:k(t)}) yields
\begin{align}
    \label{eq:beat_signal_derivation}
    \begin{split}
        s(t) &= \frac{\sigma}{R_0^2} e^{j 2 \pi (f_0 \tau_0 + K \tau_0 t)}, \\
        &= \frac{\sigma}{R_0^2} e^{j 2 \pi \tau_0 (f_0 + K t)}, \\
        &= \frac{\sigma}{R_0^2} e^{j 2 \pi \frac{2R_0}{c} (f_0 + K t)}, \\
        &= \frac{\sigma}{R_0^2} e^{j 2 \pi \frac{2R_0}{c} f(t)}, \\
        &= \frac{\sigma}{R_0^2} e^{j 2 R_0 \frac{2 \pi}{c} f(t)}, \\
        s(k) &= \frac{\sigma}{R_0^2} e^{j 2 R_0 k}, \quad 0 \leq t \leq T.
    \end{split}
\end{align}

More commonly, the beat signal is expressed as 
\begin{equation}
    \label{eq:beat_signal}
    s(k) = \frac{\sigma}{R_0^2} e^{j2 k R_0}
\end{equation}

We have derived a compact representation of the FMCW beat signal, $s(k)$. 
It is clear from (\ref{eq:beat_signal}) that the frequency of the beat signal corresponds directly with the radial distance, known as the ``range'', $R_0$. 
For this single point scatter case, the radar beat signal $s(k)$ is a single tone sinusoid whose frequency corresponds with $R_0$.
Hence, the Fourier transform of $s(k)$ would have a single peak at a position corresponding to the range $R_0$.

This section detailed the derivation of the radar beat signal. 
However, now that the beat signal has been derived, it can be applied simply by the definition of $s(k)$ in (\ref{eq:beat_signal}) without needing to go through all steps every time.

\subsection{Multiple Targets}
Suppose $N$ point scatterers are in the radar FOV such that the $n$-th point scatterer is located at $(x_n,y_n,z_n)$ and has reflectivity $\sigma_n$. 
In this case, the transmit signal is the same, but the received signal is now a sum (by superposition) of the received signals from each of the point scatterers. 
As a result, the radar beat signal can be written as
\begin{equation}
    \label{eq:multiple_discrete_targets}
    s(k) = \sum_{n = 1}^{N} \frac{\sigma_n}{R_n^2} e^{j2 k R_n}, \quad 0 \leq t \leq T,
\end{equation}
which is clearly a sum of sinusoidal signals whose frequencies depend on the distances $R_n$.
Hence, the Fourier transform of (\ref{eq:multiple_discrete_targets}) would result in multiple peaks such that the location of the $n$-th peak corresponds with the distance $R_n$.

Alternatively, if we model the target as a continuous set of point targets, rather than a discrete set as in (\ref{eq:multiple_discrete_targets}), where the target is located in a volume $V$ inside $(x,y,z)$ space, we express the reflectivity as a continuous function $p(x,y,z)$ and the summation expressed in (\ref{eq:multiple_discrete_targets}) becomes an integral as
\begin{equation}
    \label{eq:continuous_targets}
    s(k) = \iiint_V \frac{p(x,y,z)}{R^2}e^{j2kR} dx dy dz,
\end{equation}
where
\begin{equation}
    \label{eq:R}
    R = \sqrt{(x-x')^2 + (y-y')^2 + (z-z')^2},
\end{equation}
recall the position of the antenna is $(x',y',z')$. 

\section{Range Resolution and Maximum Resolvable Range}
\label{sec:range_resolution}
Knowing the target range information is present in the frequency of the beat signal (\ref{eq:beat_signal}), we examine the minimum resolvable distance between two targets in the scene and the maximum resolvable range.

The simplest method of identifying the frequency content of a given signal is the Fourier transform. According to Fourier transform theory, frequency components can be resolved if separated by at least the reciprocal of the observation time ($T$) as

\begin{equation}
	\label{eq:deltahat_1}
	\Delta {f} > \frac{1}{T},
\end{equation}
where $\Delta {f}$ is the change in frequency of the beat signal

Considering a monostatic radar with two targets separated by a distance $\Delta R$, the difference in frequency between the two beat signals, $\Delta {f}$, is expressed as

\begin{equation}
	\label{eq:deltahat_2}
	\Delta {f} = \frac{2K \Delta R}{c}.
\end{equation}

Combining (\ref{eq:deltahat_1}) with (\ref{eq:deltahat_2}) yields

\begin{gather}
	\label{eq:range_resolution}
	\frac{2K \Delta R}{c} > \frac{1}{T} \Rightarrow \Delta R > \frac{c}{2KT}, \\
	\Delta R > \frac{c}{2B}.
\end{gather}

This minimum resolvable distance is commonly known as the radar range resolution for ultra-wide-band (UWB) systems. 
For an automotive radar with a several GHz of bandwidth, the range resolution will be on the order of centimeters. 
For example, a common $4$ GHz chirp yields a range resolution of 3.75 cm. 

Similar analysis can be performed to compute the maximum resolvable range of a given set of chirp parameters. 
The maximum range $R_{max}$ yields an IF frequency of ${f}_{max} = 2KR_{max}/c$. 
Assuming a complex baseband IF signal, the maximum frequency is limited by the ADC sampling rate $f_S$ as

\begin{equation}
	f_S > \frac{2KR_{max}}{c}.
\end{equation}

The expression above can be rearranged yielding the maximum range as a function of the chirp slope, the sampling frequency, and the speed of light as

\begin{equation}
	\label{eq:maximum_range}
	R_{max} < \frac{f_S c}{2K}.
\end{equation}

\chapter{Spatial Fourier Transform and Relations}
\label{app:spatial_ft}
Neglecting amplitude terms, the \mbox{1-D}, \mbox{2-D} and \mbox{3-D} spatial Fourier transforms can be defined as \cite{yanik2019sparse}
\begin{equation}
\label{eq:ft1D}
    \text{FT}_{\text{1D}}^{(u)} \left[ s(u) \right] = S(k_u) = \int s(u) e^{-jk_u u}du,
\end{equation}

\begin{equation}
\label{eq:ft2D}
    \text{FT}_{\text{2D}}^{(u,v)} \left[ s(u,v) \right] = S(k_u,k_v) = \iint s(u,v) e^{-j(k_u u + k_v v)}du dv,
\end{equation}

\begin{equation}
\label{eq:ft3D}
    \text{FT}_{\text{3D}}^{(u,v,w)} \left[ s(u,v,w) \right] = S(k_u,k_v,k_w) = \iiint s(u,v,w) e^{-j(k_u u + k_v v + k_w w)}du dv dw.
\end{equation}

Similarly, the \mbox{1-D}, \mbox{2-D} and \mbox{3-D} inverse spatial Fourier transforms can be expressed as
\begin{equation}
\label{eq:ift1D}
    \text{IFT}_{\text{1D}}^{(k_u)} \left[ S(k_u) \right] = s(u) = \int S(k_u) e^{jk_u u}du,
\end{equation}

\begin{equation}
\label{eq:ift2D}
    \text{IFT}_{\text{2D}}^{(k_u,k_v)} \left[ S(k_u,k_v) \right] = s(u,v) = \iint S(k_u,k_v) e^{j(k_u u + k_v v)}du dv,
\end{equation}

\begin{equation}
\label{eq:ift3D}
    \text{IFT}_{\text{3D}}^{(k_u,k_v,k_w)} \left[ S(k_u,k_v,k_w) \right] = s(u,v,w) = \iiint S(k_u,k_v,k_w) e^{j(k_u u + k_v v + k_w w)}du dv dw.
\end{equation}

A shift in the spatial domain results in a corresponding phase shift in the spatial spectral domain. The example given here is in the \mbox{3-D} spatial domain but holds true for the \mbox{2-D} and \mbox{1-D} cases also:
\begin{equation}
\label{eq:ft_shiftForward}
    \text{FT}_{\text{3D}}^{(u,v,w)} \left[ s(u - u_0,v - v_0,w - w_0) \right] = e^{-j(k_u u_0 + k_v v_0 + k_w w_0)}S(k_u,k_v,k_w).
\end{equation}

Similarly, a shift in the spatial spectral domain results in a phase shift in the spatial domain:

\begin{equation}
\label{eq:ft_shiftInverse}
    \text{IFT}_{\text{3D}}^{(k_u,k_v,k_w)} \left[ S(k_u - k^u_0,k_v - k^v_0,k_w - k^w_0) \right] = e^{j(k_u u_0 + k_v v_0 + k_w w_0)}s(u,v,w).
\end{equation}

These spatial Fourier transform definitions and relations are useful in deriving the reconstruction algorithms discussed in the subsequent appendices.
\chapter{Method of Stationary Phase}
\label{ch:msp}
In this chapter, we discuss the important topic of the Method of Stationary Phase (MSP) required for many of the subsequent imaging algorithms detailed in following chapters. 
A highly recommended exercise to the reader is to follow the example derivation in Section \ref{subsec:msp_example} closely and derive the helpful approximations given in (\ref{eq:mspLinear})--(\ref{eq:mspCylindrical}) showing every step.
Additionally, a careful review of the spatial Fourier relationships in Appendix \ref{app:spatial_ft} is recommended. 

As discussed in \cite{cook2012radar,papoulis1968systems,mcclure2016multidimensional}, the general form of the \mbox{$n$-dimensional} Method of Stationary Phase (MSP) can be expressed as the following. 
A rigorous mathematical perspective is offered in \cite{mcclure2016multidimensional}, whereas our discussion does not comprehensively address the underlying assumptions and constraints.
Rather, this section is meant to serve as a resource to researchers and engineers to apply the results of the MSP approximation to near-field spherical wave decomposition problems.

Given an oscillatory integral with a wide phase variation of the form
\begin{equation}
    I(\mathbf{x}) = \int g(\mathbf{x}) e^{jf(\mathbf{x})}d\mathbf{x}, \quad \mathbf{x} \in \mathbb{R}^n,
\end{equation}
where $f(\mathbf{x})$ is assumed to be twice-continuously differentiable, the major contribution to the quantity $I(\mathbf{x})$ is from the stationary points, $\mathbf{x}_0$, which are calculated by
\begin{equation}
    \nabla f(\mathbf{x}) |_{\mathbf{x} = \mathbf{x}_0} = 0
\end{equation}

Thus, the integral can be approximated by
\begin{equation}
    I(\mathbf{x}) \approx \frac{g(\mathbf{x}_0)}{\sqrt{\det\mathbf{A}}} e^{j f(\mathbf{x}_0)},
\end{equation}
where $\mathbf{x}_0$ is the set of stationary points and $\mathbf{A}$ is the Hessian matrix of $f(\mathbf{x})$ evaluated at $\mathbf{x}_0$ and defined as
\begin{equation}
    \mathbf{A} = \left( \frac{\partial^2f(\mathbf{x})}{\partial x_i \partial x_j} \right) \biggr\rvert_{\mathbf{x} = \mathbf{x}_0}.
\end{equation}

For the derivations required in this article, we can limit $n$ to 1 or 2 dimensions. 
\section{\mbox{1-D} Method of Stationary Phase}
\label{subsec:msp_1D}
The \mbox{1-D} MSP can be written as the following. The following integral with the same assumptions as the MSP,
\begin{equation}
    I(u) = \int g(u) e^{j f(u)} du,
\end{equation}
can be approximated as 
\begin{equation}
\label{eq:msp_1D}
    I(u) \approx \frac{g(u_0)}{\sqrt{f''(u_0)}} e^{j f(u_0)},
\end{equation}
where $u_0$ is the stationary point calculated by
\begin{equation}
\label{eq:msp_1D_delf}
    \frac{\partial f(u)}{\partial u} \biggr\rvert_{u = u_0} = 0,
\end{equation}
and $f''(u_0)$ is the second derivative of $f(u)$ evaluated at the stationary point $u_0$.

\section{\mbox{2-D} Method of Stationary Phase}
\label{subsec:msp_2D}
Similarly, for the \mbox{2-D} case, the integral,
\begin{equation}
    I(u,v) = \iint g(u,v) e^{j f(u,v)} du dv,
\end{equation}
can be approximated by
\begin{equation}
\label{eq:msp_2D}
    I(u,v) \approx \frac{g(u_0,v_0)}{\sqrt{f_{uu}f_{vv}-f^2_{uv}}} e^{j f(u_0,v_0)},
\end{equation}
where the stationary points $u_0$, $v_0$ are calculated by
\begin{gather}
\label{eq:msp_2D_delf}
    \frac{\partial f(u,v)}{\partial u} \biggr\rvert_{(u=u_0,v=v_0)} = 0, \\
    \frac{\partial f(u,v)}{\partial v} \biggr\rvert_{(u=u_0,v=v_0)} = 0,
\end{gather}
and $f_{uu}$, $f_{vv}$, $f_{uv}$ are the second partial derivatives of $f(u,v)$ evaluated at the stationary points. 

\section{Useful MSP Identities}
\label{subsec:useful_msp}
Using the aforementioned method for the \mbox{1-D} and \mbox{2-D} cases, the MSP is applied to several integrals and the corresponding approximations are provided for reference in this section.

We will demonstrate the steps for the approximation below which have been applied to the other spherical wavefronts to yield the relations in  (\ref{eq:mspLinear})--(\ref{eq:mspCylindrical}). 

\subsection{Example MSP Derivation}
\label{subsec:msp_example}
We consider the linear array case with a monostatic single antenna array being scanned along the $x$-axis at the positions labeled $x'$. Further, we consider a \mbox{1-D} target at some line $z_0$ in the $x$-$z$ plane, where the $x$ and $x'$ coordinate systems are coincident. Thus, the radar beat signal can be modeled, neglecting path loss, as
\begin{equation}
\label{eq:msp_ex1}
    s(x',k) = \int p(x) e^{j2kR} dx,
\end{equation}
where $R$ is the radial distance from each of the antenna locations $(x',0)$ to the target locations $(x,z)$ and is expressed as
\begin{equation}
    R = \sqrt{(x-x')^2 + z_0^2}.
\end{equation}

It is desired to approximate the spherical wavefront term in (\ref{eq:msp_ex1}), $e^{j2kR}$, as a more tractable expression. Thus, the MSP is exploited. For generality, the following substitutions are made $u = x'$, $r = 2k$, $w = z_0$. The \mbox{1-D} spatial Fourier transform (\ref{eq:ft1D}) is performed over the $u$ dimension of the spherical wave term and the spatial translation property (\ref{eq:ft_shiftForward}) is applied as
\begin{equation}
\label{eq:msp_ex2}
    \text{FT}_{\text{1D}}^{(u)} \left[ e^{jr\sqrt{(x-u)^2 + w^2}} \right] = e^{-jk_u x} \int e^{jr\sqrt{u^2 + w^2} - jk_u u}du.
\end{equation}

The MSP will be applied to the Fourier integral in (\ref{eq:msp_ex2}), implying for this example
\begin{gather}
    g(u) = 1, \\
    f(u) = r\sqrt{u^2 + w^2} - k_u u.
\end{gather}

Using (\ref{eq:msp_1D_delf}), the stationary point $u_0$ can be computed as 
\begin{gather}
    \frac{\partial f(u)}{\partial u} \biggr\rvert_{u = u_0} = \frac{r u_0}{\sqrt{u_0^2 + w^2}} - k_u = 0, \\
    u_0 = \frac{k_u w}{\sqrt{r^2 - k_u^2}}, \\
    f(u_0) = w\sqrt{r^2 - k_u^2}
\end{gather}

Finally, $u_0$ can be substituted into (\ref{eq:msp_1D}) ignoring the factor of $1/f''(u_0)$ as 
\begin{equation}
\label{eq:msp_ex3}
    \int e^{jr\sqrt{u^2 + w^2} - jk_u u}du \approx e^{jw\sqrt{r^2 - k_u^2}}.
\end{equation}

Substituting (\ref{eq:msp_ex3}) into (\ref{eq:msp_ex2}) yields

\begin{equation}
\label{eq:msp_ex4}
    \text{FT}_{\text{1D}}^{(u)} \left[ e^{jr\sqrt{(x-u)^2 + w^2}} \right] = e^{-jk_u x + jw\sqrt{r^2 - k_u^2}}.
\end{equation}

Taking the \mbox{1-D} inverse spatial Fourier transform of (\ref{eq:msp_ex4}) results in (\ref{eq:mspLinear}), labeled Approximation 1 below. This example illustrates the key steps of the spherical wave decomposition using the method of stationary phase. Similar analysis has been employed on the other examples below yielding the corresponding approximations using the MSP.

\subsection{Useful MSP Approximations}
\label{subsec:msp_useful}

\textbf{Approximation 1:}
\begin{equation}
\label{eq:mspLinear}
    \begin{split}
        e^{jr\sqrt{(x-u)^2 + w^2}} \approx \int e^{jk_u(u-x) + j k_w w} dk_u,
    \end{split}
\end{equation}
where
\begin{equation}
    k_w^2 = r^2 - k_u^2.
\end{equation}
\textbf{Approximation 2:}
\begin{equation}
\label{eq:mspRectilinear}
    \begin{split}
        \frac{e^{jr\sqrt{(x-u)^2 + (y-v)^2 + w^2}}}{\sqrt{(x-u)^2 + (y-v)^2 + w^2}} \approx \iint \frac{1}{k_w} e^{jk_u(u-x) + jk_v(v-y) + jk_w w} dk_u dk_v,
    \end{split}
\end{equation}
where
\begin{equation}
    k_w^2 = r^2 - k_u^2 - k_v^2.
\end{equation}
\textbf{Approximation 3:}
\begin{equation}
\label{eq:mspCircular}
    \begin{split}
        e^{jr\sqrt{(x-u)^2 + (z-w)^2}} \approx \iint e^{jk_u(u-x) + j k_w (w-z)} dk_u dk_w.
    \end{split}
\end{equation}
\textbf{Approximation 4:}
\begin{equation}
\label{eq:mspCylindrical}
    \begin{split}
        e^{jr\sqrt{(x-u)^2 + (y-v)^2 + (z-w)^2}}  \approx \iiint e^{jk_u(u-x) + jk_v(v-y) + jk_w(w-z)} dk_u dk_v dk_w.
    \end{split}
\end{equation}
\chapter{Efficient Near-Field SAR Image Reconstruction Algorithms for Various Geometries}
\label{ch:reconstruction_algos}

In this chapter, we detail the efficient image reconstruction algorithms for several synthetic aperture radar (SAR) scanning geometries. 
The algorithms in this section have been derived elsewhere and are included for the benefit of the reader.

\section{\mbox{1-D} Linear Synthetic Array \mbox{1-D} Imaging - Fourier-based}
\label{sec:linear_fft}
In this section, we derive the image reconstruction algorithm for recovering a \mbox{1-D} reflectivity function from a \mbox{1-D} linear SAR scenario in the near-field \cite{paul2021systematic,maisto2021sensor,soumekh1998wide,soumekh1999synthetic,smith2021An}. 
Given a \mbox{1-D} linear SISO synthetic array whose elements are located at the points $(y',Z_0)$ in the $y$-$z$ plane and a \mbox{1-D} target with reflectivity function $p(y)$ located at the points $(y,z_0)$, the isotropic beat signal can be written as
\begin{equation}
\label{eq:linear_fft1}
    s(y',k) = \int \frac{p(y)}{R^2} e^{j2kR} dy,
\end{equation}
where
\begin{equation}
    R = \sqrt{(y-y')^2 + (z_0 - Z_0)^2}.
\end{equation}
Ignoring amplitude terms, applying the MSP derived in (\ref{eq:mspLinear}), the spherical phase term in (\ref{eq:linear_fft1}) can be substituted yielding
\begin{equation}
\label{eq:linear_fft2}
    s(y',k) = \iint p(y) e^{j(k_y'(y'-y) + k_z(z_0-Z_0))}dy dk_y', 
\end{equation}
where
\begin{equation}
    k_z = \sqrt{4k^2 - k_y^2}.
\end{equation}
Rearranging the phase terms in (\ref{eq:linear_fft2}), a forward spatial Fourier transform on $y$ and inverse spatial Fourier transform on $y'$ become evident as
\begin{gather}
    s(y',k) = \int  \left[ \int p(y)e^{-jk_y'y}dy \right] e^{j(k_y'y'+ k_z(z_0-Z_0))} dk_y'.
\end{gather}
The term inside the brackets can be rewritten as the spatial-spectral representation of the target reflectivity function, $P(k_y)$. Then, performing a forward Fourier transform along $y'$ on both sides simplifies the expression as the following. Note that the distinction between the primed and unprimed domains can be dropped in the spatial Fourier domain as they coincide.
\begin{gather}
    s(y',k) = \int \left[ P(k_y)  e^{jk_z(z_0-Z_0)} \right] e^{jk_y'y'} dk_y', \\
    S(k_y,k) = P(k_y)e^{jk_z(z_0-Z_0)}, \\
    P(k_y) = S(k_y,k)e^{-jk_z(z_0-Z_0)}.
    \label{eq:linear_fft3}
\end{gather}

For wideband waveforms, (\ref{eq:linear_fft3}) is evaluated at multiple wavenumbers thus coherent summation is performed over $k$. 
Hence, the complete expression for the Fourier-based \mbox{1-D} image reconstruction algorithm for a \mbox{1-D} linear SISO synthetic array is
\begin{equation}
\label{eq:linear_fft_final}
    p(y) = \int \text{IFT}_{\text{1D}}^{(k_y)} \left[ \text{FT}_{\text{1D}}^{(y')} [s(y',k)] e^{-jk_z(z_0-Z_0)} \right]dk.
\end{equation}

\section{\mbox{1-D} Linear Synthetic Array \mbox{2-D} Imaging - Range Migration Algorithm}
\label{sec:linear_rma}
In this section we derive the image reconstruction algorithm for recovering a \mbox{2-D} reflectivity function from a \mbox{1-D} linear SAR scenario in the near-field \cite{paul2021systematic,maisto2021sensor,soumekh1998wide,soumekh1999synthetic,smith2021An}. 
Given a \mbox{1-D} linear SISO synthetic array whose elements are located at the points $(y',Z_0)$ in the $y$-$z$ plane and a \mbox{2-D} target with reflectivity function $p(y,z)$ located at the points $(y,z)$, the isotropic beat signal can be written as
\begin{equation}
\label{eq:linear_rma1}
    s(y',k) = \iint \frac{p(y,z)}{R^2} e^{j2kR} dy dz,
\end{equation}
where
\begin{equation}
    R = \sqrt{(y-y')^2 + (z - Z_0)^2}.
\end{equation}
Ignoring amplitude terms, applying the MSP derived in (\ref{eq:mspLinear}), the spherical phase term in (\ref{eq:linear_rma1}) can be substituted yielding
\begin{equation}
\label{eq:linear_rma2}
    s(y',k) = \iiint p(y,z) e^{j(k_y'(y'-y) + k_z(z-Z_0))}dy dz dk_y', 
\end{equation}
where
\begin{equation}
    k_z = \sqrt{4k^2 - k_y^2}.
\end{equation}
Leveraging conjugate symmetry of the spherical wavefront, (\ref{eq:linear_rma2}) can be rewritten in the following form to exploit the spatial Fourier transform on $z$
\begin{equation}
\label{eq:linear_rma3}
    s^*(y',k) = \iiint p(y,z) e^{j(k_y'(y'-y) - k_z(z-Z_0))}dy dz dk_y', 
\end{equation}
where $(\cdot)^*$ is the complex conjugate operation.

Rearranging the phase terms in (\ref{eq:linear_rma3}), a forward spatial Fourier transform on $y$-$z$ and inverse spatial Fourier transform on $y'$ become evident as
\begin{equation}
    \begin{split}
        s^*(y',k) &= \int  \left[ \iint p(y,z)e^{-j(k_y'y+k_z z)}dydz \right]  e^{j(k_y'y'+ k_z Z_0)} dk_y'.
    \end{split}
\end{equation}
The term inside the brackets can be rewritten as the spatial-spectral representation of the target reflectivity function, $P(k_y,k_z)$. Then, performing a forward Fourier transform along $y'$ on both sides simplifies the expression as the following. Note that the distinction between the primed and unprimed domains can be dropped in the spatial Fourier domain as they coincide.
\begin{gather}
    s^*(y',k) = \int \left[ P(k_y,k_z)  e^{jk_z Z_0} \right] e^{jk_y'y'} dk_y', \\
    \Tilde{S}(k_y,k) = \text{FT}_{\text{1D}}^{(y')} [s^*(y',k)], \\
    \Tilde{S}(k_y,k) = P(k_y,k_z)e^{jk_z Z_0}, \\
    P(k_y,k_z) = \Tilde{S}(k_y,k)e^{-jk_z Z_0}.
    \label{eq:linear_rma4}
\end{gather}

The direct relationship between $P(k_y,k_z)$ and $\Tilde{S}(k_y,k)$ is now obvious in (\ref{eq:linear_rma4}); however, $P(k_y,k_z)$ is sampled on a uniform $k_y$-$k_z$ grid and $\Tilde{S}(k_y,k)$ is sampled on a uniform $k_y$-$k$ grid. Before the reflectivity function can be recovered using in inverse Fourier transform, $\Tilde{S}(k_y,k)e^{-jk_z Z_0}$ must be interpolated to a uniform $k_y$-$k_z$ grid using Stolt interpolation, represented by the $\mathcal{S}[\cdot]$ operator, to account for the curvature of the wavefront \cite{lopez20003}.
\begin{equation}
    S(k_y,k_z) = \mathcal{S} \left[ \Tilde{S}(k_y,k)e^{-jk_z Z_0} \right].
\end{equation}

Finally, the complete expression for the Fourier-based \mbox{2-D} image reconstruction algorithm for a \mbox{1-D} linear SISO synthetic array can be written as
\begin{equation}
\label{eq:linear_rma_final}
    p(y,z) = \text{IFT}_{\text{2D}}^{(k_y,k_z)}  \left[ \mathcal{S} \left[ \text{FT}_{\text{1D}}^{(y')} [s^*(y',k)] e^{-jk_z Z_0} \right] \right].
\end{equation}

\section{\mbox{2-D} Rectilinear Array \mbox{2-D} Imaging - Fourier-based}
\label{sec:rectilinear_fft}
In this section, we derive the image reconstruction algorithm for recovering a \mbox{2-D} reflectivity function from a \mbox{2-D} rectilinear SAR scenario in the near-field \cite{guo2019millimeter,yanik2018millimeter}. 
Given a \mbox{2-D} rectilinear SISO synthetic array whose elements are located at the points $(x',y',Z_0)$ in $x$-$y$-$z$ space and a \mbox{2-D} target with reflectivity function $p(x,y)$ located at the points $(x,y,z_0)$, the isotropic beat signal can be written as
\begin{equation}
\label{eq:rectilinear_fft1}
    s(x',y',k) = \iint \frac{p(x,y)}{R^2} e^{j2kR} dx dy,
\end{equation}
where
\begin{equation}
    R = \sqrt{(x-x')^2 + (y-y')^2 + (z_0 - Z_0)^2}.
\end{equation}
Assuming the points of the target scene are closely located, the $R^{-2}$ factor in (\ref{eq:rectilinear_fft1}) can be approximated as $R^{-1}$ \cite{yanik2019sparse}. Applying the MSP derived in (\ref{eq:mspRectilinear}), the spherical phase term in (\ref{eq:rectilinear_fft1}) can be substituted yielding
\begin{equation}
\label{eq:rectilinear_fft2}
    \begin{split}
        s(x',y',k) &= \iiiint \frac{p(x,y)}{k_z} e^{j(k_x'(x'-x) + k_y'(y'-y))}  e^{jk_z(z_0-Z_0)}dx dy dk_x' dk_y', 
    \end{split}
\end{equation}
where
\begin{equation}
    k_z = \sqrt{4k^2 - k_x^2 - k_y^2}.
\end{equation}
Rearranging the phase terms in (\ref{eq:rectilinear_fft2}), a forward spatial Fourier transform on $x$-$y$ and inverse spatial Fourier transform on $x'$-$y'$ become evident as
\begin{equation}
    \begin{split}
        s(x'y',k) &= \iint  \left[ \iint \frac{p(x,y)}{k_z} e^{-j(k_x'x + k_y'y)}dx dy \right]  e^{j(k_x'x' + k_y'y')+ jk_z(z_0-Z_0)} dk_x' dk_y'.
    \end{split}
\end{equation}
The term inside the brackets can be rewritten as the spatial-spectral representation of the target reflectivity function. Then, performing a forward Fourier transform along $x'$-$y'$ on both sides simplifies the expression as the following. Note that the distinction between the primed and unprimed domains can be dropped in the spatial Fourier domain as they coincide.
\begin{equation}
    \begin{split}
        s(x',y',k) &= \int \left[ \frac{P(k_x,k_y)}{k_z}  e^{jk_z(z_0-Z_0)} \right]  e^{j(k_x'x' + k_y'y')}dk_x' dk_y', 
    \end{split}
\end{equation}
writing

\begin{gather}
    S(k_x,k_y,k) = \frac{P(k_x,k_y)}{k_z}e^{jk_z(z_0-Z_0)}, \\
    P(k_x,k_y) = S(k_y,k)k_z e^{-jk_z(z_0-Z_0)}.
    \label{eq:rectilinear_fft3}
\end{gather}

For wideband waveforms, (\ref{eq:rectilinear_fft3}) is evaluated at multiple wavenumbers thus coherent summation is performed over $k$. 
Hence, the complete expression for the Fourier-based \mbox{2-D} image reconstruction algorithm for a \mbox{2-D} rectilinear SISO synthetic array is
\begin{equation}
\label{eq:rectilinear_fft_final}
    \begin{split}
        p(x,y) &= \int \text{IFT}_{\text{2D}}^{(k_x,k_y)} \biggr[ \text{FT}_{\text{2D}}^{(x',y')} [s(x',y',k)]  k_z e^{-jk_z(z_0-Z_0)} \biggr] dk.
    \end{split}
\end{equation}

\section{\mbox{2-D} Rectilinear Array \mbox{3-D} Imaging -  Range Migration Algorithm}
\label{sec:rectilinear_rma}
In this section we derive the image reconstruction algorithm for recovering a \mbox{3-D} reflectivity function from a \mbox{2-D} rectilinear SAR scenario in the near-field \cite{lopez20003,zhuge2010sparse,savelyev2010comparison,fan2020linearMIMOArbitraryTopologies,mohammadian2019sar,zhuge2012three,zhu2017range,qiao2015compressive,yanik2019cascaded,yanik2019sparse,yanik2020development,smith2021sterile,sheen2016three,sheen2010near,sheen2001three,sheen2018simulation}. 
Given a \mbox{2-D} rectilinear SISO synthetic array whose elements are located at the points $(x',y',Z_0)$ in $x$-$y$-$z$ space and a \mbox{3-D} target with reflectivity function $p(x,y,z)$ located at the points $(x,y,z)$, the isotropic beat signal can be written as
\begin{equation}
\label{eq:rectilinear_rma1}
    s(x',y',k) = \iiint \frac{p(x,y,z)}{R^2} e^{j2kR} dx dy dz,
\end{equation}
where
\begin{equation}
    R = \sqrt{(x-x')^2 + (y-y')^2 + (z - Z_0)^2}.
\end{equation}
Assuming the points of the target scene are closely located, the $R^{-2}$ factor in (\ref{eq:rectilinear_rma1}) can be approximated as $R^{-1}$ \cite{yanik2019sparse}. 
Applying the MSP derived in (\ref{eq:mspRectilinear}), the spherical phase term in (\ref{eq:rectilinear_rma1}) can be substituted yielding
\begin{equation}
\label{eq:rectilinear_rma2}
    \begin{split}
        s(x',y',k) &= \iint \biggr[ \iiint \frac{p(x,y,z)}{k_z} e^{j(k_x'(x'-x) + k_y'(y'-y))}  e^{jk_z(z-Z_0)}dx dy dz \biggr] dk_x' dk_y', 
    \end{split}
\end{equation}
where
\begin{equation}
    k_z = \sqrt{4k^2 - k_x^2 - k_y^2}.
\end{equation}
Leveraging conjugate symmetry of the spherical wavefront, (\ref{eq:rectilinear_rma2}) can be rewritten in the following form to exploit the spatial Fourier transform on $z$
\begin{equation}
\label{eq:rectilinear_rma3}
    \begin{split}
        s^*(x',y',k) &= \iint \biggr[ \iiint \frac{p(x,y,z)}{k_z} e^{j(k_x'(x'-x) + k_y'(y'-y))}  e^{-jk_z(z-Z_0)}dx dy dz \biggr] dk_x' dk_y', 
    \end{split}
\end{equation}
where $(\cdot)^*$ is the complex conjugate operation.

Rearranging the phase terms in (\ref{eq:rectilinear_rma3}), a forward spatial Fourier transform on $x$,$y$,$z$ and inverse spatial Fourier transform on $x'$,$y'$ become evident as
\begin{equation}
    \begin{split}
        s^*(x',y',k) &= \iint  \biggr[ \iiint \frac{p(x,y,z)}{k_z} e^{-(jk_x' x + jk_y' y + jk_z z)}  dxdydz \biggr] e^{j(k_x'x'+k_y'y')+ jk_z Z_0} dk_x' dk_y'.
    \end{split}
\end{equation}
The term inside the brackets can be rewritten as the spatial-spectral representation of the target reflectivity function. Then, performing a forward Fourier transform along $x'$, $y'$ on both sides simplifies the expression as the following. Note that the distinction between the primed and unprimed domains can be dropped in the spatial Fourier domain as they coincide.
\begin{equation}
    \begin{split}
        s^*(x',y',k) &= \int \left[ P(k_x,k_y,k_z)  e^{jk_z Z_0} \right]  e^{j(k_x'x'+k_y'y')} dk_x' dk_y', 
    \end{split}
\end{equation}
writing

\begin{gather}
    \Tilde{S}(k_x,k_y,k) = \text{FT}_{\text{2D}}^{(x',y')} [s^*(x',y',k)], \\
    \Tilde{S}(k_x,k_y,k) = \frac{P(k_x,k_y,k_z)}{k_z} e^{jk_z Z_0}, \\
    P(k_x,k_y,k_z) = \Tilde{S}(k_x,k_y,k) k_z e^{-jk_z Z_0}.
    \label{eq:rectilinear_rma4}
\end{gather}

The direct relationship between $P(k_x,k_y,k_z)$ and $\Tilde{S}(k_x,k_y,k)$ is now obvious in (\ref{eq:rectilinear_rma4}); however, $P(k_x,k_y,k_z)$ is sampled on a uniform $k_x$-$k_y$-$k_z$ grid and $\Tilde{S}(k_x,k_y,k)$ is sampled on a uniform $k_x$-$k_y$-$k$ grid. 
Before the reflectivity function can be recovered using an inverse Fourier transform, $\Tilde{S}(k_x,k_y,k)e^{-jk_z Z_0}$ must be interpolated to a uniform $k_x,k_y$-$k_z$ grid using the Stolt interpolation, represented by the $\mathcal{S}[\cdot]$ operator, to account for the curvature of the wavefront \cite{lopez20003}.
\begin{equation}
    S(k_x,k_y,k_z) = \mathcal{S} \left[ \Tilde{S}(k_x,k_y,k) k_z e^{-jk_z Z_0} \right].
\end{equation}

Finally, the complete expression for the Fourier-based \mbox{3-D} image reconstruction algorithm for a \mbox{2-D} rectilinear SISO synthetic array can be written as
\begin{equation}
\label{eq:rectilinear_rma_final}
    \begin{split}
        p(x,y,z) &= \text{IFT}_{\text{3D}}^{(k_x,k_y,k_z)}  \biggr[ \mathcal{S} \biggr[ \text{FT}_{\text{2D}}^{(x',y')} [s^*(x',y',k)] k_z e^{-jk_z Z_0} \biggr] \biggr].
    \end{split}
\end{equation}

\section{\mbox{1-D} Circular Synthetic Array \mbox{2-D} Imaging - Polar Formatting Algorithm}
\label{sec:circular_pfa}
In this section, we derive the image reconstruction algorithm for recovering a \mbox{2-D} reflectivity function from a \mbox{1-D} circular SAR scenario in the near-field \cite{demirci2011back,jia2014modifiedBPA,gao2016terahertz}. 
Given a \mbox{1-D} circular SISO synthetic array whose elements are located at the points $(R_0\cos\theta,R_0\sin\theta)$ in the $x$-$z$ plane at $y = 0$, where $R_0$ and $\theta$ are the constant radial distance from the antenna elements to the origin and the angular dimension, respectively, and a \mbox{2-D} target with reflectivity function $p(x,z)$ located at the points $(x,z)$, the isotropic beat signal can be written as
\begin{equation}
\label{eq:circular_pfa1}
    s(\theta,k) = \iint \frac{p(x,z)}{R^2} e^{j2kR} dx dz,
\end{equation}
where
\begin{equation}
    R = \sqrt{(x-R_0\cos\theta)^2 + (z-R_0\sin\theta)^2}.
\end{equation}

The MSP derived in (\ref{eq:mspCircular}) can be applied to the spherical phase term in (\ref{eq:circular_pfa1}) after the following substitutions
\begin{gather}
\label{eq:circular_pfa_sub1}
    x' = R_0\cos\theta, \\
    z' = R_0\cos\theta, \\
    k_x' = k_r \cos\alpha, \\
    k_z' = k_r \cos\alpha, \\
    k_r^2 = k_x'^2 + k_z'^2,
    \label{eq:circular_pfa_subEnd}
\end{gather}
yielding
\begin{gather}
\label{eq:circular_pfa2}
    e^{j2kR} \approx \iint e^{j(k_x'(x'-x) + k_z'(z'-z))} dk_x' dk_z'.
\end{gather}

Neglecting path loss, (\ref{eq:circular_pfa1}) and (\ref{eq:circular_pfa2}) can be combined as
\begin{equation}
\label{eq:circular_pfa3}
    s(\theta,k) = \iiiint p(x,z) e^{j(k_x'(x'-x) + k_z'(z'-z))} dx dz dk_x' dk_z'.
\end{equation}

Rearranging the phase terms in (\ref{eq:circular_pfa3}), a forward spatial Fourier transform on $x$-$z$ and inverse spatial Fourier transform on $x'$-$z'$ become evident as
\begin{equation}
\label{eq:circular_pfa4}
    \begin{split}
        s(\theta,k) &= \iint  \left[ \iint p(x,z) e^{-j(k_x'x + k_z'z)}dx dz \right]  e^{j(k_x'x' + k_z'z')} dk_x' dk_z'.
    \end{split}
\end{equation}

The term inside the brackets can be rewritten as the spatial spectral representation of the target reflectivity function, $P(k_x,k_z)$. Then using the relations (\ref{eq:circular_pfa_sub1})-(\ref{eq:circular_pfa_subEnd}), the expression in (\ref{eq:circular_pfa4}) can be rewritten as
\begin{equation}
    \begin{split}
        s(\theta,k) &= \iint  P(k_x,k_z) e^{j(k_r\cos\theta R_0\cos\alpha + k_r\sin\theta R_0\sin\alpha)}  k_r dk_r d\alpha.
    \end{split}
\end{equation}

Rewriting the spectral $P(k_x,k_z)$ as its equivalent spectral polar form $P(\alpha,k_r)$ and simplifying the phase term
\begin{equation}
\label{eq:circular_pfa5}
    s(\theta,k) = \int \biggr[ \int P(\alpha,k_r) e^{jk_r R_0 cos(\theta - \alpha)} d\alpha \biggr] k_r dk_r.
\end{equation}

The term inside the brackets in (\ref{eq:circular_pfa5}) is a convolution operation in the $\theta$ domain, where the $\theta$ and $\alpha$ domains are coincident and can be exploited using Fourier relations by taking a Fourier transform across $\theta$ on both sides of the equation as 
\begin{equation}
\label{eq:circular_pfa6}
    S(k_\theta,k) = \int P(k_\theta,k_r) \text{FT}_{\text{1D}}^{(\theta)} \left[ e^{jk_r R_0 \cos\theta} \right] k_r dk_r.
\end{equation}

Considering only the values lying on the Ewald sphere, $k_r^2 = 4k^2$ imposes a $\delta$-function behavior of the integrand in (\ref{eq:circular_pfa6}) with respect to $k_r$ \cite{amineh2019real}.
As such, (\ref{eq:circular_pfa6}) can be simplified as such, substituting $k_r = 2k$,
\begin{equation}
\label{eq:circular_pfa7}
    P(k_\theta,k_r) = S(k_\theta,k)G^*(k_\theta,k),
\end{equation}
where
\begin{equation}
\label{eq:circular_pfa_G}
    G(k_\theta,k) = \text{FT}_{\text{1D}}^{(\theta)} \left[ e^{j2 k R_0 \cos\theta} \right].
\end{equation}

The spatial spectral reflectivity function in polar coordinates can be recovered from (\ref{eq:circular_pfa7}) as 
\begin{equation}
    P(\theta,k_r) = \text{IFT}_{\text{1D}}^{(k_\theta)} \left[ S(k_\theta,k)G^*(k_\theta,k) \right].
\end{equation}

Finally, the reflectivity function $p(x,z)$ can be recovered using a nonuniform FFT (NUFFT) \cite{gao2016efficient} or via interpolation to the rectangular spatial Fourier domain $k_x$-$k_z$ followed by a uniform IFFT. This interpolation operation, known as the polar formatting algorithm (PFA), is denoted by $\mathcal{P}[\cdot]$. Thus, the final step in the image recovery process is (prime notation will be ignored for the remainder of this derivation as the primed and unprimed coordinate systems are coincident)
\begin{equation}
    p(x,z) = \text{IFT}_{\text{2D}}^{(k_x,k_z)} \left[ \mathcal{P}[P(\theta,k_r)] \right].
\end{equation}

Finally, the complete expression for the Fourier-based \mbox{2-D} image reconstruction algorithm for a \mbox{1-D} circular SISO synthetic array can be written as
\begin{equation}
\label{eq:circular_pfa_final}
    \begin{split}
        p(x,z) &= \text{IFT}_{\text{2D}}^{(k_x,k_z)} \biggr[ \mathcal{P}\biggr[ \text{IFT}_{\text{1D}}^{(k_\theta)} \biggr[ S(k_\theta,k)  \text{FT}_{\text{1D}}^{(\theta)} \biggr[ e^{j2 k R_0 \cos\theta} \biggr]^* \biggr] \biggr] \biggr].
    \end{split}
\end{equation}

\section{\mbox{2-D} Cylindrical Synthetic Array \mbox{3-D} Imaging - Polar Formatting Algorithm}
\label{sec:cylindrical_pfa}
In this section, we derive the image reconstruction algorithm for recovering a \mbox{3-D} reflectivity function from a \mbox{2-D} cylindrical SAR (also known as ECSAR) scenario in the near-field \cite{amineh2019real,fortuny2001extension,detlefsen2005effective,laviada2017multiview,gao2018cylindricalMIMO,smith2020nearfieldisar}. 
Given a \mbox{2-D} cylindrical SISO synthetic array whose elements are located at the points $(R_0\cos\theta,y',R_0\sin\theta)$ in $x$-$y$-$z$ space, where $R_0$ and $\theta$ are the constant radial distance from the antenna elements to the origin and the angular dimension, respectively, and a \mbox{3-D} target with reflectivity function $p(x,y,z)$ located at the points $(x,y,z)$, the isotropic beat signal can be written as
\begin{equation}
\label{eq:cylindrical_pfa1}
    s(\theta,y',k) = \iiint \frac{p(x,y,z)}{R^2} e^{j2kR} dx dy dz,
\end{equation}
where
\begin{equation}
    R = \sqrt{(x-R_0\cos\theta)^2 + (y-y')^2 + (z-R_0\sin\theta)^2}.
\end{equation}

The MSP derived in (\ref{eq:mspCylindrical}) can be applied to the spherical phase term in (\ref{eq:cylindrical_pfa1}) after the following substitutions
\begin{gather}
\label{eq:cylindrical_pfa_sub1}
    x' = R_0\cos\theta, \\
    z' = R_0\cos\theta, \\
    k_x' = k_r \cos\alpha, \\
    k_z' = k_r \cos\alpha, \\
    k_r^2 = k_x'^2 + k_z'^2 = 4k^2 - k_y'^2,
    \label{eq:cylindrical_pfa_subEnd}
\end{gather}
yielding
\begin{gather}
\label{eq:cylindrical_pfa2}
    e^{j2kR} \approx \iint e^{j(k_x'(x'-x) + k_y'(y'-y) + k_z'(z'-z))} dk_x' dk_y' dk_z'.
\end{gather}

Neglecting path loss, (\ref{eq:cylindrical_pfa1}) and (\ref{eq:cylindrical_pfa2}) can be combined as
\begin{equation}
\label{eq:cylindrical_pfa3}
    \begin{split}
        s(\theta,y',k) &= \iiint \biggr[ \iiint p(x,y,z) e^{j(k_x'(x'-x)+ k_y'(y'-y) + k_z'(z'-z))} dx dy dz \biggr] dk_x' dk_y' dk_z'.
    \end{split}
\end{equation}

Rearranging the phase terms in (\ref{eq:cylindrical_pfa3}), a forward spatial Fourier transform on $x$-$y$-$z$ and inverse spatial Fourier transform on $x'$-$y'$-$z'$ become evident as
\begin{equation}
\label{eq:cylindrical_pfa4}
    \begin{split}
        s(\theta,y',k) &= \iiint  \left[ \iiint p(x,y,z) e^{-j(k_x'x + k_y'y + k_z'z)}dx dy dz \right]  e^{j(k_x'x' + k_y'y' + k_z'z')} dk_x' dk_y' dk_z'.
    \end{split}
\end{equation}

The term inside the brackets can be rewritten as the spatial spectral representation of the target reflectivity function, $P(k_x,k_y,k_z)$. Then using the relations (\ref{eq:cylindrical_pfa_sub1})-(\ref{eq:cylindrical_pfa_subEnd}), the expression in (\ref{eq:cylindrical_pfa4}) can be rewritten as
\begin{equation}
    \begin{split}
        s(\theta,y',k) &= \iiint  P(k_x,k_y,k_z) e^{j(k_r\cos\theta R_0\cos\alpha} e^{j(k_r\sin\theta R_0\sin\alpha) + k_y'y')} k_r dk_y' dk_r d\alpha.
    \end{split}
\end{equation}

Taking a Fourier transform on both side with respect to $y'$, rewriting the spectral $P(k_x,k_y,k_z)$ as its equivalent spectral polar form $P(\alpha,k_y,k_r)$, and simplifying the phase term yields (prime notation will be dropped for the remainder of this derivation as the primed and unprimed coordinate systems are coincident)
\begin{equation}
\label{eq:cylindrical_pfa5}
    s(\theta,k_y,k) = \int \biggr[ \int P(\alpha,k_y,k_r) e^{jk_r R_0 cos(\theta - \alpha)} d\alpha \biggr] k_r dk_r.
\end{equation}

The term inside the brackets in (\ref{eq:cylindrical_pfa5}) is a convolution operation in the $\theta$ domain, where the $\theta$ and $\alpha$ domains are coincident and can be exploited using Fourier relations by taking a Fourier transform across $\theta$ on both sides of the equation as 
\begin{equation}
\label{eq:cylindrical_pfa6}
    S(k_\theta,k_y,k) = \int P(k_\theta,k_y,k_r) \text{FT}_{\text{1D}}^{(\theta)} \left[ e^{jk_r R_0 \cos\theta} \right] k_r dk_r.
\end{equation}

Considering only the values lying on the Ewald sphere, $k_r^2 = 4k^2 - k_y^2$ imposes a $\delta$-function behavior of the integrand in (\ref{eq:cylindrical_pfa6}) with respect to $k_r$ \cite{amineh2019real}.
As such, (\ref{eq:cylindrical_pfa6}) can be simplified as such, substituting $k_r = \sqrt{4k^2 - k_y^2}$,
\begin{equation}
\label{eq:cylindrical_pfa7}
    P(k_\theta,k_y,k_r) = S(k_\theta,k_y,k)G^*(k_\theta,k_y,k),
\end{equation}
where
\begin{equation}
\label{eq:cylindrical_pfa_G}
    G(k_\theta,k_y,k) = \text{FT}_{\text{1D}}^{(\theta)} \left[ e^{j \sqrt{4k^2 - k_y^2} R_0 \cos\theta} \right].
\end{equation}

The spatial spectral reflectivity function in polar coordinates can be recovered from (\ref{eq:cylindrical_pfa7}) as 
\begin{equation}
    P(\theta,k_y,k_r) = \text{IFT}_{\text{1D}}^{(k_\theta)} \left[ S(k_\theta,k_y,k)G^*(k_\theta,k_y,k) \right].
\end{equation}

Finally, the reflectivity function $p(x,y,z)$ can be recovered using a nonuniform FFT (NUFFT) \cite{gao2016efficient} or via interpolation to the rectangular spatial Fourier domain $k_x$-$k_y$-$k_z$ followed by a uniform IFFT. This interpolation operation, known as the polar formatting algorithm (PFA), is denoted by $\mathcal{P}[\cdot]$. Thus, the final step in the image recovery process is
\begin{equation}
    p(x,y,z) = \text{IFT}_{\text{3D}}^{(k_x,k_y,k_z)} \left[ \mathcal{P}[P(\theta,k_y,k_r)] \right].
\end{equation}

Finally, the complete expression for the Fourier-based \mbox{3-D} image reconstruction algorithm for a \mbox{2-D} cylindrical SISO synthetic array can be written as
\begin{equation}
\label{eq:cylindrical_pfa_final}
    \begin{split}
        p(x,y,z) &= \text{IFT}_{\text{3D}}^{(k_x,k_y,k_z)} \biggr[ \mathcal{P}\biggr[ \text{IFT}_{\text{1D}}^{(k_\theta)} \biggr[ S(k_\theta,k_y,k) \text{FT}_{\text{1D}}^{(\theta)} \biggr[ e^{j \sqrt{4k^2 - k_y^2} R_0 \cos\theta} \biggr]^* \biggr] \biggr] \biggr].
    \end{split}
\end{equation}


\printbibliography

@inproceedings{sheen2016three,
	title        = {Three-dimensional radar imaging techniques and systems for near-field applications},
	author       = {David M. Sheen and Thomas E. Hall and Douglas L. McMakin and A. Mark Jones and Jonathan R. Tedeschi},
	year         = 2016,
	month        = may,
	booktitle    = {Proc. SPIE},
	address      = {Baltimore, MD, USA},
	volume       = 9829,
	pages        = {98290V}
}

@article{sheen2010near,
	title        = {Near-field three-dimensional radar imaging techniques and applications},
	author       = {Sheen, David and McMakin, Douglas and Hall, Thomas},
	year         = 2010,
	month        = jun,
	journal      = {Appl. Opt.},
	volume       = 49,
	number       = 19,
	pages        = {E83--E93}
}

@article{sheen2001three,
	title        = {Three-dimensional millimeter-wave imaging for concealed weapon detection},
	author       = {Sheen, David M and McMakin, Douglas L and Hall, Thomas E},
	year         = 2001,
	month        = sep,
	journal      = {IEEE Trans. Microw. Theory Techn.},
	volume       = 49,
	number       = 9,
	pages        = {1581--1592}
}

@inproceedings{sheen2018simulation,
	title        = {Simulation of active cylindrical and planar millimeter-wave imaging systems},
	author       = {Sheen, David M and Jones, A Mark and Hall, Thomas E},
	year         = 2018,
	month        = may,
	booktitle    = {Proc. SPIE},
	address      = {Orlando, FL, USA},
	volume       = 10634,
	pages        = 1063408
}

@inproceedings{yanik2018millimeter,
	title        = {Millimeter-wave near-field imaging with two-dimensional {SAR} data},
	author       = {M. E. Yanik and M. Torlak},
	year         = 2018,
	month        = sep,
	booktitle    = {Proc. SRC Techcon},
	address      = {Austin, TX, USA},
	number       = {P093929}
}

@article{yanik2020development,
	title        = {Development and demonstration of {MIMO-SAR mmWave} imaging testbeds},
	author       = {Yanik, Muhammet Emin and Wang, Dan and Torlak, Murat},
	year         = 2020,
	month        = jul,
	journal      = {IEEE Access},
	volume       = 8,
	pages        = {126019--126038}
}

@article{yanik2019sparse,
	title        = {Near-field {MIMO-SAR} millimeter-wave imaging with sparsely sampled aperture data},
	author       = {Yanik, Muhammet Emin and Torlak, Murat},
	year         = 2019,
	month        = mar,
	journal      = {IEEE Access},
	volume       = 7,
	pages        = {31801--31819}
}

@inproceedings{yanik2019cascaded,
	title        = {{3-D MIMO-SAR} imaging using multi-chip cascaded millimeter-wave sensors},
	author       = {Yanik, Muhammet Emin and Wang, Dan and Torlak, Murat},
	year         = 2019,
	month        = nov,
	booktitle    = {Proc. IEEE Global Conf. Signal Inf. Process. (GlobalSIP)},
	address      = {Ottawa, ON, Canada},
	pages        = {1--5}
}

@article{smith2021An,
	title        = {An {FCNN}-Based Super-Resolution {mmWave} Radar Framework for Contactless Musical Instrument Interface},
	author       = {Smith, J. W. and Furxhi, O. and Torlak, M.},
	year         = 2021,
	month        = may,
	journal      = {IEEE Trans. Multimedia},
	volume       = 24,
	pages        = {2315--2328}
}

@inproceedings{smith2020nearfieldisar,
	title        = {Near-Field {MIMO-ISAR} Millimeter-Wave Imaging},
	author       = {Smith, J. W. and Yanik, M. E. and Torlak, M.},
	year         = 2020,
	month        = sep,
	booktitle    = {Proc. IEEE Radar Conf. (RadarConf)},
	address      = {Florance, Italy},
	pages        = {1--6}
}

@article{smith2021sterile,
	title        = {Improved Static Hand Gesture Classification on Deep Convolutional Neural Networks Using Novel Sterile Training Technique},
	author       = {Smith, J. W. and Thiagarajan, S. and Willis, R. and Makris, Y. and Torlak, M.},
	year         = 2021,
	month        = jan,
	journal      = {IEEE Access},
	volume       = 9,
	pages        = {10893--10902}
}

@article{gao2016efficient,
	title        = {Efficient terahertz wide-angle {NUFFT}-based inverse synthetic aperture imaging considering spherical wavefront},
	author       = {Gao, Jingkun and Deng, Bin and Qin, Yuliang and Wang, Hongqiang and Li, Xiang},
	year         = 2016,
	month        = dec,
	journal      = {Sensors},
	volume       = 16,
	number       = 12,
	pages        = 2120
}

@article{gao2016terahertz,
	title        = {Terahertz Wide-Angle Imaging and Analysis on Plane-wave Criteria Based on Inverse Synthetic Aperture Techniques},
	author       = {Gao, Jing Kun and Qin, Yu Liang and Deng, Bin and Wang, Hong Qiang and Li, Jin and Li, Xiang},
	year         = 2016,
	month        = jan,
	journal      = {J. Infrared Millim. Terahertz Waves},
	volume       = 37,
	number       = 4,
	pages        = {373--393}
}

@article{gao2018cylindricalMIMO,
	title        = {An Efficient Algorithm for {MIMO} Cylindrical Millimeter-Wave Holographic {3-D} Imaging},
	author       = {J. {Gao} and B. {Deng} and Y. {Qin} and H. {Wang} and X. {Li}},
	year         = 2018,
	month        = aug,
	journal      = {IEEE Trans. Microw. Theory Techn.},
	volume       = 66,
	number       = 11,
	pages        = {5065--5074}
}

@inproceedings{demirci2011back,
	title        = {Back-projection algorithm for {ISAR} imaging of near-field concealed objects},
	author       = {Demirci, Sevket and Cetinkaya, Harun and Tekbas, Mustafa and Yigit, Enes and Ozdemir, Caner and Vertiy, Alexey},
	year         = 2011,
	month        = aug,
	booktitle    = {Proc. 30th URSI Gen. Assem. Sci. Symp. (URSI GASS)},
	address      = {Istanbul, Turkey},
	pages        = {1--4}
}

@inproceedings{jia2014modifiedBPA,
	title        = {Modified back projection reconstruction for circular {FMCW SAR}},
	author       = {Jia, Gaowei and Chang, Wenge},
	year         = 2014,
	month        = oct,
	booktitle    = {Proc. Inter. Radar Conf.},
	address      = {Lille, France},
	pages        = {1--5}
}

@book{amineh2019real,
	title        = {Real-Time Three-Dimensional Imaging of Dielectric Bodies Using Microwave/Millimeter Wave Holography},
	author       = {Amineh, Reza K and Nikolova, Natalia K and Ravan, Maryam},
	year         = 2019,
	publisher    = {John Wiley \& Sons}
}

@article{fortuny2001extension,
	title        = {Extension of the {3-D} range migration algorithm to cylindrical and spherical scanning geometries},
	author       = {Fortuny-Guasch, Joaquim and Lopez-Sanchez, Juan M},
	year         = 2001,
	month        = oct,
	journal      = {IEEE Trans. Antennas Propag.},
	volume       = 49,
	number       = 10,
	pages        = {1434--1444}
}

@inproceedings{detlefsen2005effective,
	title        = {Effective reconstruction approaches to millimeter-wave imaging of humans},
	author       = {Detlefsen, J and Dallinger, A and Huber, S and Schelkshorn, S and und Schaltungen, Fachgebiet Hochfrequente Felder},
	year         = 2005,
	month        = oct,
	booktitle    = {Proc. XXVIIIth URSI Gen. Assem. Sci. Symp. (URSI GASS)},
	address      = {New Delhi, India},
	pages        = {23--29}
}

@article{laviada2017multiview,
	title        = {Multiview three-dimensional reconstruction by millimetre-wave portable camera},
	author       = {Laviada, Jaime and Arboleya-Arboleya, Ana and {\'A}lvarez, Yuri	and Gonz{\'a}lez-Vald{\'e}s, Borja and Las-Heras, Fernando},
	year         = 2017,
	month        = jul,
	journal      = {Sci. Rep.},
	volume       = 7,
	number       = 1,
	pages        = {6479--6479}
}

@article{paul2021systematic,
	title        = {A Systematic Comparison of Near-Field Beamforming and {Fourier}-Based Backward-Wave Holographic Imaging},
	author       = {Paul, Sebastian and Schwartau, Fabian and Krueckemeier, Markus and Caspary, Reinhard and Monka-Ewe, Carsten and Schoebel, Joerg and Kowalsky, Wolfgang},
	year         = 2021,
	month        = aug,
	journal      = {IEEE Open J. Antennas Propag.},
	volume       = 2,
	pages        = {921--931}
}

@article{maisto2021sensor,
	title        = {Sensor Arrangement in Monostatic Subsurface Radar Imaging},
	author       = {Maisto, Maria Antonia and Pierri, Rocco and Solimene, Raffaele},
	year         = 2020,
	month        = nov,
	journal      = {IEEE Open J. Antennas Propag.},
	volume       = 2,
	pages        = {3--13}
}

@article{guo2019millimeter,
	title        = {Millimeter-wave imaging with accelerated super-resolution range migration algorithm},
	author       = {Guo, Qijia and Liang, Jie and Chang, Tianying and Cui, Hong-Liang},
	year         = 2019,
	month        = jul,
	journal      = {IEEE Trans. Microw. Theory Techn.},
	volume       = 67,
	number       = 11,
	pages        = {4610--4621}
}

@article{lopez20003,
	title        = {{3-D} radar imaging using range migration techniques},
	author       = {Lopez-Sanchez, Juan M and Fortuny-Guasch, Joaquim},
	year         = 2000,
	month        = may,
	journal      = {IEEE Trans. Antennas Propag.},
	volume       = 48,
	number       = 5,
	pages        = {728--737}
}

@article{zhuge2010sparse,
	title        = {A sparse aperture {MIMO-SAR}-based {UWB} imaging system for concealed weapon detection},
	author       = {Zhuge, Xiaodong and Yarovoy, Alexander G},
	year         = 2010,
	month        = jul,
	journal      = {IEEE Trans. Geosci. Remote Sens.},
	volume       = 49,
	number       = 1,
	pages        = {509--518}
}

@article{savelyev2010comparison,
	title        = {Comparison of 10-18 {GHz SAR and MIMO}-based short-range imaging radars},
	author       = {Savelyev, Timofey and Zhuge, Xiaodong and Yang, Bill and Aubry, Pascal and Yarovoy, Alexander and Ligthart, Leo and Levitas, Boris},
	year         = 2010,
	month        = aug,
	journal      = {Int. J. Microw. Wireless Techn.},
	volume       = 2,
	number       = {3--4},
	pages        = 369
}

@article{fan2020linearMIMOArbitraryTopologies,
	title        = {Near-Field {3D SAR} Imaging Using a Scanning Linear {MIMO} Array With Arbitrary Topologies},
	author       = {B. {Fan} and J. {Gao} and H. {Li} and Z. {Jiang} and Y. {He}},
	year         = 2019,
	month        = dec,
	journal      = {IEEE Access},
	volume       = 8,
	pages        = {6782--6791}
}

@inproceedings{mohammadian2019sar,
	title        = {{SAR} millimeter wave imaging systems},
	author       = {Mohammadian, Nafiseh and Furxhi, Orges and Short, Robert and Driggers, Ronald},
	year         = 2019,
	month        = may,
	booktitle    = {Proc. SPIE},
	address      = {Baltimore, MD, USA},
	volume       = 10994,
	pages        = {109940A}
}

@article{zhuge2012three,
	title        = {Three-dimensional near-field {MIMO} array imaging using range migration techniques},
	author       = {Zhuge, Xiaodong and Yarovoy, Alexander G},
	year         = 2012,
	month        = feb,
	journal      = {IEEE Trans. Image Process.},
	volume       = 21,
	number       = 6,
	pages        = {3026--3033}
}

@article{zhu2017range,
	title        = {Range migration algorithm for near-field {MIMO-SAR} imaging},
	author       = {Zhu, Rongqiang and Zhou, Jianxiong and Jiang, Ge and Fu, Qiang},
	year         = 2017,
	month        = nov,
	journal      = {IEEE Geosci. Remote Sens. Lett.},
	volume       = 14,
	number       = 12,
	pages        = {2280--2284}
}

@article{qiao2015compressive,
	title        = {Compressive sensing for direct millimeter-wave holographic imaging},
	author       = {Lingbo Qiao and Yingxin Wang and Zongjun Shen and Ziran Zhao and Zhiqiang Chen},
	year         = 2015,
	month        = apr,
	journal      = {Appl. Opt.},
	volume       = 54,
	number       = 11,
	pages        = {3280--3289}
}

@inproceedings{soumekh1998wide,
	title        = {Wide-bandwidth continuous-wave monostatic/bistatic synthetic aperture radar imaging},
	author       = {Soumekh, Mehrdad},
	year         = 1998,
	month        = oct,
	booktitle    = {Proc. IEEE Int. Conf. Image Process. (ICIP)},
	address      = {Chicago, IL, USA},
	pages        = {361--365}
}

@book{soumekh1999synthetic,
	title        = {Synthetic aperture radar signal processing},
	author       = {Soumekh, Mehrdad},
	year         = 1999,
	publisher    = {New York: Wiley},
	volume       = 7
}

@book{cook2012radar,
	title        = {Radar signals: an introduction to theory and application},
	author       = {Cook, Charles},
	year         = 2012,
	publisher    = {Elsevier}
}

@book{papoulis1968systems,
	title        = {Systems and transforms with applications in optics},
	author       = {Papoulis, Athanasios},
	year         = 1968,
	publisher    = {McGraw-Hill Series in System Science}
}

@article{mcclure2016multidimensional,
	title        = {Multidimensional stationary phase approximation: boundary stationary point},
	author       = {J.P. McClure and R. Wong},
	year         = 1990,
	month        = may,
	journal      = {J. Comput. Appl. Mathematics},
	volume       = 30,
	number       = 2,
	pages        = {213--225}
}

@article{li2008mft,
	title        = {Data-Level Fusion of Multilook Inverse Synthetic Aperture Radar Images},
	author       = {Li, Zhixi and Papson, Scott and Narayanan, Ram M.},
	year         = 2008,
	month        = apr,
	journal      = {IEEE Trans. Geosci. Remote Sens.},
	volume       = 46,
	number       = 5,
	pages        = {1394--1406}
}

@article{wang2018wavenumber,
	title        = {Wavenumber-domain multiband signal fusion with matrix-pencil approach for high-resolution imaging},
	author       = {Wang, Jianping and Aubry, Pascal and Yarovoy, Alexander},
	year         = 2018,
	month        = apr,
	journal      = {IEEE Trans. Geosci. Remote Sens.},
	volume       = 56,
	number       = 7,
	pages        = {4037--4049}
}

\end{document}